%% file: ms.tex
\documentclass[useAMS,twocolumn,usenatbib]{mn2e}
\usepackage{graphicx}
\usepackage{color}
\usepackage{times}
\usepackage{amsmath}
\usepackage{url}
\usepackage{soul}

\include{JournalAbbr}

\newcommand{\Msun}{\rm M_{\odot}}
\newcommand{\Msunh}{h^{-1}{\rm M}_{\odot}}
\newcommand{\kms}{{\rm km\,s^{-1}}}
\newcommand{\kpch}{h^{-1}{\rm kpc}}

\newcommand{\gsim}{\lower.7ex\hbox{$\;\stackrel{\textstyle>}{\sim}\;$}}
\newcommand{\lsim}{\lower.7ex\hbox{$\;\stackrel{\textstyle<}{\sim}\;$}}

\title[Baryonic impact on dark matter]{Baryonic impact on the dark matter distribution in Milky Way-size galaxies and their satellites}

\author[Q. Zhu et al.]
{\parbox{\textwidth}{Qirong Zhu,$^{1,2} \thanks{E-mail: qxz125@psu.edu}$
Federico Marinacci,$^{3}$
Moupiya Maji,$^{1,2}$
Yuexing Li,$^{1,2}$
Volker Springel$^{4,5}$ and
Lars Hernquist$^{6}$}\vspace{0.4cm}\\
\parbox{\textwidth}{
$^{1}$Department of Astronomy \& Astrophysics, The Pennsylvania State University, 525 Davey Lab, University Park, PA 16802, USA \\
$^{2}$Institute for Gravitation and the Cosmos, The Pennsylvania State University, University Park, PA 16802, USA\\
$^{3}$Department of Physics, Kavli Institute for Astrophysics and Space Research, Massachusetts Institute of Technology,
Cambridge, MA 02139, USA\\
$^{4}$Heidelberg Institute for Theoretical Studies, Schloss-Wolfsbrunnenweg 35, 69118 Heidelberg, Germany\\
$^{5}$Zentrum f\"{u}r Astronomie der Universit\"{a}t Heidelberg, ARI, M\"{o}nchhofstr. 12-14, 69120 Heidelberg, Germany\\
$^{6}$Harvard-Smithsonian Center for Astrophysics, Harvard University, 60 Garden Street, Cambridge, MA 02138, USA}}

\date{Accepted XXX. Received YYY; in original form ZZZ}
\pubyear{2015}
\begin{document}                          
\pagerange{\pageref{firstpage}--\pageref{lastpage}}

\maketitle

\label{firstpage}

\begin{abstract} 
We study the impact of baryons on the distribution of dark matter in a Milky 
Way-size halo by comparing a high-resolution, moving-mesh cosmological 
simulation with its dark matter-only counterpart. We identify three main 
processes related to baryons -- adiabatic contraction, tidal disruption and 
reionization -- which jointly shape the dark matter distribution in both the 
main halo and its subhalos.  The relative effect of each baryonic process 
depends strongly on the subhalo mass. For massive subhalos with maximum circular 
velocity $v_{\rm max} > 35~\kms$, adiabatic contraction increases the dark 
matter concentration, making these halos less susceptible to tidal disruption. For 
low-mass subhalos with $v_{\rm max} < 20~\kms$, reionization effectively reduces 
their mass on average by $\approx$ 30\% and $v_{\rm max}$ by $\approx$ 20\%. 
For intermediate 
subhalos with $20~\kms < v_{\rm max} < 35~\kms$, which share a similar mass 
range as the classical dwarf spheroidals, strong tidal truncation induced by the 
main galaxy reduces their $v_{\rm max}$. As a combined 
result of reionization and increased tidal disruption, the total number of 
low-mass subhalos in the hydrodynamic simulation is nearly halved compared to 
that of the $\textit{N-}$body simulation. We do not find dark matter cores in 
dwarf galaxies, unlike previous studies that employed bursty feedback-driven 
outflows. The substantial impact of baryons on the abundance and internal 
structure of subhalos suggests that galaxy formation and evolution models based 
on $\textit{N}$-body simulations should include these physical processes as 
major components.  
\end{abstract}


\section{Introduction}
\label{sec:introduction}

A major achievement in observational cosmology is the discovery that our 
Universe is composed of $\sim4\%$ baryons, $20\%$ dark matter (DM), and $76\%$ 
dark energy (DE) \citep{Frieman2008}. The first observational evidence for DM 
dates back to 1933 when Zwicky noted a missing mass problem in the Coma cluster 
of galaxies: the visible galaxies account for only a small fraction of the 
total mass inferred from the dynamics \citep{Zwicky1937}. More evidence came 
later from galactic rotation curves in spiral galaxies \citep{Rubin1980}, 
gravitational lensing, and the Bullet cluster which shows an offset of the 
center of the total mass from that of the baryons \citep{Clowe2006}. The first 
compelling observational evidence for DE was found later in 1998 when two teams 
studying Type Ia supernovae independently found that the expansion of the 
universe is accelerating \citep{Riess1998, Perlmutter1999}. This finding has 
been confirmed by subsequent supernovae observations, and independent evidence 
from galaxy clusters (e.g., \citealt{Vikhlinin2006, Allen2011}), large-scale 
structure (e.g., \citealt{Tegmark2006, Addison2013}), and the cosmic microwave 
background (e.g., \citealt{Spergel2007, Komatsu2011, Hinshaw2013, Planck2015}).

These observations motivate the current ``standard model'' of cosmology 
($\Lambda$CDM), where dark energy and cold dark matter shape the formation and 
evolution of cosmic structures (e.g., \citealt{Frenk2012, Kravtsov2012, 
Conselice2014, Somerville2015}). To date, numerous $\Lambda$CDM cosmological 
simulations have produced clumpy and filamentary large-scale structures as seen 
in galaxy surveys (e.g., \citealt{Navarro1997, Springel2005, Springel2006, 
Gao2012, Vogelsberger2014a}), and have confirmed that structures form through 
hierarchical assembly in CDM dominated universes. However, on small scales 
(i.e.~less than $\sim 10\,{\rm kpc}$), there appear to be a number of tensions 
between predictions from the $\Lambda$CDM model and observations, notably: (1) 
the ``missing satellites problem'', in which the abundance of subhalos produced 
by $\textit{N}$-body simulations is orders of magnitude larger than the two 
dozens of satellites observed in the MW (e.g., \citealt{Klypin1999, Moore1999, 
Kravtsov2004, Kravtsov2009}); (2) the ``too big to fail problem", in which 
$\textit{N}$-body simulations produce overly dense massive subhalos compared to 
the brightest dwarf galaxies in the MW and Local Group \citep{Kolchin2011, 
Kolchin2012, Tollerud2014, GarrisonKimmel2014}; and (3) the ``core-vs-cusp 
problem'', in which the central dark matter density profiles of DM-dominated 
dwarf spheroids are observed to apparently feature smooth cores instead of the 
cusps that are generically predicted by CDM models (e.g., \citealt{Gilmore2007, 
Evans2009, deBlok2010, Amorisco2012, Strigari2010, Martinez2013}). These 
problems have motivated alternative models such as self-interacting DM 
\citep[][]{Dave2001, Vogelsberger2012, Elbert2014, Vogelsberger2014b}, or warm 
DM (e.g., \citealt{Polisensky2014, Schneider2014, Kennedy2014}).

However, some of the discrepancies were reported in DM-only simulations in which the dynamical coupling between baryons and dark matter was ignored. While this assumption may be justified on large scales, this is no longer the case on kpc scales, where the density starts to be dominated by baryons, and the dynamics becomes governed by baryonic processes such as gas dynamics, star formation, black hole accretion, and feedback from stars and active galactic nuclei (AGN).
	
An impact of baryons on the dark matter arises from different spatial 
distributions of the two components. A well-known effect is adiabatic 
contraction \citep{Young1980, Barnes1984, Blumenthal1986, Ryden1987, Gnedin2004, 
Gnedin2011, Zemp2012, Pillepich2014}, which causes an increase of the mass 
concentration of dark matter in the center of a galaxy due to gas inflow as a 
result of cooling. The increased DM mass concentration and the presence of 
a stellar disk can produce stronger tidal forces, which have been suggested as a 
way to significantly affect the abundance and distribution of subhalos and 
satellite galaxies in the Milky Way (MW) \citep[e.g.][]{DOnghia2010, Penarrubia2010, 
Zolotov2012, Arraki2014}. 

Hydrodynamic simulations have become powerful tools to investigate the response of dark matter to baryons and vice-versa, thanks to recent progress in numerical methods \citep[e.g.][]{Springel2010, Read2012, Hopkins2013, Hu2014} and physical modeling \citep[][]{Vogelsberger2013, Stinson2013, Aumer2013, Hopkins2014b} that improved upon long-standing issues in the field \citep{Agertz2007, Sijacki2012, Torrey2012}. Recent cosmological simulations such as the Illustris \citep{Vogelsberger2014a} and {\sc eagle} \citep{Schaye2015} projects were able to reproduce different galaxy populations that resemble the observed ones both locally and in the high-redshift Universe \citep[e.g.][] {Vogelsberger2014c, Genel2014}. In particular, \cite{Vogelsberger2014a} showed that subhalos in DM-only simulation are more prone to tidal disruption than those in hydrodynamic simulations, leading to a depletion of satellites near galaxy cluster centers and a drop in the matter power spectrum on small scales.

Several solutions have been proposed to solve the ``too big to fail'' problem, including tidal effects \citep{Zolotov2012, Arraki2014}, which however might be insufficient in some cases \citep[for instance in M31, see][]{Tollerud2014},  a mass-dependent abundance of subhalos which may alleviate the problem if a 
lower total mass of the MW is assumed \citep{Wang2012, VeraCiro2013, Sawala2014a}, and strong outflows driven by supernovae explosions which can have a direct impact on the central dark matter content in dwarf galaxies, possibly leading to a cored profile \citep[e.g.][]{Navarro1996, Governato2012, Teyssier2013, Madau2014, Brooks2014, Ogiya2015}. However, \cite{Kolchin2012} and \cite{GarrisonKimmel2013} argue that the required energy from supernovae may not be sufficient given the low stellar mass 
in some of the dwarf galaxies  \citep[but see also the energy argument by][]{Madau2014}. Moreover, most hydrodynamic simulations \cite[e.g.][] {Mashchenko2008, Governato2012, Teyssier2013, Madau2014} are focused on dwarf galaxies in field environments, which may not be representative for the dwarfs in the MW or M31.  

In order to investigate baryonic effects on dark matter in dwarf galaxies of the MW, we need high-resolution, cosmological hydrodynamic simulations which produce a spiral galaxy with properties similar to those of the MW. Producing MW-like disk galaxies in cosmological simulations has been a decade-long challenge, but recently several groups have succeeded in this endeavor \citep[][]{Agertz2011, Guedes2011, Aumer2013, Okamoto2013, Hopkins2014b}. Equipped with the same implementation of baryon physics as in the Illustris simulations, \cite{Marinacci2014a} successfully produced MW-size disk galaxies in a suite of zoom-in simulations. These simulations used the same initial conditions as the Aquarius Project \citep{Springel2008}, and the highest-resolution hydrodynamical rerun (Aq-C-4) has sufficient resolution to identify and study the formation history and properties of the predicted dwarf galaxies. 

In this work, we use both DM-only and hydrodynamical simulations of Aq-C-4 by \cite{Marinacci2014a} to study the impact of baryon processes on the halo/subhalo properties and the subhalo abundance. We will not limit our study on the bright satellites alone, i.e.~those subhalos containing stars, but we will also analyze the ``dark" ones. As it turns out, even the ``dark'' subhalos are systematically affected by baryonic processes in terms of their spatial distribution and mass functions. 

This paper is organized as follows. In Section~\ref{sec:technique}, we describe the numerical technique used in our simulations and the structure identification. The impact of baryons on the smooth dark matter distribution in the main halo and the global statistics of subhalos are presented in Section~\ref{sec:results}. In Section~\ref{sec:matchedsubhaloresult}, we investigate the impact of baryons on the total mass, the DM density profiles and $v_{\rm max}$ values of objects extracted from a matched subhalo catalogue of the DM and Hydro simulations. We aim to determine the main physical processes that shape the DM content in subhalos by tracking the assembly history and evolution of bright satellites and ``dark" subhalos. We discuss the implications of our study and its limitations in Section~\ref{sec:discussion}, and summarize our main findings in Section~\ref{sec:conclusion}.

\section{Methods}
\label{sec:technique}

\subsection{The Simulations} 

In this study, we use two cosmological simulations of a MW-size halo, one being the full hydrodynamical Aq-C-4 run by \cite{Marinacci2014a} (referred to as ``Hydro" hereafter), and the other being a \textit{control} DM-only simulation of the same halo (referred to as ``DMO" hereafter). This Aq-C halo was selected as a close match to the MW for the Aquila Comparison Project \citep{Scannapieco2012}, as well as several other studies \citep{Wadepuhl2011, Okamoto2013, Sawala2012}. The hydrodynamical Aq-C-4 simulation by \cite{Marinacci2014a} was performed with the moving mesh code {\small\sc Arepo} \citep{Springel2010}. The simulation adopted a physical model for galaxy formation and evolution developed by \cite{Vogelsberger2013}, which includes supernovae feedback, metal enrichment and stellar mass return, AGN feedback \citep{Dimatteo2005, Springel2005a, Sijacki2007}, 
and a spatially uniform, redshift-dependent ionizing background by \cite{Faucher2009}, which leads to complete re-ionization of neutral hydrogen by $z = 6$. Thermal feedback from supernovae was implemented following a hybrid ISM model developed by \cite{Springel2003}, and galactic outflows were launched with a velocity scaled with the local dark matter velocity dispersion of the host halo, following a kinetic model similar to \cite{Okamoto2010} and \cite{Puchwein2013}.

The Aq-C-4 Hydro simulation has a mass resolution of $5.0 \times 10^{4}\, \Msun$ for gas and stars, and $3.0 \times 10^{5}\, \Msun$ for the DM component \citep[see Table 1 in][]{Marinacci2014a}, sufficient to simultaneously follow the main galaxy and its classical dwarf galaxies with a maximum circular velocity $v_{\rm max}$ between 12 and 24 $\kms$ \citep[][]{Kolchin2011}. The gravitational softening length in the high-resolution region was kept fixed in comoving coordinates, corresponding to a physical length of 340 pc at $z = 0$. We re-run a DM-only simulation of Aq-C-4 with the same numerical parameters controlling the force and time integration accuracy as used in \cite{Marinacci2014a}. Thus, the effects of baryonic processes on the dark matter distribution can be well studied by comparing the DM simulation and its Hydro counterpart. At $z = 0$, the properties of the central galaxy of Aq-C-4 in the Hydro run are in very good agreement with those of a typical disk-dominated galaxy in terms of the mass budget in various components, the morphology, and the star formation history \citep{Marinacci2014a}. The properties of the diffuse gas and the metal distribution are also consistent with observations \citep{Marinacci2014b}. Moreover, we note that the robustness of the results was verified by a resolution study in \cite{Marinacci2014a}.

\subsection{Structure Identification}

To identify subhalos both in the DMO and Hydro simulations, the snapshots were post-processed with the Amiga Halo Finder \citep[{\small\sc AHF},][]{Knollmann2009}.\footnote{The code is available at \url{http://popia.ft.uam.es/AHF/Download.html}.  In this study, we use the version ahf-v1.0-084. It also contains an analysis tool called {\sc MergerTree}, which we have used to construct the merger tree and to cross-match the subhalos between the DMO and Hydro simulations.} The {\small\sc AHF} algorithm identifies structures based on density estimates calculated with an adaptive refinement technique, and naturally builds a halo-subhalo-subsubhalo hierarchy. The extent of a halo is determined by its density, $\bar \rho(r_{\rm vir}) = \Delta_{\rm vir}(z) \rho_{\rm bg}$, where $\bar \rho(r_{\rm vir})$ is the mean density within the virial radius $r_{\rm vir}$, $\rho_{\rm bg}$ is the background density, and $\Delta_{\rm vir}(z)= 178$ is the adopted virial overdensity.

{\small\sc AHF} performs an iterative process to remove unbound particles until the final result converges to a set of bound particles within $r_{\rm vir}$. These sets of particles form the halo and subhalos. The code then calculates various properties of the halos and subhalos, such as the mass in different components, the maximum value of the rotation curve $v_{\rm max}$, and the spin parameter. The results of {\small\sc AHF} and other substructure finders such as {\small\sc SUBFIND} \citep{Springel2001} are generally in very good agreement~\citep[e.g.][]{Onions2012, Pujol2014}, and any residual differences are not expected to influence our results.

The IDs of collisionless particles are preserved in our simulations since there is no mass exchange between them. This allows us to construct subhalo merger trees using the built-in module {\small\sc MergerTree} of {\small\sc AHF}, which relies on tracking the membership of dark matter and star particles (identified by their IDs) within the different halos and subhalos.  The merger trees are constructed for both the DM and Hydro simulations. For each halo/subhalo, we only consider the most massive progenitor in the previous snapshot as its parent. In addition, we only consider the halos/subhalos comprised of at least 10 particles. We have carefully checked the validity of the constructed merger trees by visually comparing the evolutionary paths of each individual object in terms of its position, velocity and mass. There are rare cases when a subhalo is not detected by {\small\sc AHF} in one snapshot output when the subhalo closely passes the center of its host halo. These objects usually reappear in the next {\small\sc AHF} catalogue if they have not been disrupted at pericenter. To avoid complications, we discard such subhalos in this study.

With the {\small\sc MergerTree} analysis package, we can also cross-match the $z = 0$ snapshots of the DM and Hydro simulations using the dark matter particles. We verify that this cross-match between the Hydro and DMO simulations is able to identify the ``same'' objects by comparing their evolutionary paths. In Section~\ref{sec:physicalprocesses}, we show the orbital and mass growth histories of several of these matched objects. We note however that substantial orbital phase offsets are expected to appear in most pairs due to the inclusion of baryonic processes; thus the positions of the subhalos in the two simulations are not expected to exactly match each other. Also, some halos/subhalos identified in the DMO simulation do not have counterparts in the Hydro simulation, given that substructures are destroyed at a higher rate in the latter run.

\section{Baryonic impact on the properties of the main galaxy and its satellites}
\label{sec:results}

One of the goals of this study is to identify the main physical processes shaping the distribution of dark matter in the main galaxy and its substructures. In this section, we focus on important galaxy properties such as the spatial distribution, abundance, and mass function of satellites, as well as the DM density profiles in the DMO and Hydro simulations.

\subsection{Spatial Distribution and Abundance of Subhalos}
\label{subsec:abundance}

\begin{figure}
\begin{center}
\begin{tabular}{c}
\resizebox{\columnwidth}{!}{\includegraphics{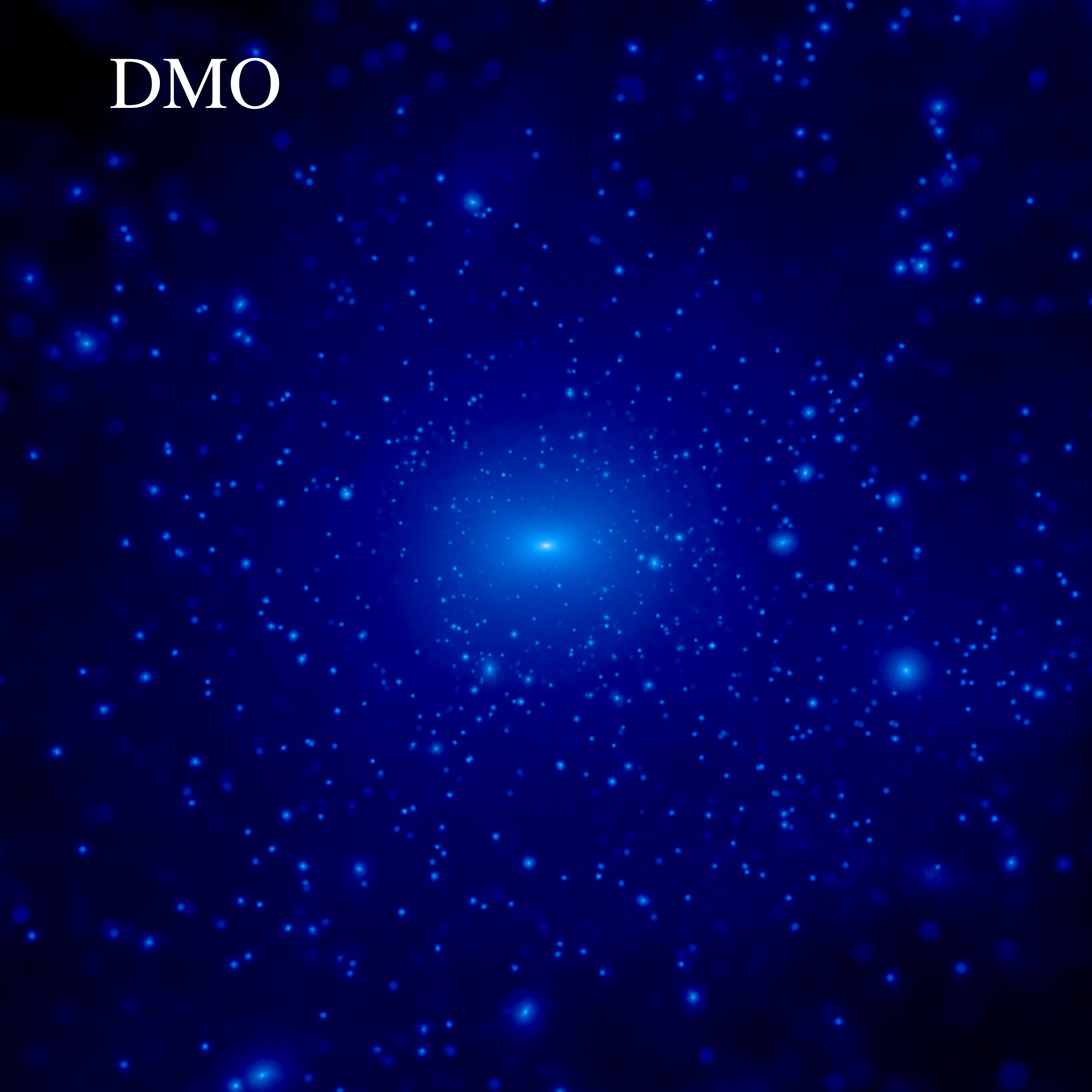}}
\end{tabular}
\begin{tabular}{c}
\resizebox{\columnwidth}{!}{\includegraphics{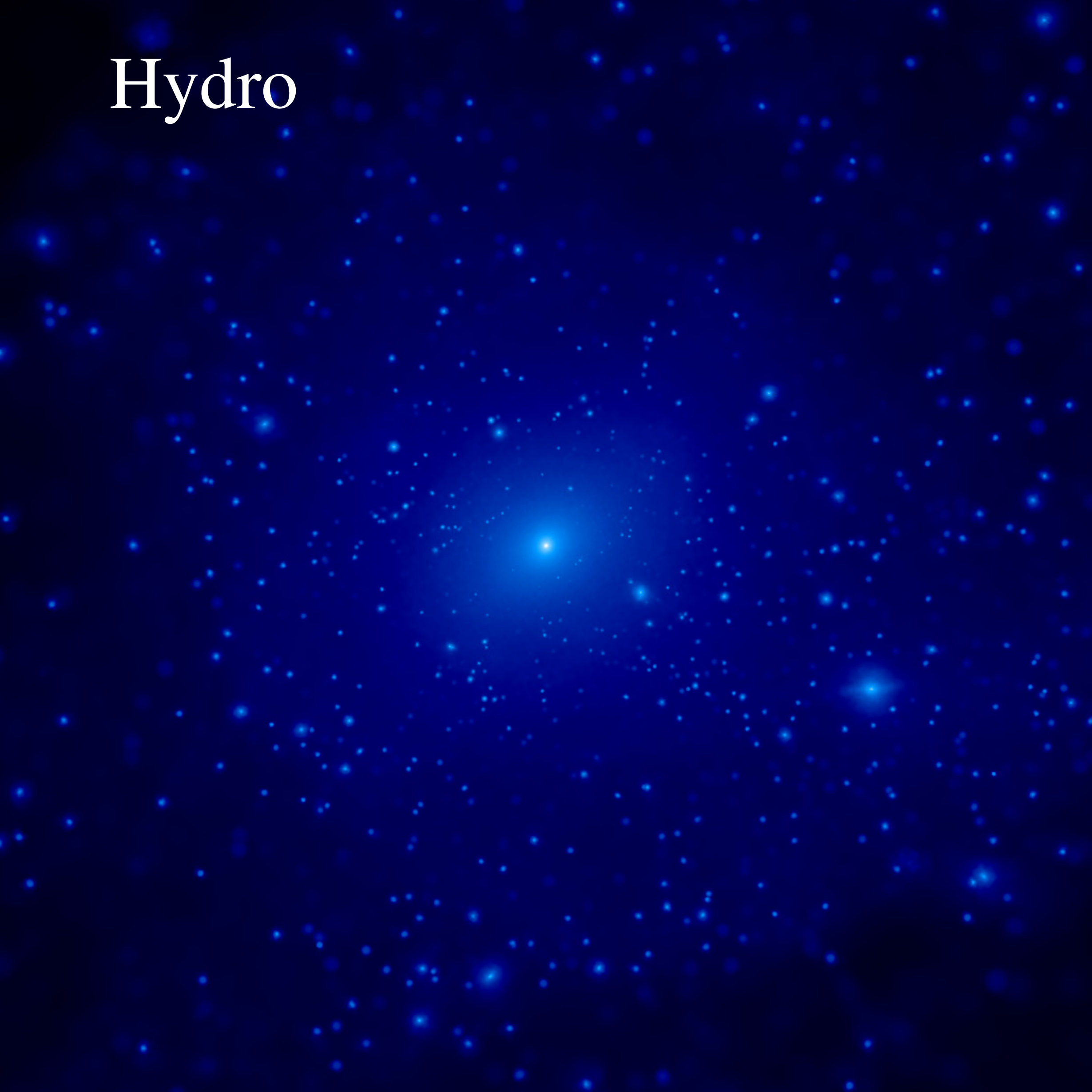}}
\end{tabular}
\caption{\label{fig:dm_distribution} Projected dark matter density maps within a $250\,h^{-1}{\rm kpc}$ slice centered on the main halo at redshift $z = 0$ in the DMO (top panel) and Hydro (bottom panel) simulations, respectively. The size of the displayed region is $0.7\,h^{-1}{\rm Mpc}$ on a side.}
\end{center}
\end{figure}

\begin{figure} \begin{center} \begin{tabular}{c} \resizebox{3.2in}{!}{\includegraphics{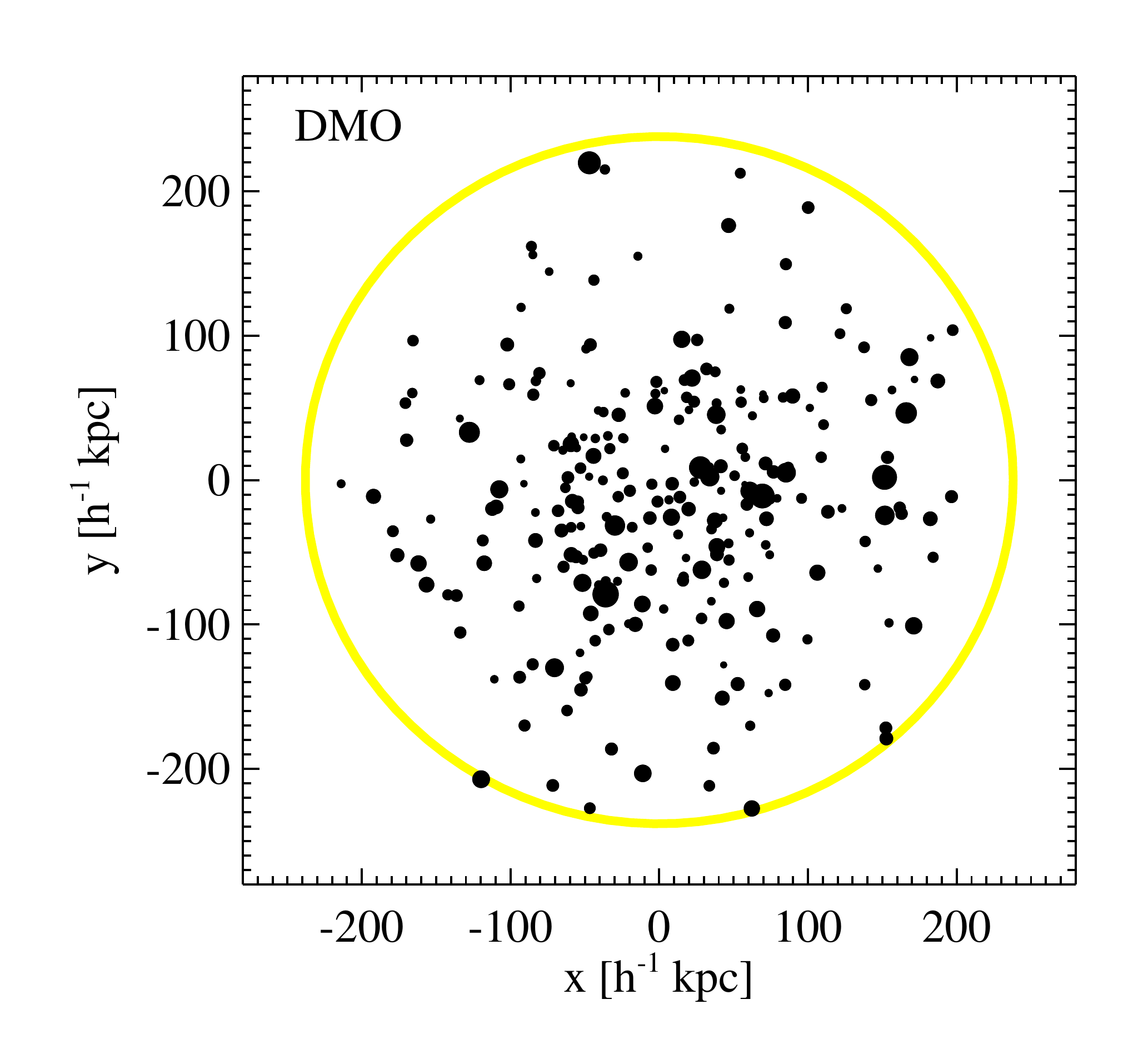}} \end{tabular} \begin{tabular}{c} \resizebox{3.2in}{!}{\includegraphics{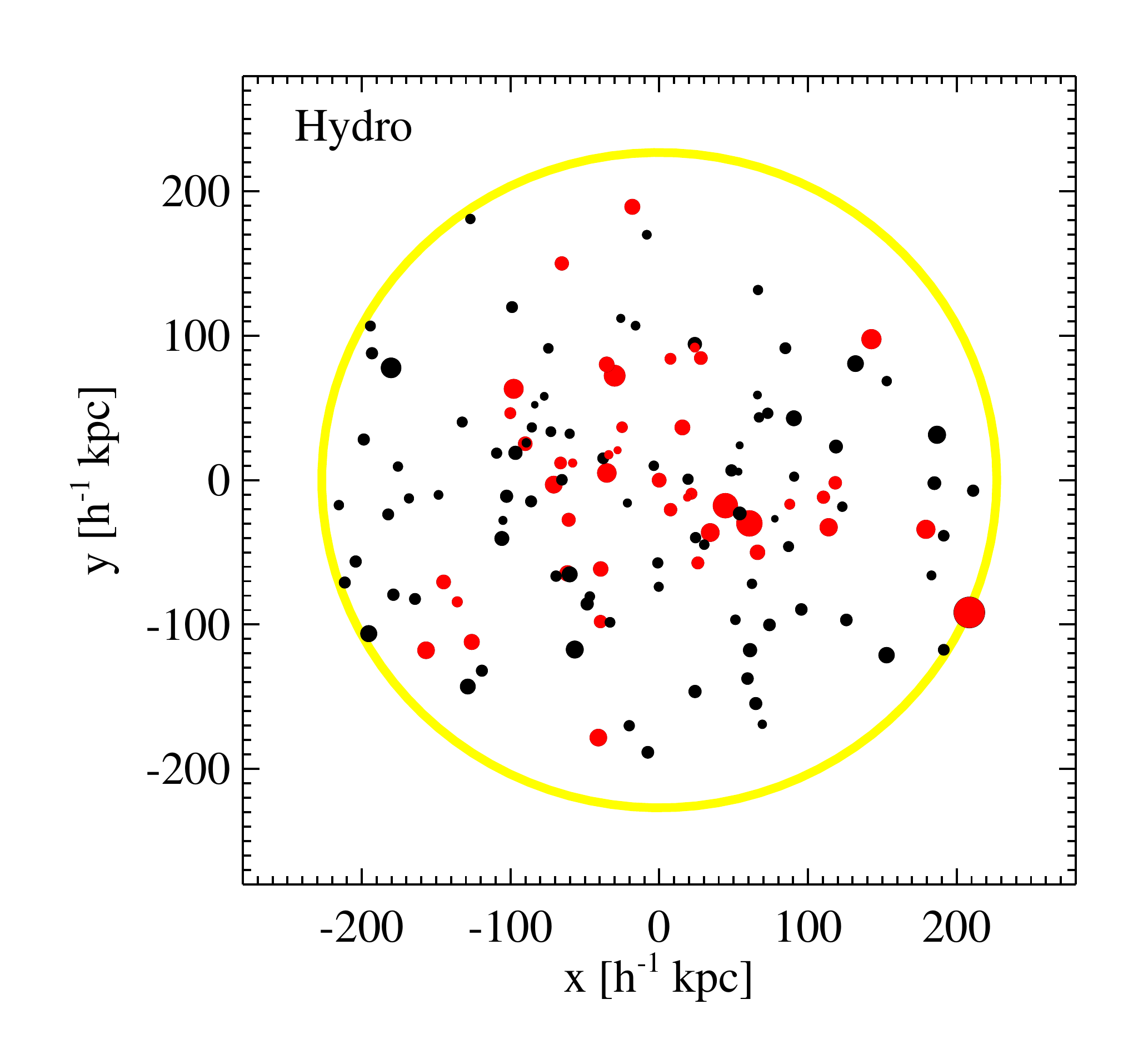}} \end{tabular} \caption{\label{fig:sub_distribution} The spatial distribution of subhalos at redshift $z = 0$ in the DMO (top panel) and Hydro (bottom panel) simulations, respectively. The black filled circles represent dark matter subhalos, while the red filled circles represent bright satellites which have formed stars. The size of the symbols is scaled with the subhalo mass. The solid yellow circle indicates the virial radius calculated by {\small\sc AHF} (with overdensity $\Delta_{\rm vir} = 178$). The Hydro simulation produces fewer subhalos than the DMO counterpart, with a pronounced depletion of low-mass subhalos near the central region. The bright satellites are only a small fraction of the entire subhalo population.}  \end{center} \end{figure}

In Figure~\ref{fig:dm_distribution}, we show projected dark matter density maps 
at $z = 0$ for a slice of thickness $250\,h^{-1}{\rm kpc}$ centered on the 
MW-size halo for the DMO (top panel) and the Hydro (bottom panel) simulations, 
respectively. Numerous substructures are clearly visible in both simulations. 
Despite the overall similarity in the morphology and size of the main halo 
between the two simulations, there are notable differences in the abundance and 
spatial distribution of the subhalos, especially in the central region, as 
demonstrated in Figure~\ref{fig:sub_distribution}, which shows the positions of 
all the DM subhalos and bright satellites (subhalos that contain stars) within 
the virial radius of the main galaxy at the present epoch. It is clear that 
there are fewer subhalos in the Hydro simulation than in the DMO one, in 
particular, the large number of low-mass subhalos found in the DMO simulation is 
clearly reduced in the Hydro case. Moreover, it is seen that only a fraction of 
the subhalos presented in the DMO simulation can be found in the Hydro simulation. 
Note that not necessarily the most massive ones are able to host bright satellites that form stars.

\begin{figure}
\begin{center}
\includegraphics[scale=0.33]{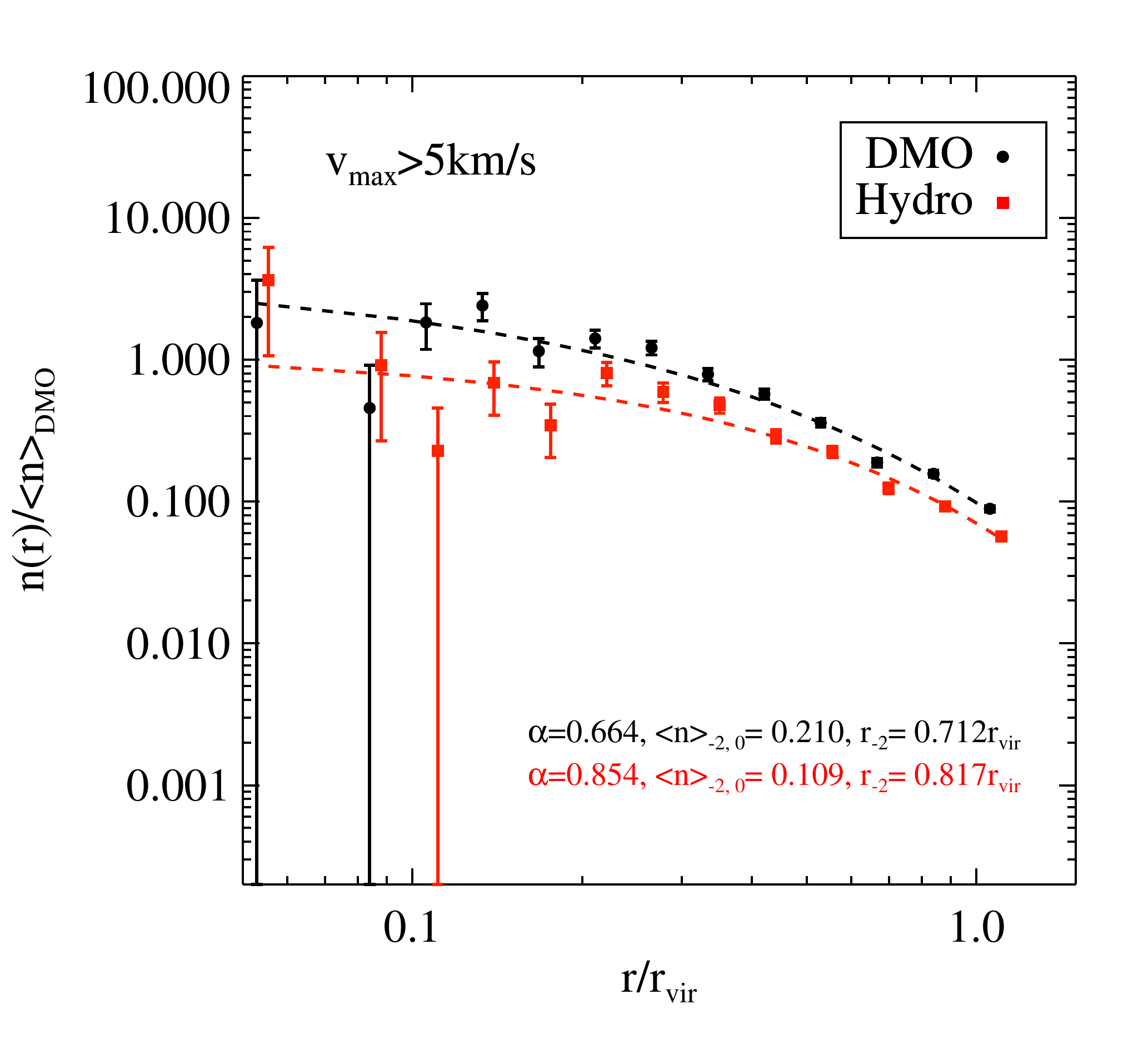}
\includegraphics[scale=0.33]{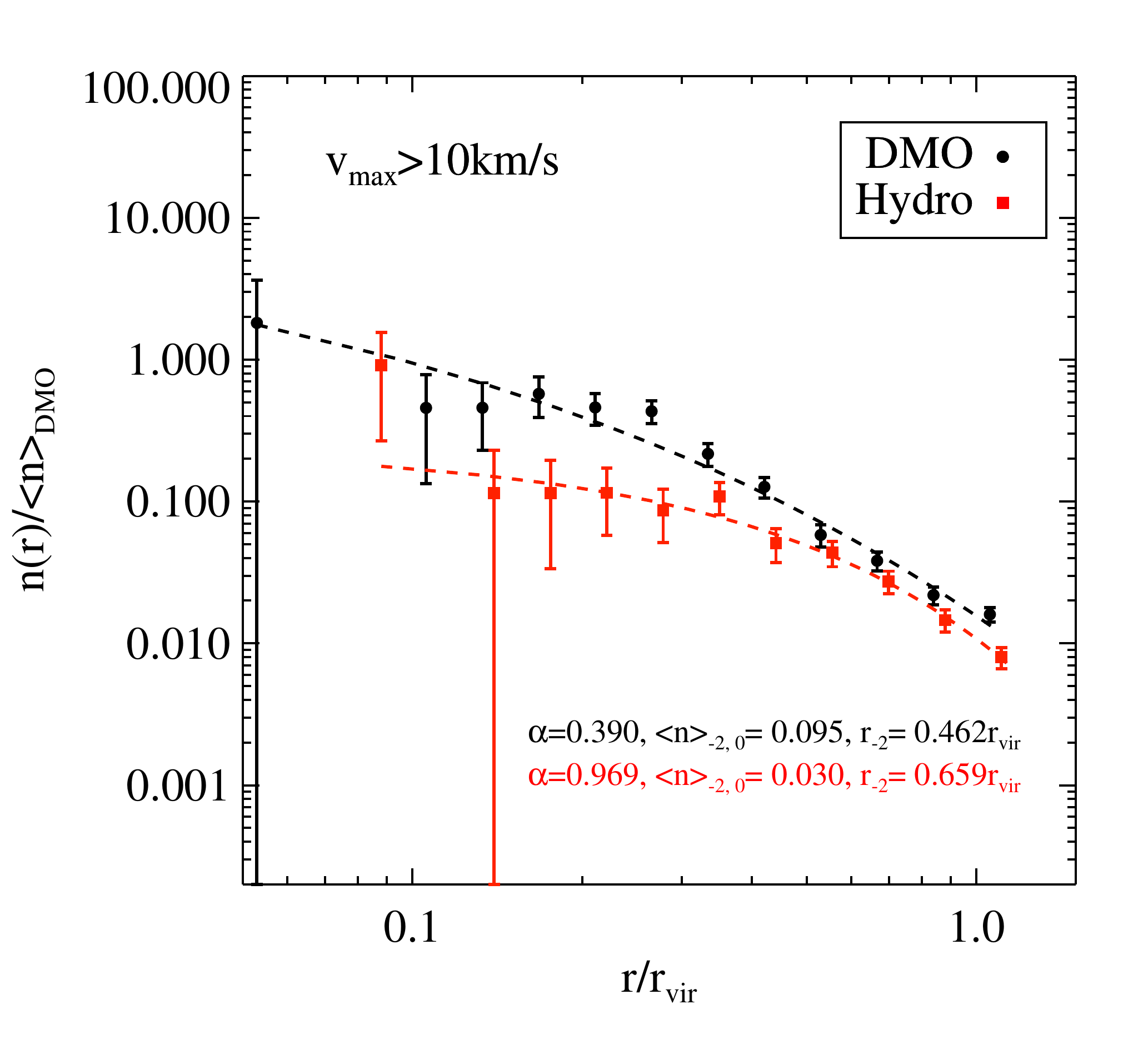}
\includegraphics[scale=0.33]{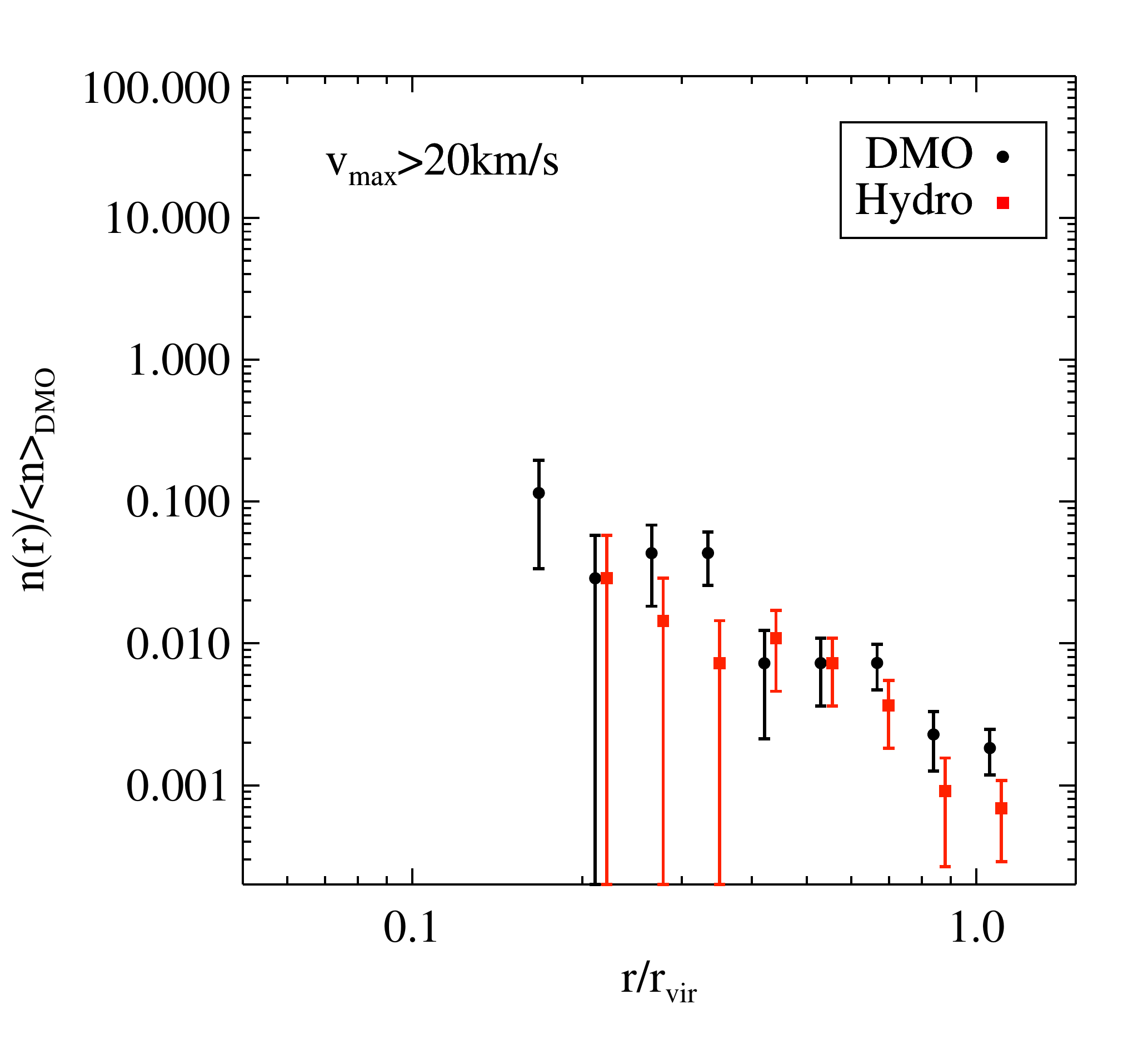}
\caption{\label{fig:sub_number_density} The number density of subhalos in 
different mass ranges as a function of distance to the center of the 
main halo, both for the DMO (black symbols) and Hydro (red symbols) simulations. 
The distance $r$ is normalized to the virial radius of the main halo, while the 
subhalo abundance is normalized to $\langle n \rangle_{\rm DMO}$, the total 
number of DM subhalos identified in the DMO simulation divided by the entire 
volume enclosed by $r_{\rm vir}$. The top, middle and bottom panels show 
subhalos in three different mass ranges, as indicated by the maximum circular
velocity $v_{\rm max} > 5\, \kms$, $v_{\rm max} > 10\, \kms$, and $v_{\rm max} > 
20\,  \kms$, respectively. The error bars are computed using the Poisson error 
$\sqrt{N_{r}}$, where $N_{r}$ is the number of subhalos within each radial bin. 
The dashed lines are fits to the Einasto profile, as given by 
Eqn.~(\ref{eq:einasto}). }
\end{center}
\end{figure}

To quantify the spatial distribution of subhalos, we compare the radial number density of subhalos in different mass ranges from both the DMO and Hydro simulations in Figure~\ref{fig:sub_number_density}. It has been shown that in \textit{N-}body simulations the spatial distribution of subhalos follows a universal function which is less concentrated than the density profiles of dark matter halos. It can be parameterized by the Einasto profile \citep{Gao2012}: 
\begin{equation}
n(r)/\langle n\rangle_{\rm{DMO}} = \langle n \rangle_{-2,0} \exp \left\{-\frac{2}{\alpha}\left[\left(\frac{r}{r_{-2}}\right)^{\alpha} - 1\right]\right\}.
\label{eq:einasto}
\end{equation}

\begin{figure}
\begin{center}
\begin{tabular}{c}
\resizebox{\columnwidth}{!}{\includegraphics{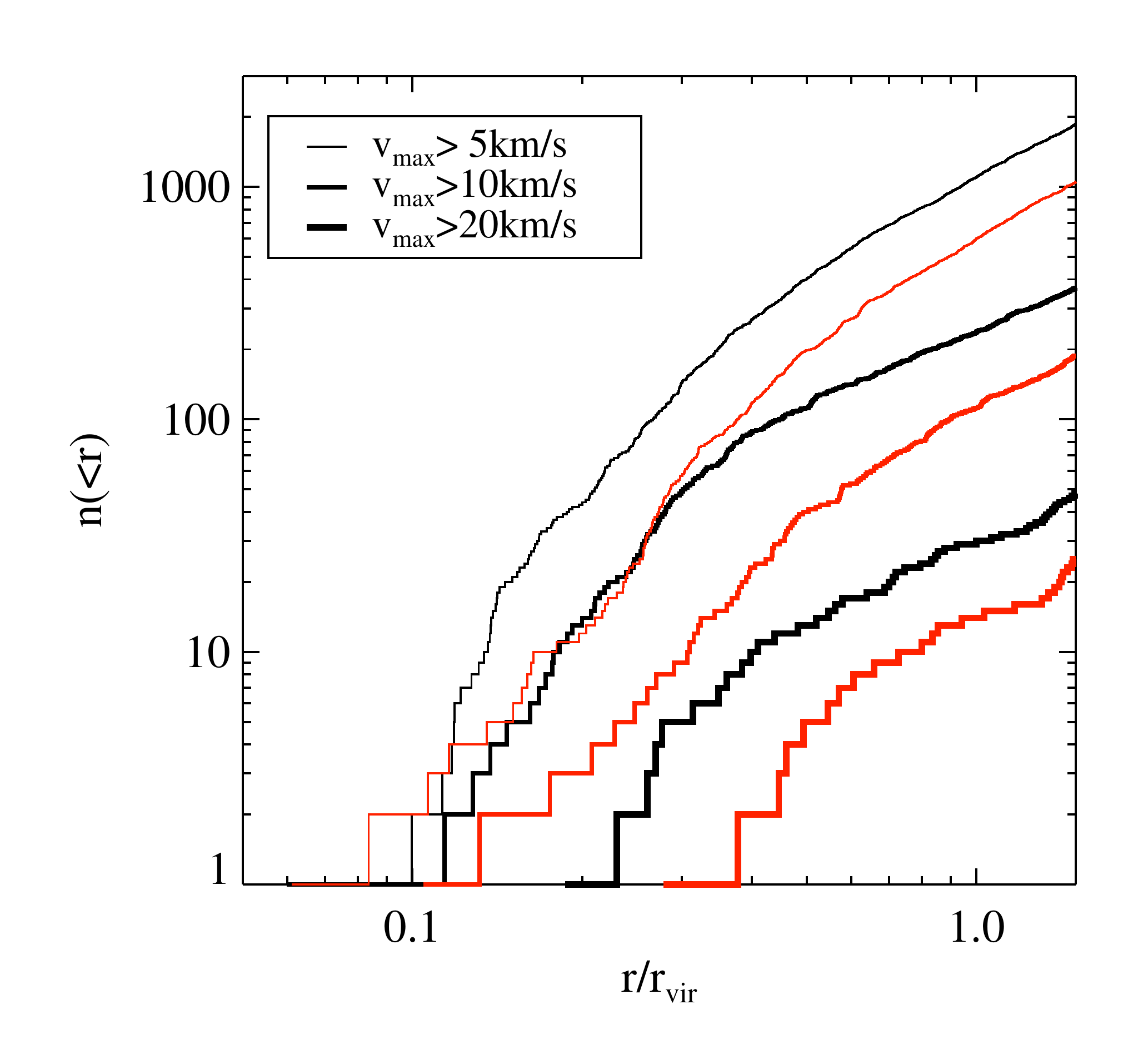}}
\end{tabular}
\caption{\label{fig:sub_cumulative} The cumulative number of subhalos in different mass ranges as a function of distance to the center of the 
main halo, both for the DMO (black lines) and Hydro simulations (red lines). The three mass ranges are the same as in Figure~\ref{fig:sub_number_density}: $v_{\rm max} > 5\, \kms$, $v_{\rm max} > 10\, \kms$, and $v_{\rm max} > 20\,  \kms$, respectively.}
\end{center}
\end{figure}

We fit our data with the Einasto profile for subhalos in two mass ranges, 
delineated by maximum circular velocities $v_{\rm max} > 5\, \kms$ and
$v_{\rm max} > 10\, \kms$ as shown  in Figure~\ref{fig:sub_number_density}. 
The radial abundance of subhalos in the Hydro simulation 
is consistently lower than that in the DMO one, and the effect increases towards 
the center for the most massive subhalos. There are not enough data points within 
$0.2 r_{\rm vir}$ for  $v_{\rm max} > 20\, \kms$ to perform reliable fitting with an Einasto 
profile.  The cumulative radial distribution of subhalos in Figure~\ref{fig:sub_cumulative} 
confirms this trend. The total number of subhalos at 
any given radius, with the exception of the innermost regions 
($r/r_{\rm vir} \lsim 0.1$) in the $v_{\rm max} > 5\, \kms$ cut,
is consistently lower in the Hydro simulation than in the DMO 
run, and this is particularly evident for the most massive subhalos 
in the central regions. 

Both Figure~\ref{fig:sub_number_density} and Figure~\ref{fig:sub_cumulative} suggest 
that subhalos are subject to being disrupted more easily in the Hydro simulation. 
A similar radial distribution can also be found in \cite{DOnghia2010} and 
\cite{Yurin2015}, in which it was suggested that the reduction of subhalos was 
due to enhanced tidal effects and accelerated disruption rates from a combination 
of DM contraction and the presence of the stellar disk. In addition, enhanced dynamical friction from the 
adiabatically contracted dark matter distribution of the main halo would cause the subhalos to sink more rapidly. 
The combination of these factors results in fewer massive subhalos in the central region in the Hydro simulation 
than in the DMO one. We will address the impact of the central galaxy on the abundance of its satellites in Section 4.

\begin{figure} \begin{center} 
\resizebox{3.15in}{!}{\includegraphics[scale=0.375]{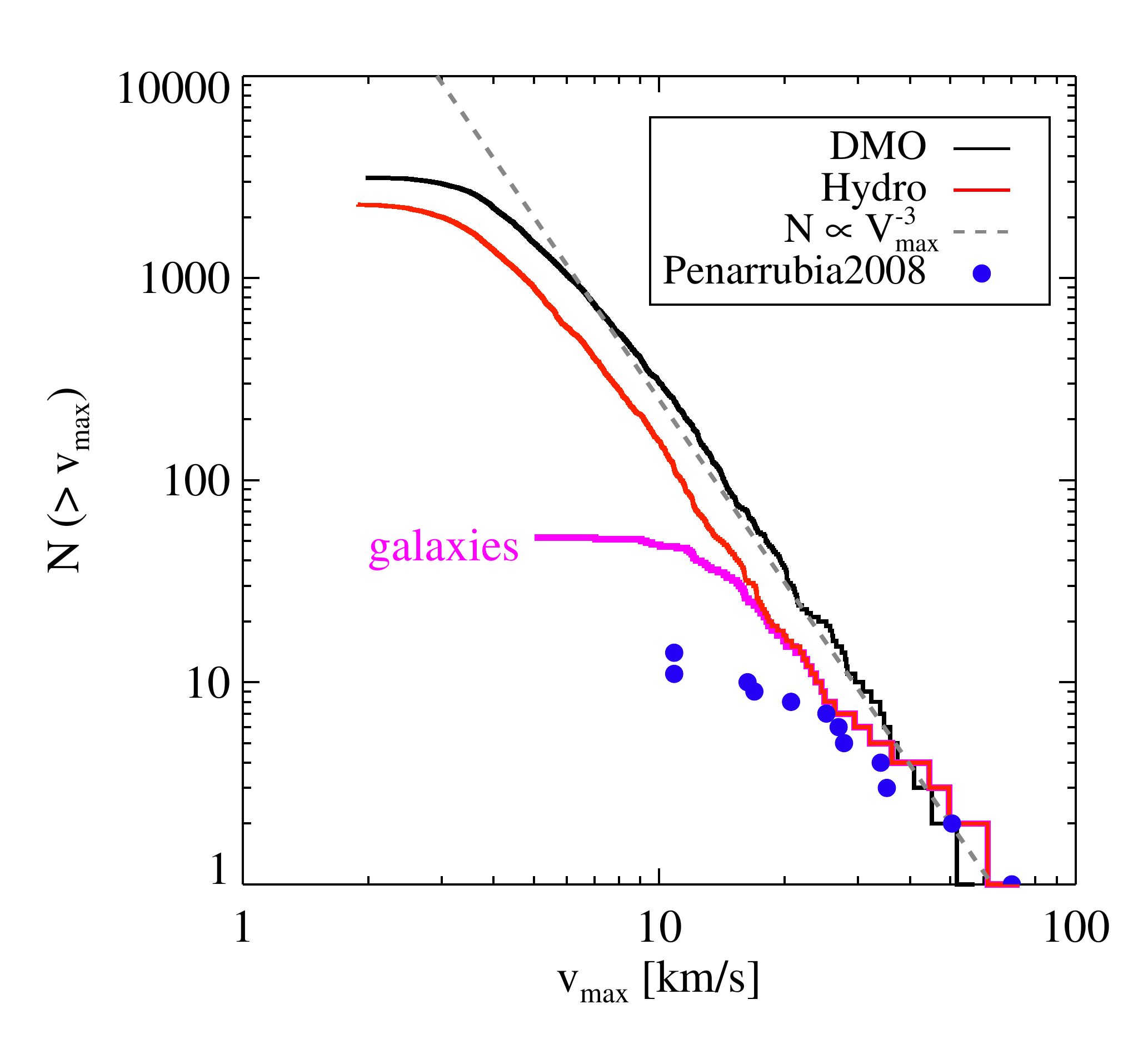}} \resizebox{3.15in}{!}{\includegraphics[scale=0.375]{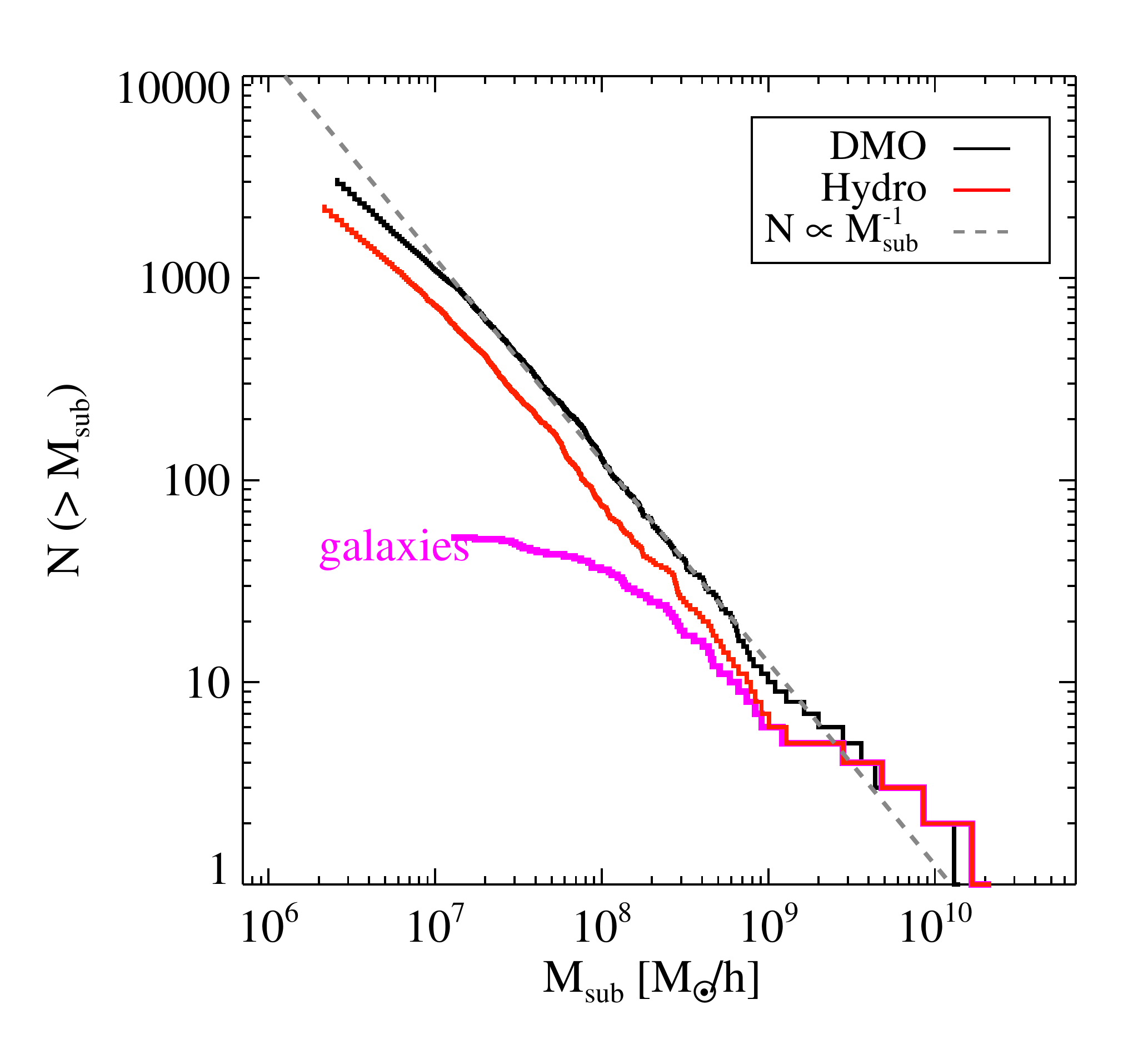}} 
\caption{\label{fig:sub_density_cumulative}. The cumulative distribution of the number of subhalos from the DMO (black solid line) and Hydro (red solid line) simulations, as a function of the maximum circular velocity $v_{\rm max}$ (top panel) and the subhalo mass ${M_{\rm sub}}$ (bottom panel). The gray dashed lines are fits from the literature, $N(>v_{\rm max}) \propto v_{\rm max}^{-3}$ (top panel), or $N(>M_{\rm sub}) \propto M_{\rm sub}^{-1}$ (bottom panel). Bright satellites (subhalos that have stars) are represented by the pink solid curve, while observations by \protect\cite{Penarrubia2008} are shown with blue dots, for comparison.}
\end{center}
\end{figure}

A comparison of the cumulative distribution of DM subhalos between the DMO and 
Hydro simulations is shown in Figure~\ref{fig:sub_density_cumulative}. Both 
simulations show a power-law distribution of the subhalo abundance, $N(>v_{\rm 
max}) \propto v_{\rm max}^{-3}$ in terms of maximum circular velocity, and 
$N(>M_{\rm sub}) \propto M_{\rm sub}^{-1}$ in terms of mass, similar to the 
relations reported for DM subhalos based on the Aquarius simulations
\citep{Springel2008} and the Phoenix simulations \citep{Gao2012}. The slopes of 
both distribution functions are similar in our DMO and Hydro simulations. 
However, the total number of subhalos in the Hydro simulation is consistently 
lower by $\sim 50\%$ than in the DMO case, except for the range where $v_{\rm 
max} > 35~\kms$ (or equivalently ${M_{\rm sub}} > 4\times10^{9}\, \Msun$ in 
terms of mass).
	
The bright satellites, which are here defined as subhalos containing stars in the Hydro simulation, show a different distribution from the DM subhalos at a critical point of $v_{\rm max} \sim 20~\kms$ (corresponding to a mass of ${M_{\rm sub}} \sim 10^{9}\, \Msun$). At the low-mass end, the probability of a subhalo hosting stars steadily decreases as $v_{\rm max}$ decreases. The ``missing satellite problem'' appears clearly striking if we simply compare the number of DM subhalos at $v_{\rm max} < 10~\kms$ with observations of \cite{Penarrubia2008}, because the former is more than two orders of magnitude higher. However, the number of satellites (i.e.~subhalos with stars) is much closer to the observations, and the discrepancy between the two becomes even smaller when detection and completeness limits of current surveys are accounted for.

At the massive end, $v_{\rm max} > 20~\kms$, the number of bright satellites agrees well with observations and it matches that of DM subhalos. The value of $v_{\rm max} \sim 20 ~\kms$ marks a transition in dwarf galaxy formation shaped by reionization, similar to previous studies \citep[][]{Okamoto2008, Okamoto2009}. The total number of massive dwarf galaxies with $v_{\rm max} > 30~\kms$ within the virial radius $r_{\rm vir}$ of the central galaxy is 6 in our Hydro simulation, which is almost half the value (11) of massive subhalos found in the DMO simulation. Note that this corresponds to the mass range of the ``massive failures'' considered in \cite{Kolchin2011, Kolchin2012}. Still, our result is slightly higher than the total number (4) of massive satellites in the Milky Way, including LMC and SMC, which have $v_{\rm max}$ above $30~\kms$ \citep[][]{Penarrubia2008}.  Moreover, detailed variations from one main galaxy to another could, in principle, resolve the residual discrepancy.

The sharp contrast in the number of dwarf galaxies between the DMO and Hydro simulations highlights the critical role of baryonic physics in galaxy formation, and it points to a potentially viable solution of the ``missing satellite'' and the ``too big to fail'' problems, in agreement with suggestions by some previous studies (e.g., \citealt{Brooks2013, Sawala2014a, Mollitor2015}).

\subsection{Mass Functions of Subhalos}
\label{sec:properties_of_matched_subhalos}
	
\begin{figure} \begin{center} \includegraphics[width=\linewidth]{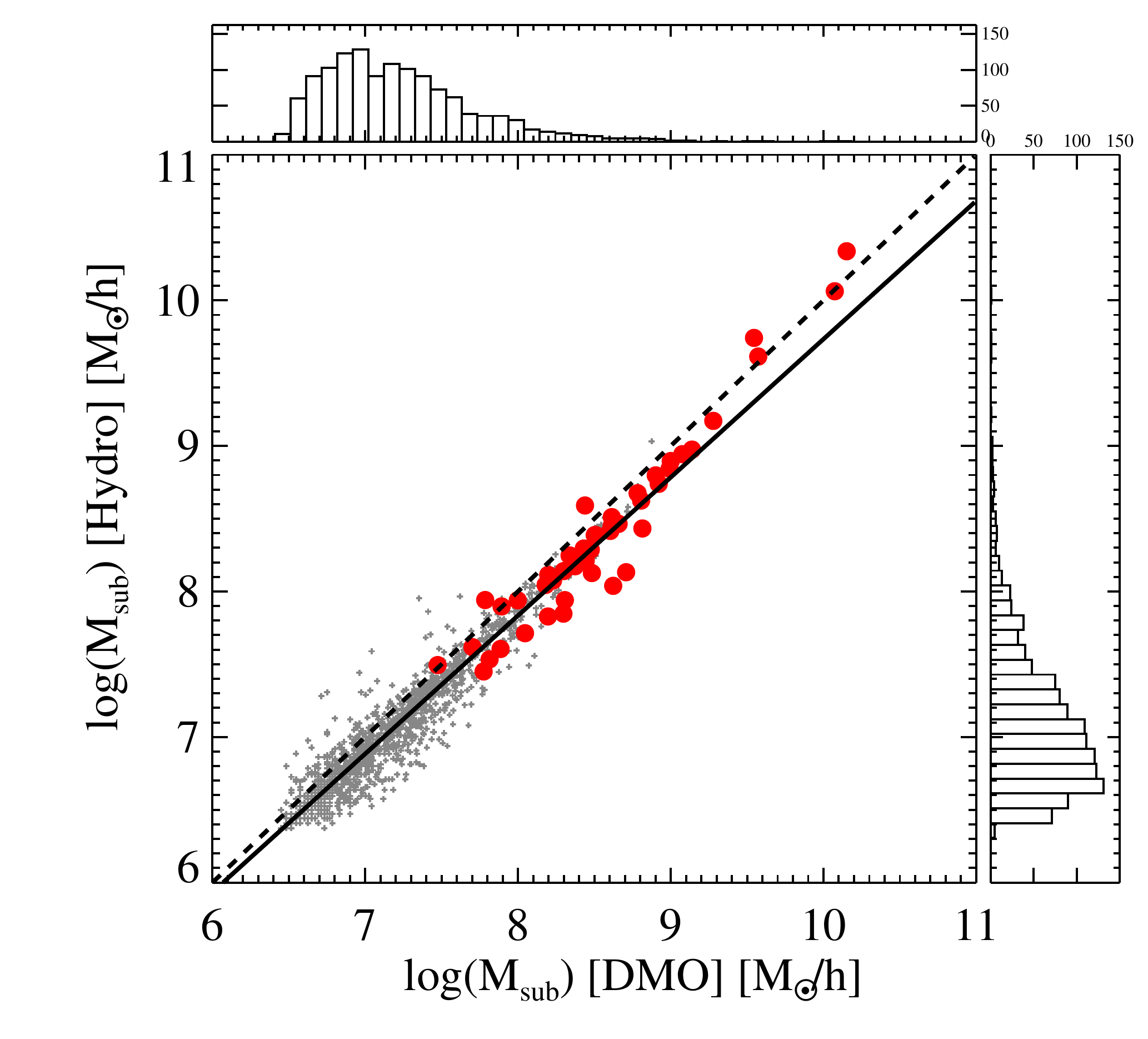} \includegraphics[width=\linewidth]{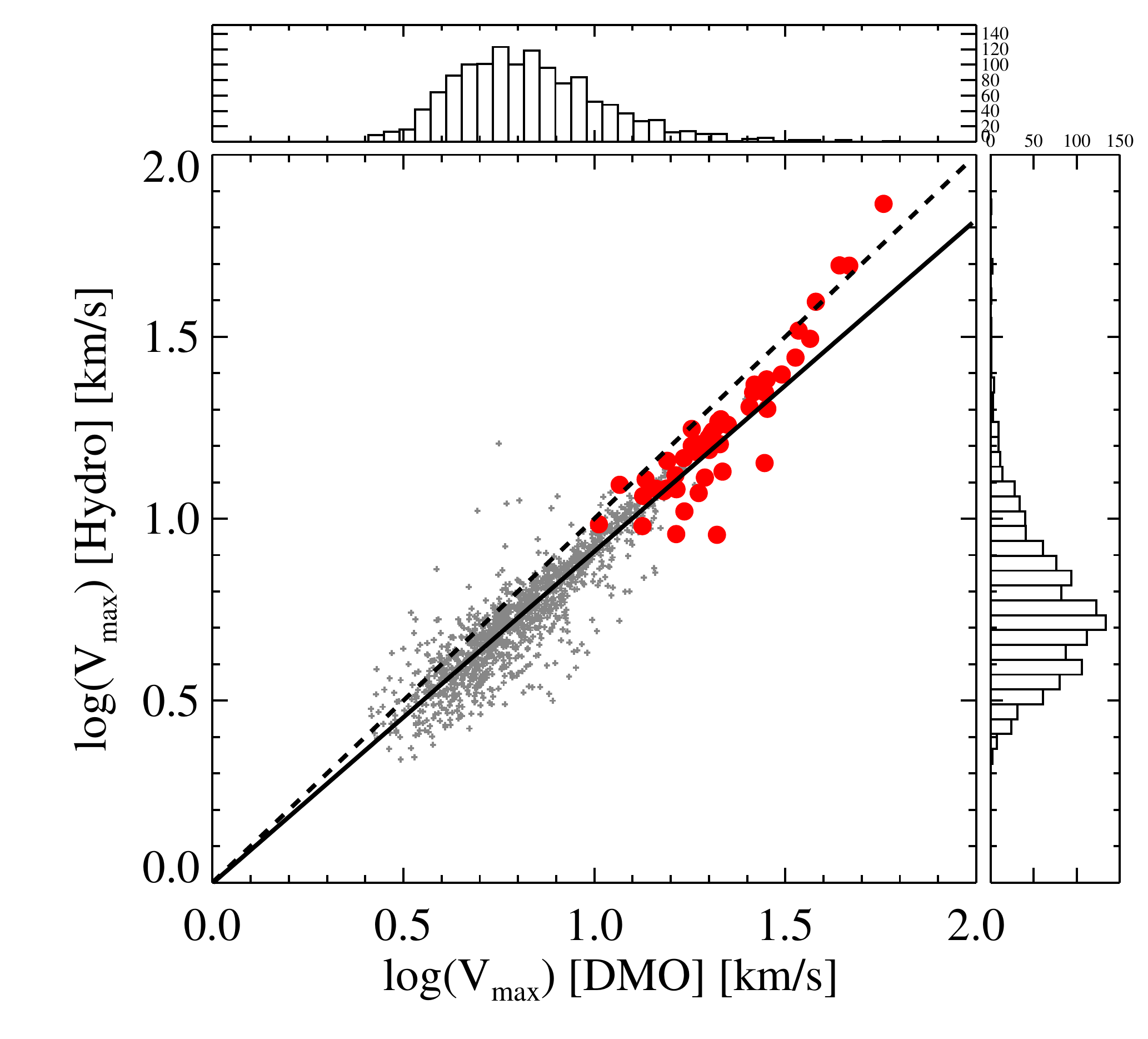} \caption{\label{fig:matched_properties_compare} Comparison of subhalo properties of the matched pairs at $z = 0$ in both the DMO and Hydro simulations. The top panel shows the subhalo mass $M_{\rm sub}$ distribution function, while the bottom panel shows the maximum circular velocity $v_{\rm max}$ distribution function. The grey dots represent all dark subhalos (subhalos that did not form stars), while the filled red circles represent bright satellites (subhalos that contain stars). The solid black line is the fitting curve for all the subhalos (including both dark and star-forming ones), while the dashed line indicates $M_{\rm sub} ({\rm Hydro}) = M_{\rm sub} ({\rm DMO})$ in the top panel, and $v_{\rm max} ({\rm Hydro}) = v_{\rm max} ({\rm DMO})$ in the bottom panel.}
\end{center}
\end{figure}

In order to investigate effects of baryons on the subhalo mass, in Figure~\ref{fig:matched_properties_compare} we compare the subhalo mass $M_{\rm sub}$ (top panel) and the maximum circular velocity $v_{\rm max}$ (bottom panel)  at $z = 0$ of the matched pairs between the two simulations. As the fitting curve (black solid line) is below the diagonal dashed line ($M_{\rm sub} ({\rm Hydro}) = M_{\rm sub} ({\rm DMO})$, or $v_{\rm max} ({\rm Hydro}) = v_{\rm max} ({\rm DMO})$), it is clearly seen that the majority of subhalos in the Hydro simulation are less massive than their counterparts in the DMO simulation, similar to the subhalo abundance findings in Section~\ref{subsec:abundance}. The subhalo mass function of the Hydro simulation peaks at $\sim 5 \times 10^6\,\Msunh$, which is about a factor of 2 lower than the peak of the DMO subhalo mass function at $\sim 10^7\, \Msunh$.

In the Hydro simulation, only massive subhalos can form stars. The minimum mass 
for subhalos to host star formation is $\log ({M_{\rm sub}}) = 7.5$ (or $v_{\rm 
max} = 10~\kms$), although it may be affected by the resolution of the 
simulation. In the mass range between ${10^8}\, \Msunh$ and ${10^9}\, \Msunh$ 
where we have sufficient mass and spatial resolution, there is a mixture of 
``dark'' subhalos and bright satellites (subhalos that contain stars). Such a 
co-existence of dark subhalos and bright satellites implies that a linear 
${M_{\rm halo}-M_{*}}$ correlation, as commonly assumed in semi-analytical 
galaxy models and abundance matching techniques 
\citep[e.g.][]{Guo2010,Moster2013}, may not hold in the dwarf galaxy regimes, 
since some massive halos do not host galaxies with stars. This would complicate 
the application of the abundance matching to dwarf galaxies 
\citep[][]{GarrisonKimmel2014b,Guo2014} and the assignment of galaxies to dark 
matter halos in $\textit{N}-$body simulations.

\begin{figure}
\begin{center}
\includegraphics[width=\linewidth]{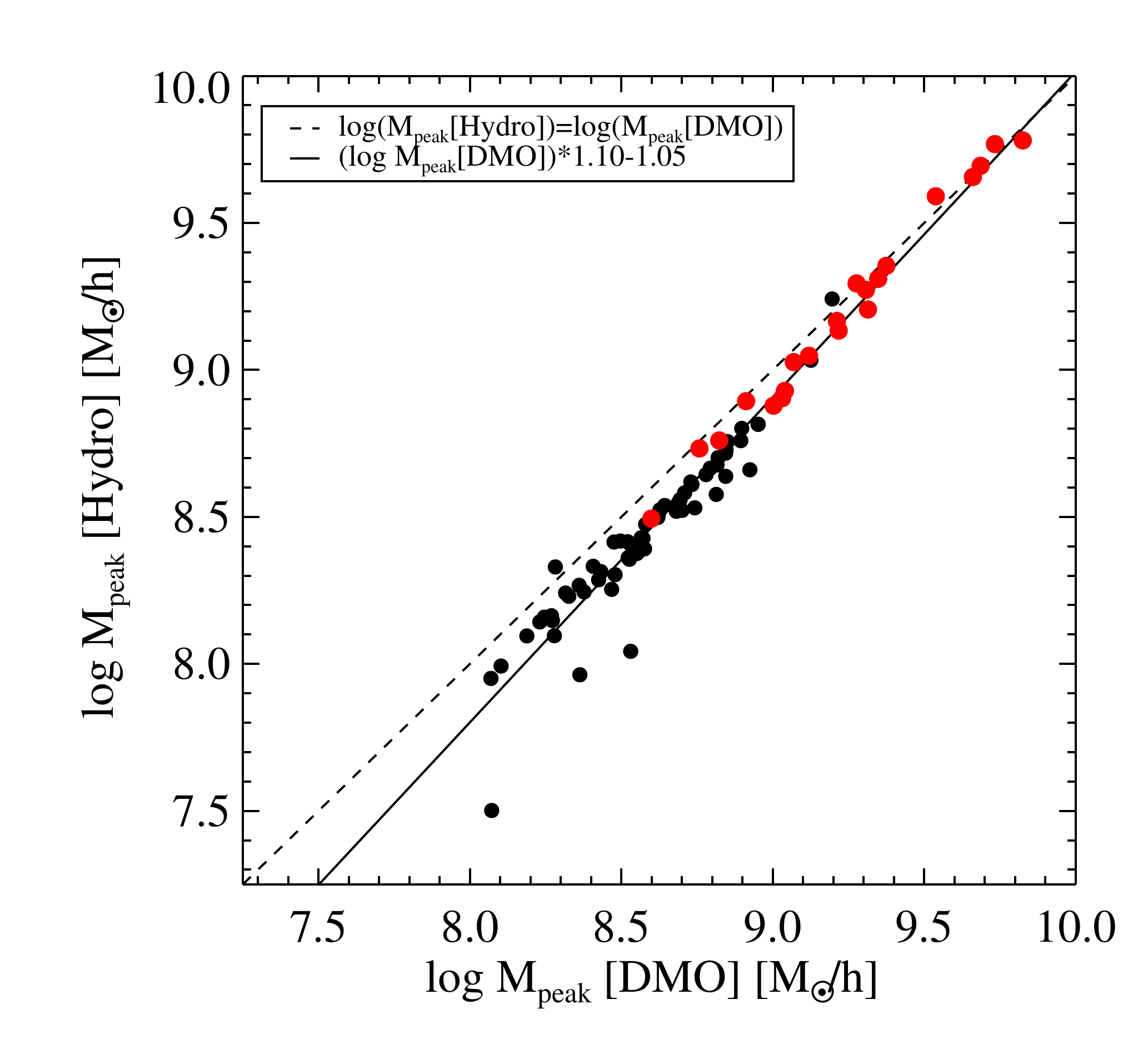}
\caption{\label{fig:peak_mass_comparison} A comparison of the subhalo peak mass ${M_{\rm peak}}$ (the maximum mass attained by the progenitor before it was accreted by its host) of matched pairs in both the DMO and Hydro simulations. The black filled circles are the ``dark'' subhalos and the red filled circles represent bright satellites. The black solid line is the fitting curve of all subhalos, while the dashed line indicates $M_{\rm peak} ({\rm Hydro}) =  M_{\rm peak} ({\rm DMO})$.} 
\end{center}
\end{figure}

Another important parameter is the peak mass of each subhalo, $M_{\rm peak}$,
defined as the maximum mass attained by the 
progenitor before it was accreted by its host. Using the peak mass is currently 
the standard method in abundance matching or semi-analytical modeling when 
dealing with subhalos \citep[e.g.][]{Guo2014, GarrisonKimmel2014b}, since this 
quantity represents a physical state unmodified by the subsequent interaction 
between the subhalo and the host. Figure~\ref{fig:peak_mass_comparison} shows a 
comparison of $M_{\rm peak}$ from the DMO and Hydro simulations. We find that 
subhalos below $10^{9}\, \Msunh$ generally have lower $M_{\rm peak}$ in the 
Hydro simulation than in the DMO simulation, and that subhalos with peak mass 
higher than ${10^{9}\, \Msunh}$ are able to form stars. However, there are a few 
``outliers": two subhalos with peak mass above  ${10^{9}\, \Msunh}$ remain 
completely dark, while three subhalos with peak mass below ${10^{9}\, \Msunh}$ 
actually contain stars. The fitting of the data shows that, 
\begin{equation}
\log {M_{\rm peak}}[{\rm Hydro}] = 1.10\log {M_{\rm peak}}[{\rm DMO}] - 1.05,
\end{equation}
which means that the subhalo peak mass is $\sim$30\% lower in the Hydro simulation than in the DMO simulation for $M_{\rm peak} \sim {10^{9}}\, \Msunh$, and $\sim$44\% lower for $M_{\rm peak} \sim {10^{8}}\, \Msunh$.

Our results suggest that the impact of hydrodynamics on the halo mass could be easily amplified in the early growth stages when the halos increase their mass exponentially \citep[][]{vandenBosch2014b, Correa2015}, and that the assumption of a monotonic relation between stellar mass of a galaxy and its peak mass in the abundance matching technique is not valid. Hydrodynamic simulations should hence be employed for a more reliable study of the properties of dwarf galaxies, also as suggested by previous work \citep[e.g.][]{Sawala2014a, Velliscig2014}.

\subsection{Dark Matter Distribution and Density Profile of the Main Host}

\begin{figure} \begin{center} \includegraphics[width=\linewidth]{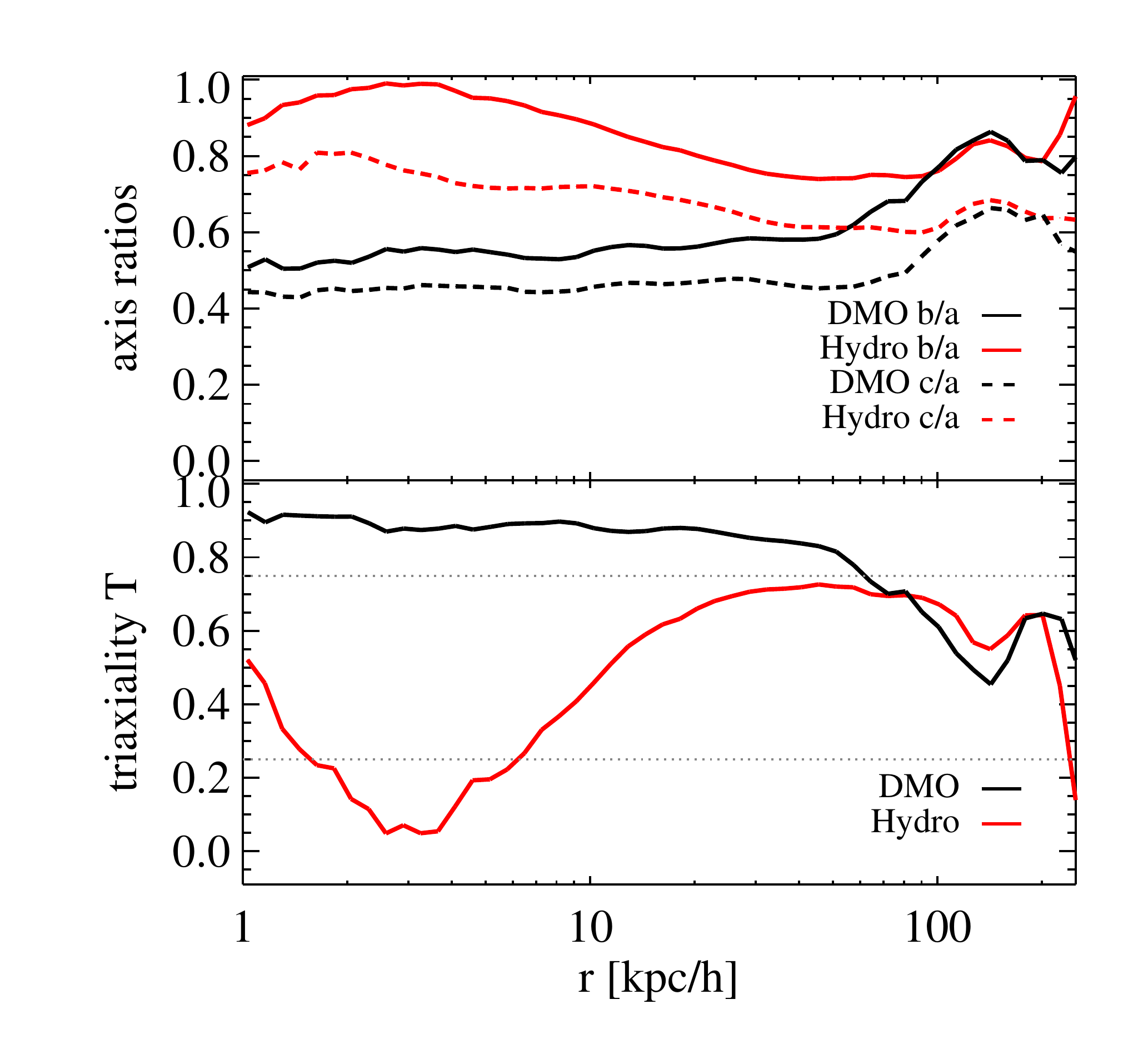} \caption{\label{fig:central_halo_shape} Comparison of the shape of the main dark matter halo from the DMO (in black) and Hydro (in red) simulations. Top panel: the intermediate to major axis ratio ${b/a}$ (solid lines) and the minor to major axis ratio ${c/a}$ (dashed lines) as a function of distance to the galactic center.  Bottom panel: the triaxiality $T$, defined as $T=(a^2-b^2)/(a^2-c^2)$, of the main dark matter halo as a function of galactic radius. The dotted lines indicate $T=0.25$ and $T=0.75$, marking the transitions from oblate/prolate to triaxial halo shapes. The plot shows that in the inner region the DM halo is more spheroidal in the Hydro simulation, but it is triaxial in the DMO simulation. }
\end{center}
\end{figure}

To study the effects of baryons on the DM distribution of the main host, we 
compare the DM shape and density profile of the main host in both simulations. We 
apply a principal component analysis to the DM halo and compute the three axis 
parameters ${a}$, ${b}$ and ${c}$ based on the eigenvalues of the moment of 
inertia tensor for all the DM particles within a given shell \citep[following 
the method by][]{Zemp2011}. The halo shape can be quantified by the intermediate 
to major axis ratio, ${b/a}$, and the minor to major axis ratio, ${c/a}$, as 
shown in Figure~\ref{fig:central_halo_shape} (top panel), which compares the 
axis ratios at different distances from the galactic center in both the DMO and 
Hydro simulations. The shape of the DM halo differs significantly between the 
two simulations. In the inner region within $10\,h^{-1}{\rm kpc}$, it is 
triaxial with the triaxiality parameter  (defined as $T=[a^2-b^2]/[a^2-c^2]$ as 
in \citealt{Zemp2011}) $T \sim 0.9$ in the DMO simulation, but in the Hydro 
simulation it is close to an oblate  spheroid with $b/a\sim 1$ and $c/a>0.7$. 
The difference in halo shape between the DMO and Hydro simulations continues 
towards larger galactic distance out to $\sim 100\, \kpch$, but they converge at 
the virial radius $r \sim  200\,\kpch$.  These results are in good agreement 
with previous studies \citep[][]{Springel2004, DOnghia2010, Vera-Ciro2011, 
Zemp2011, Bryan2013}, and demonstrate that the impact of baryons on the dark 
matter halo shape is significant up to the virial radius, a spatial scale much 
larger than that of the stellar disk or the central galaxy. We note that a 
similar effect is seen in gas-rich mergers of galaxies, where gas inflows 
\citep{Barnes1991} and nuclear starbursts \citep{Mihos1996} can significantly 
modify the shapes and orbital properties of remnants \citep{Barnes1996}.

\begin{figure} \begin{center} \includegraphics[width=\linewidth]{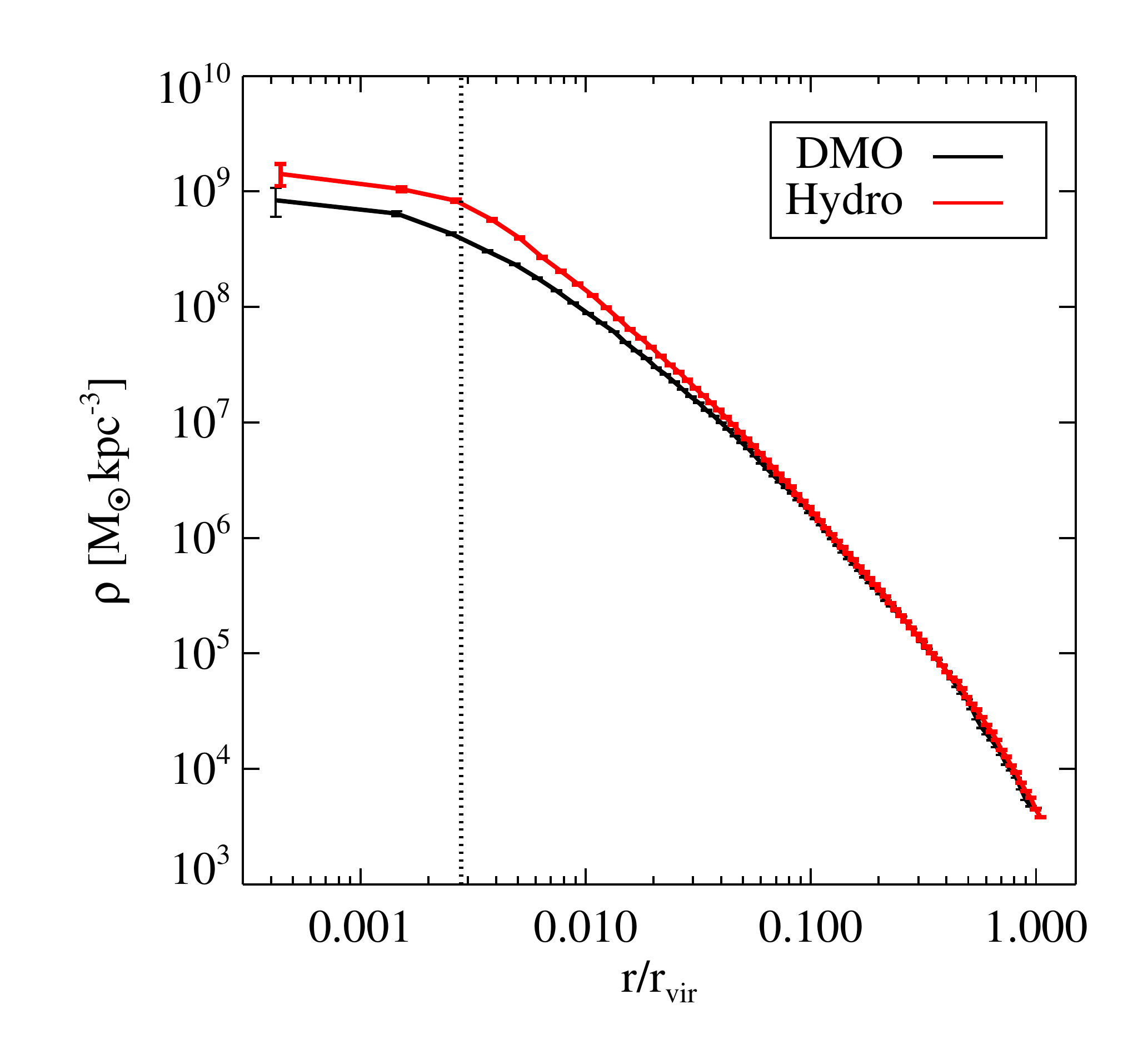} \includegraphics[width=\linewidth]{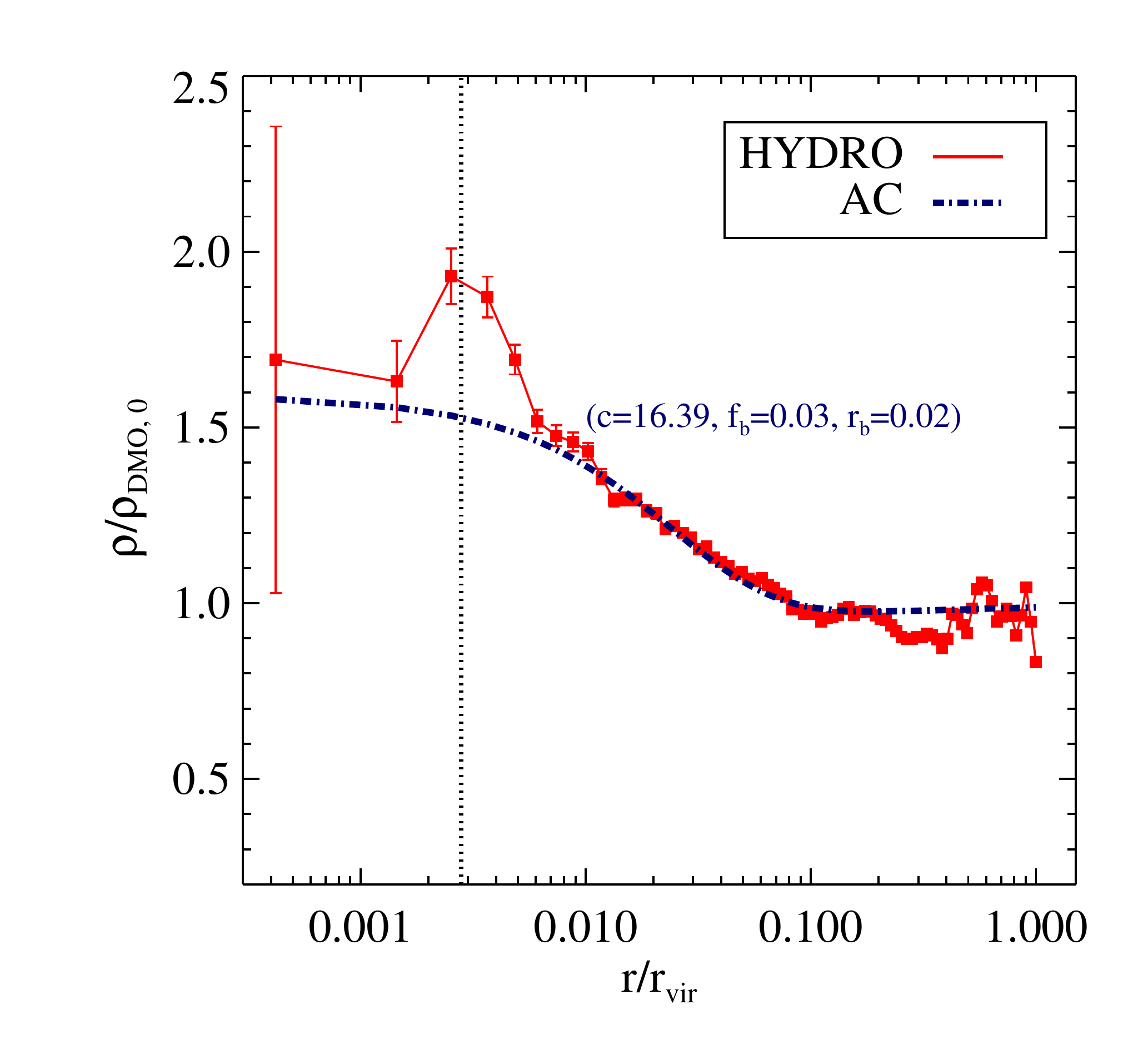} \caption{\label{fig:adiabatic_contraction} Comparison of the radial dark matter density profile of the main halo in our different simulations. \textit{Top panel:} The halo DM density profiles from the DMO (black solid curve) and Hydro (red solid curve) simulations, respectively. A factor of $\rm{\frac{\Omega_{m} - \Omega_{b}}{\Omega_{m}}}$ is applied to the density profile from the DMO simulation. \textit{Bottom panel:} The enhancement of DM from the Hydro simulation (red solid curve) in comparison with the adiabatic contraction calculation (blue dashed-dotted curve) based on the DMO density profile using the {\small\sc contra} code \citep{Gnedin2011}. The error bars are based on the Poisson error $\sqrt{N}$, where $N$ is the number of DM particles in each radial bin. The vertical dotted line in both plots marks the position at which the gravity force equals its exact Newtonian form, $r=2.8\,\epsilon$, where $\epsilon$ is the gravitational softening length. An enhancement of the DM concentration due to adiabatic contraction is present in the Hydro simulation in the inner region up to $r \sim 0.1 r_{\rm vir}$.}
\end{center}
\end{figure}

In addition to the halo shape, remarkable differences between the Hydro and DMO simulations are also present in the DM density profile of the main host. Adiabatic contraction of the DM distribution by baryonic processes is a well known effect. Historically, adiabatic contraction effects were calculated analytically based on the assumption of circular orbits and conservation of angular momentum. \cite{Gnedin2004} refined the calculation by considering the eccentricities of the orbits in a more realistic cosmological context. Even with such a modification, the density profile of DM in the inner region could be overestimated without the inclusion of more baryon physics other than cooling and star formation. Recently, \cite{Marinacci2014a} showed that an enhancement of the DM density in the inner region was present for most of the eight Milky Way-size halos studied in their simulations (however at a lower resolution than that used here).

To illustrate the effect of adiabatic contraction on the DM distribution, we present in Figure~\ref{fig:adiabatic_contraction} the spherically-averaged DM density profile of the main halo both from  the DMO and Hydro simulations (top panel), as well as a comparison with the adiabatic contraction calculation (bottom panel) based on the DMO density profile using the {\small\sc contra}\footnote{\url{http://dept.astro.lsa.umich.edu/~ognedin/contra}. } code \citep{Gnedin2004, Gnedin2011}. This code calculates the response of a DM distribution to the condensation of baryons. We input the DM distribution at $z = 0$ from the DMO simulation as the state before contraction, and the baryonic mass distribution at $z = 0$  from the Hydro simulation as the source of adiabatic contraction. Since the contraction of DM  is naturally followed in the Hydro simulation, we can compare the DM density profile from the Hydro simulation with the theoretical expectation from {\small\sc contra}. The lower panel of Figure~\ref{fig:adiabatic_contraction} shows the expected enhancement of DM from the DMO simulation (dash-dotted line) as if it would host the same galaxy produced by the Hydro simulation. The red solid symbols show the DM density profile measured in the Hydro simulation.This plot shows a significant enhancement of the DM in the inner region out to $r \sim 0.1\, r_{\rm vir}$ in the Hydro simulation, which is consistent with the expected adiabatic contraction. Interestingly, in the very central region, we see more DM enhancement in the Hydro simulation compared to the result from {\small\sc contra}. While it is possible that the DM distribution in the Hydro simulation follows a much more complex evolution than the simplified analytical calculation in {\small\sc contra}, we also note that the ``bump" in the lower panel of Figure~\ref{fig:adiabatic_contraction} may be subject to numerical uncertainty as it occurs on a scale very close to the spatial resolution of the simulations.

In our Hydro simulation, the response of DM due to gas cooling and condensation is consistent with the expectation of adiabatic contraction. As demonstrated in \cite{Marinacci2014a}, this holds true for the majority of the MW-size halos in the hydrodynamic simulations. The agreement between our simulation and {\small\sc contra} fitting suggests that the latter provides a good description of the DM distribution in $L_\star$ galaxies. Since {\small\sc contra} has been calibrated using a suite of different hydrodynamic simulations \citep{Gnedin2011}, the agreement between our result and {\small\sc contra} therefore reflects a consistent response of dark matter with respect to gas cooling and condensation, despite substantial differences in how feedback was modeled. 

Interestingly, no dark matter core is formed in our simulated halo.  We note that recent simulations of halos more massive than  ${10^{12}}\, \Msun$ by \cite{Maccio2012} and \cite{Mollitor2015} show flattened DM distributions within the inner 5 kpc. Compared to our simulations, these studies employed a different feedback model featuring a more bursty stellar feedback. While feedback processes such as energy and momentum from SNe explosions may hence change the contraction of DM in the central region, such effects sensitively depend also on how the feedback injection is modeled in detail.

\subsection{Dark Matter Distributions and Density Profiles of Subhalos}

\begin{figure}
\begin{center}
\begin{tabular}{cc}
\resizebox{1.6in}{!}{\includegraphics{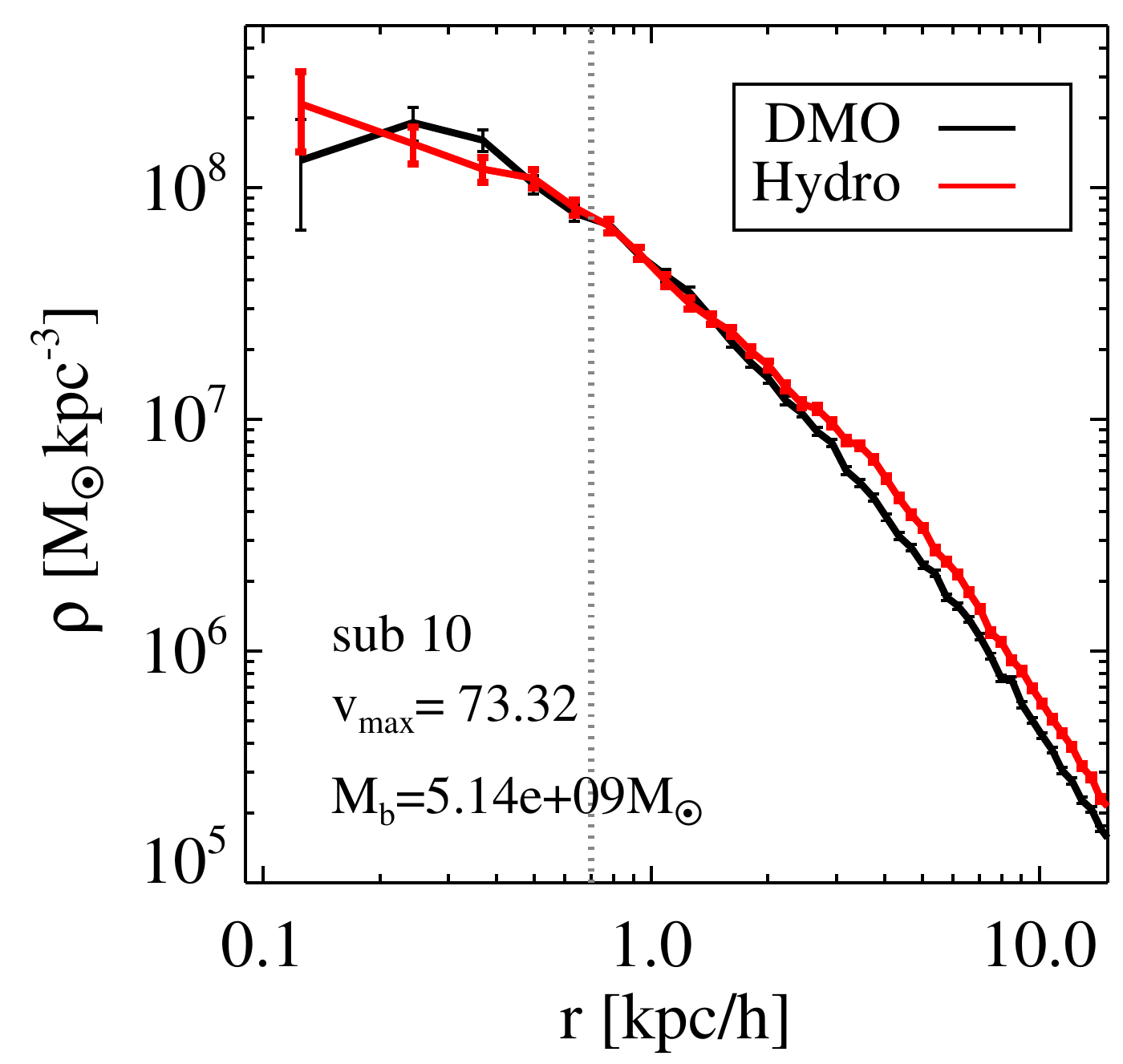}}
\resizebox{1.6in}{!}{\includegraphics{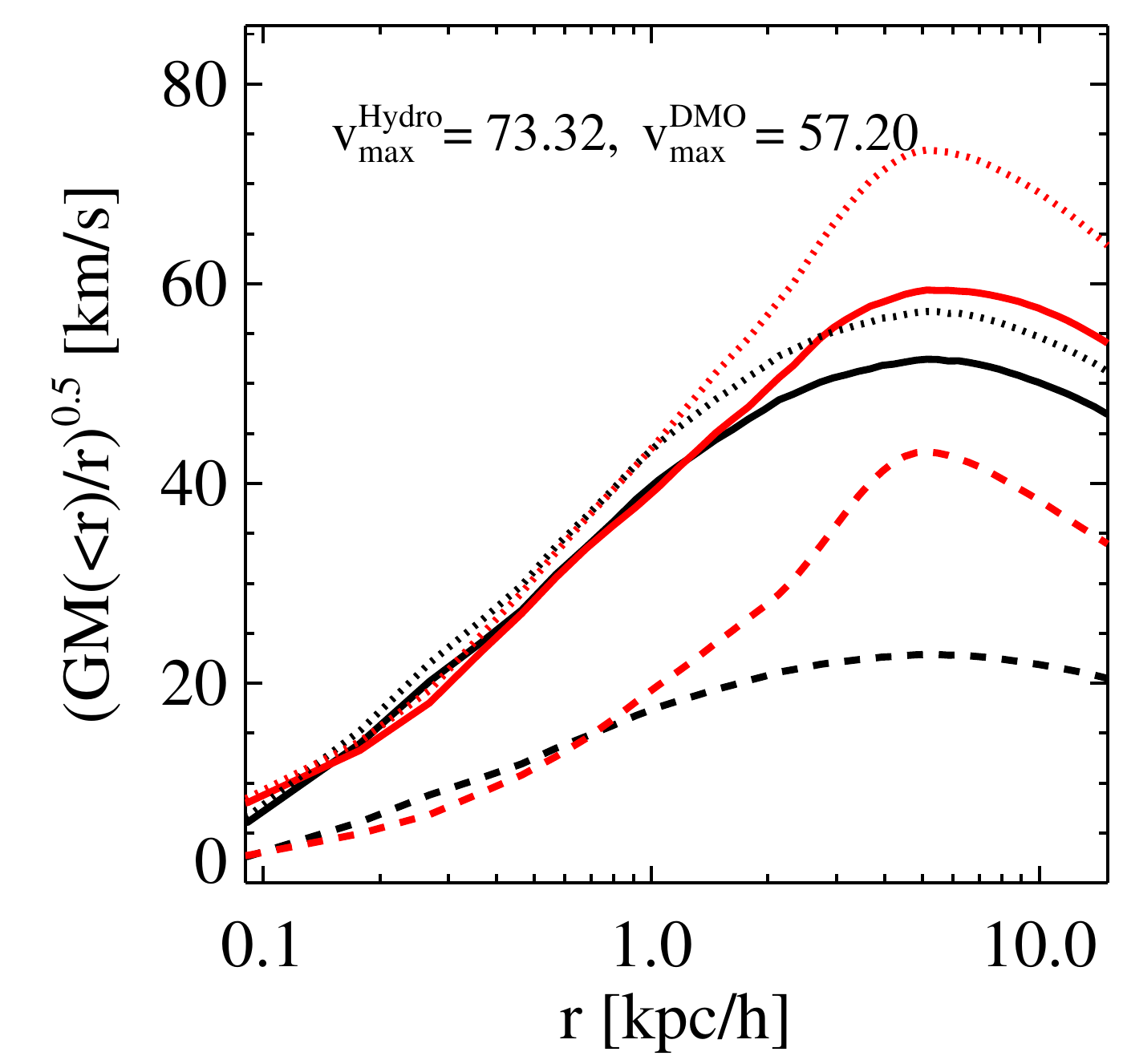}} \\
\resizebox{1.6in}{!}{\includegraphics{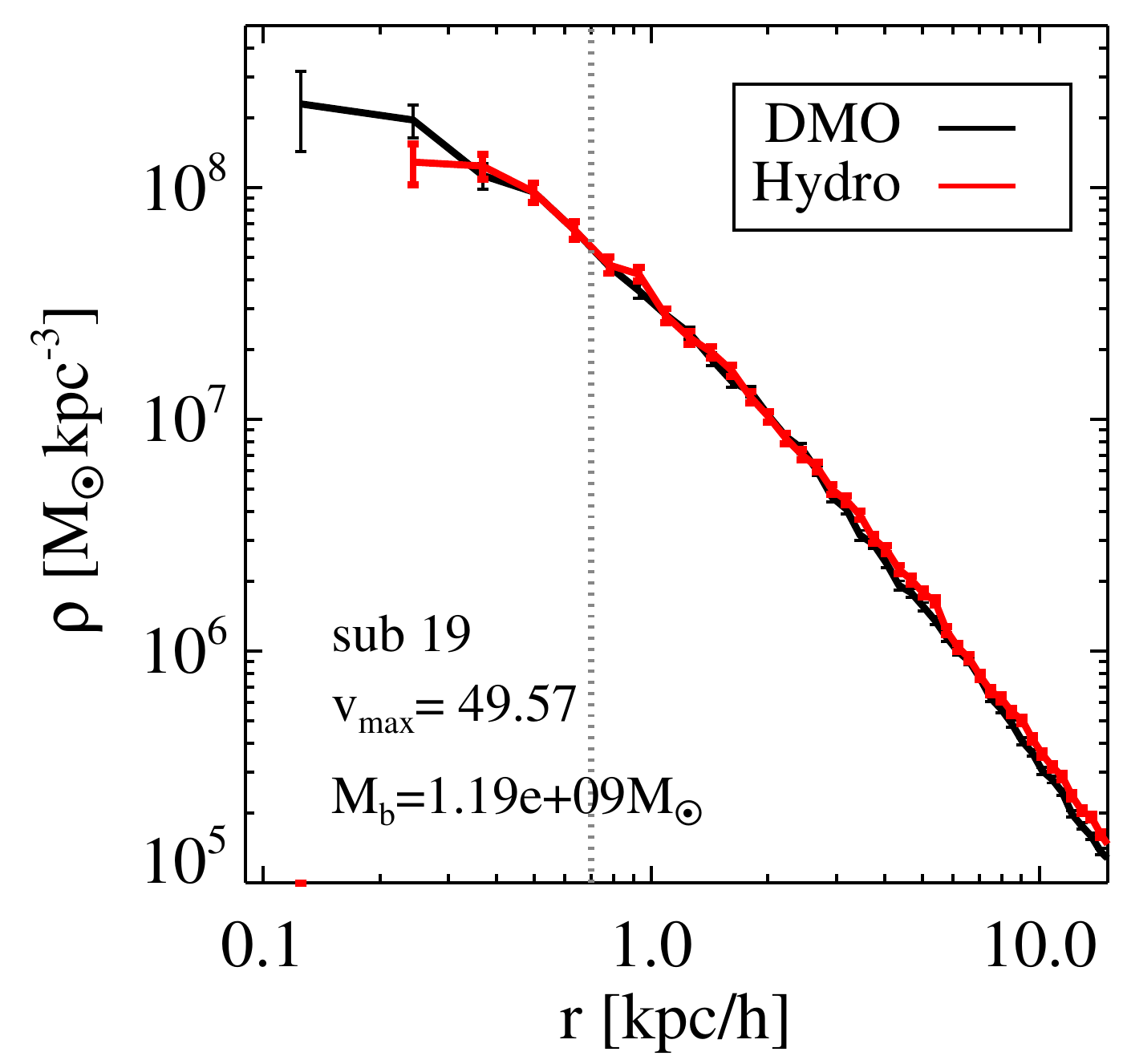}}
\resizebox{1.6in}{!}{\includegraphics{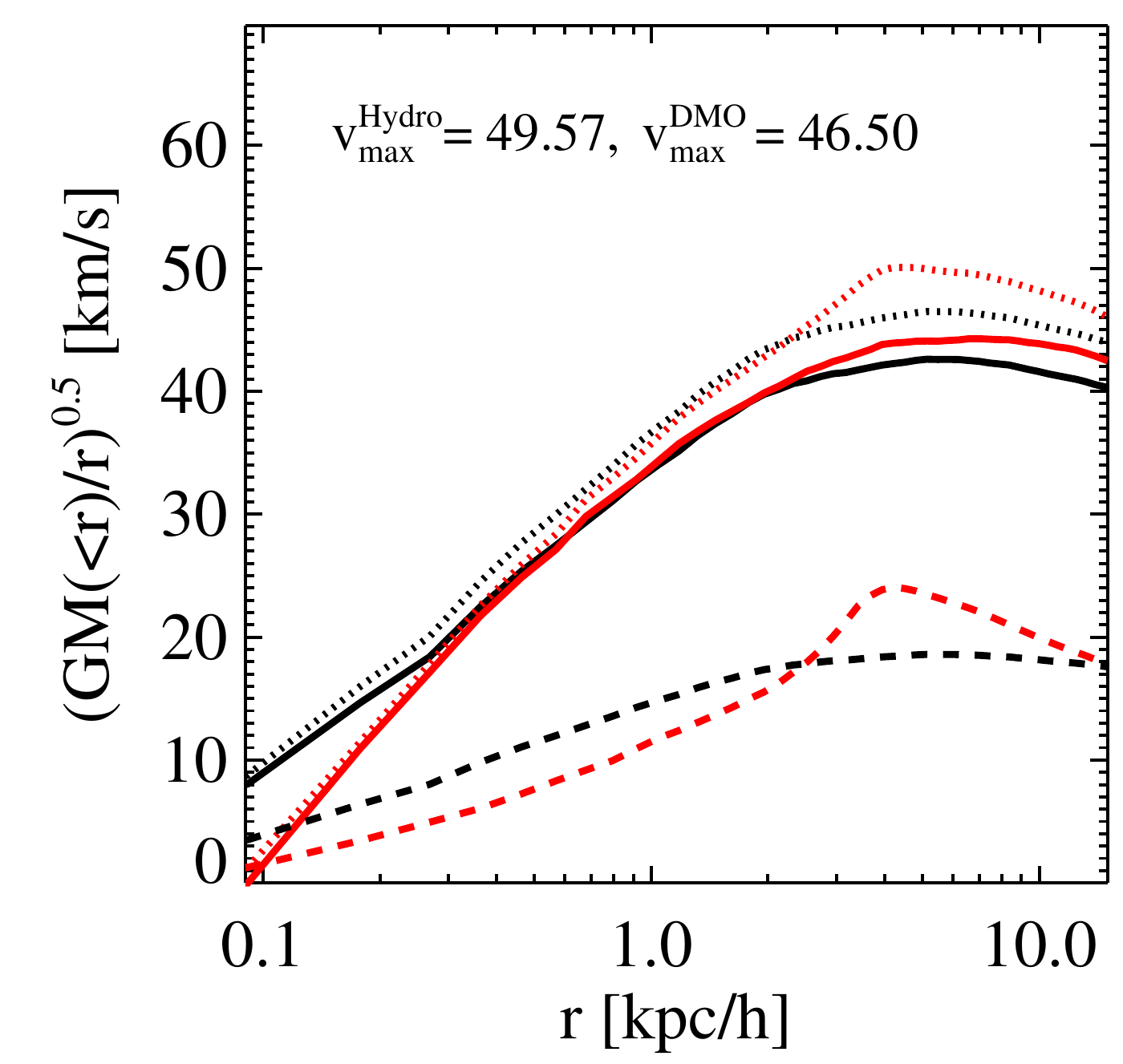}} \\
\resizebox{1.6in}{!}{\includegraphics{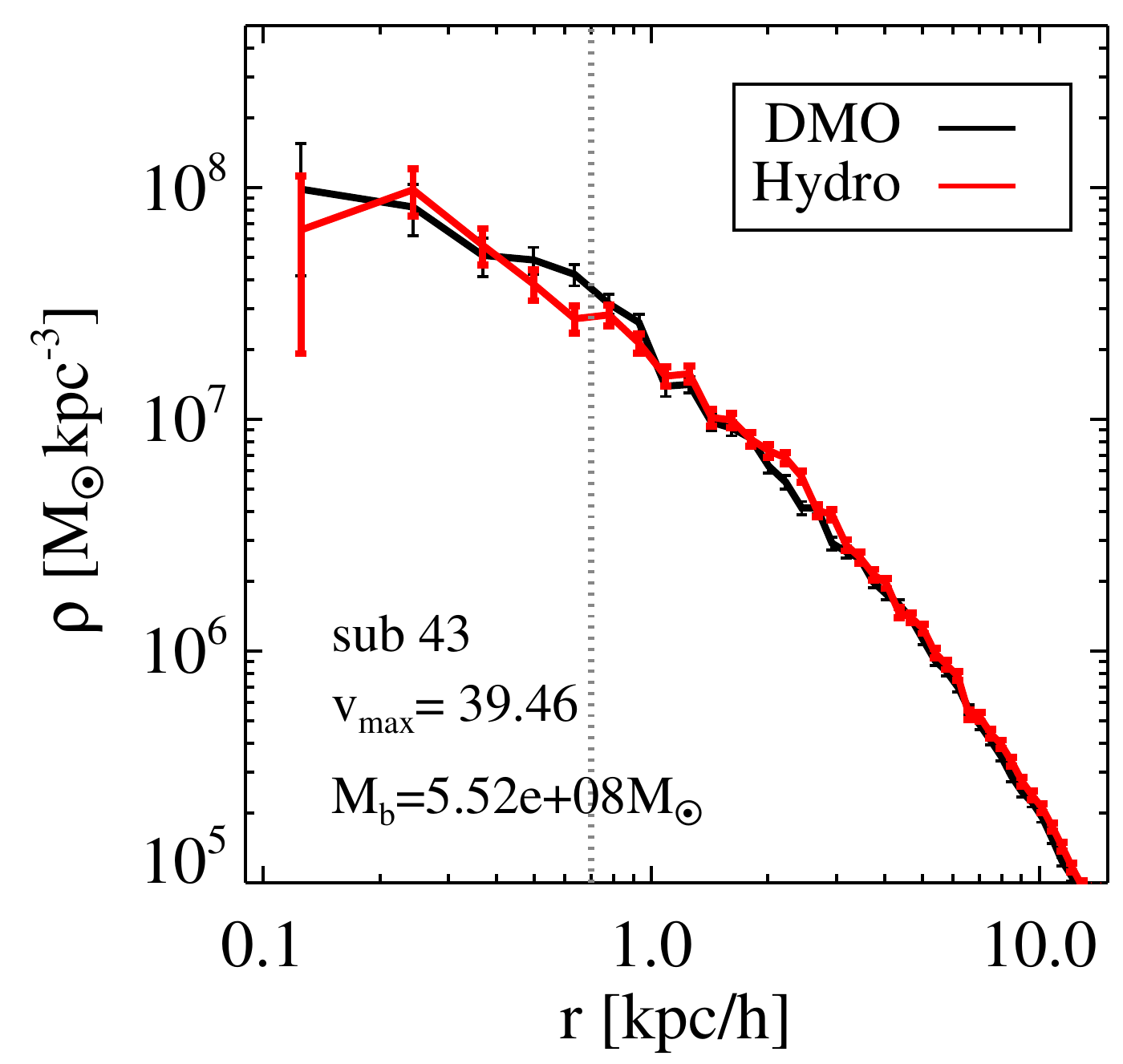}}
\resizebox{1.6in}{!}{\includegraphics{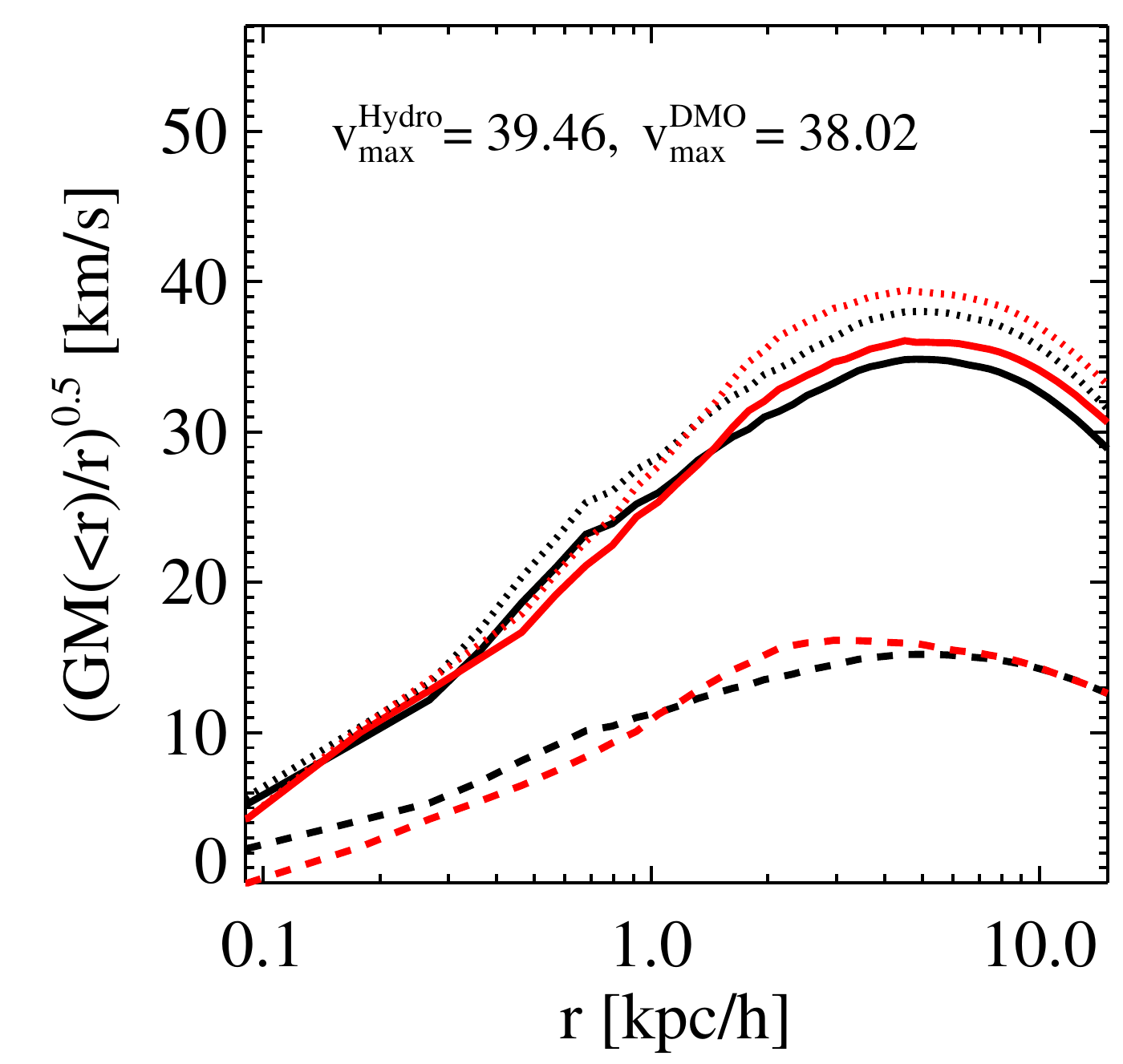}} \\
\resizebox{1.6in}{!}{\includegraphics{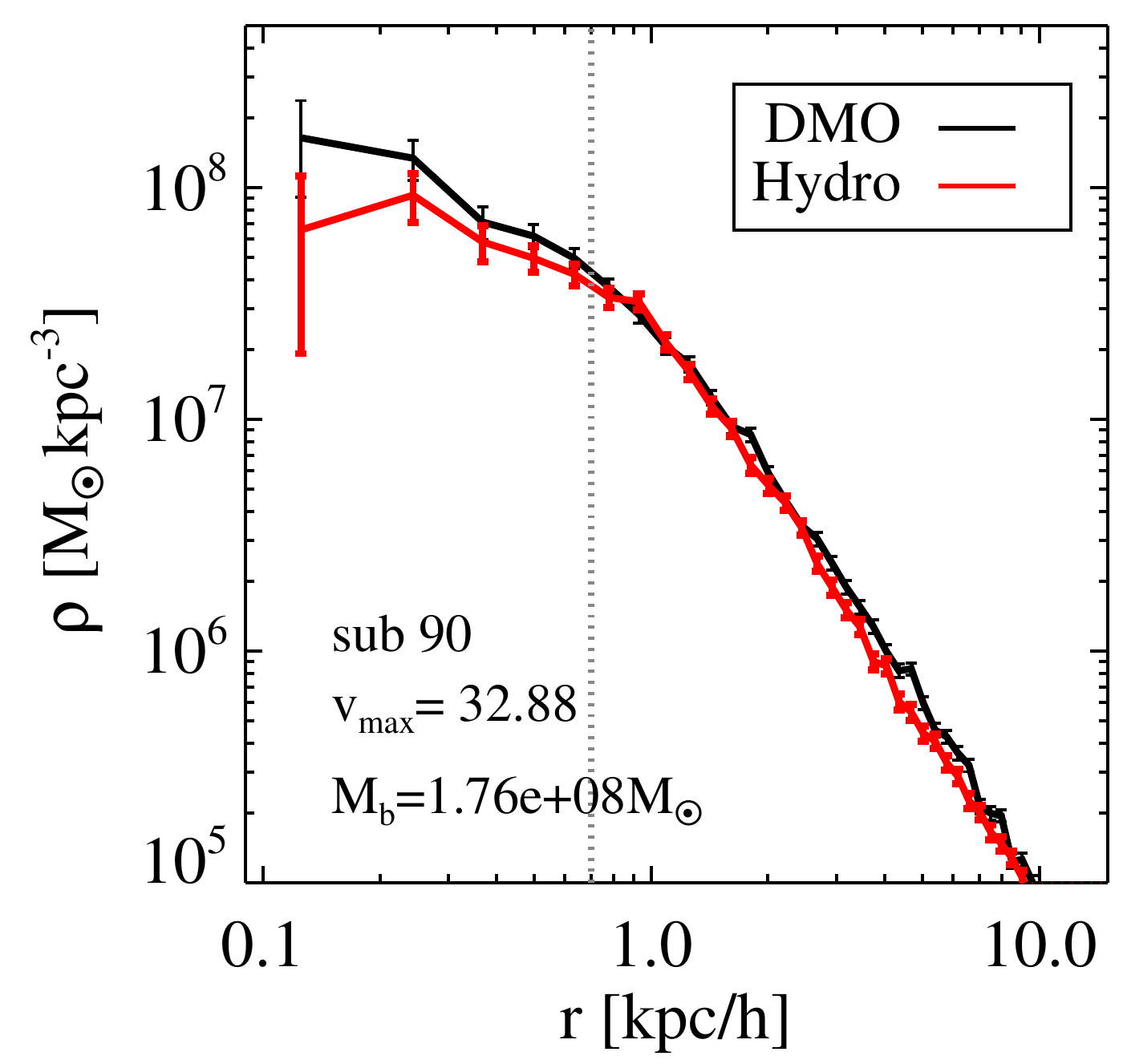}}
\resizebox{1.6in}{!}{\includegraphics{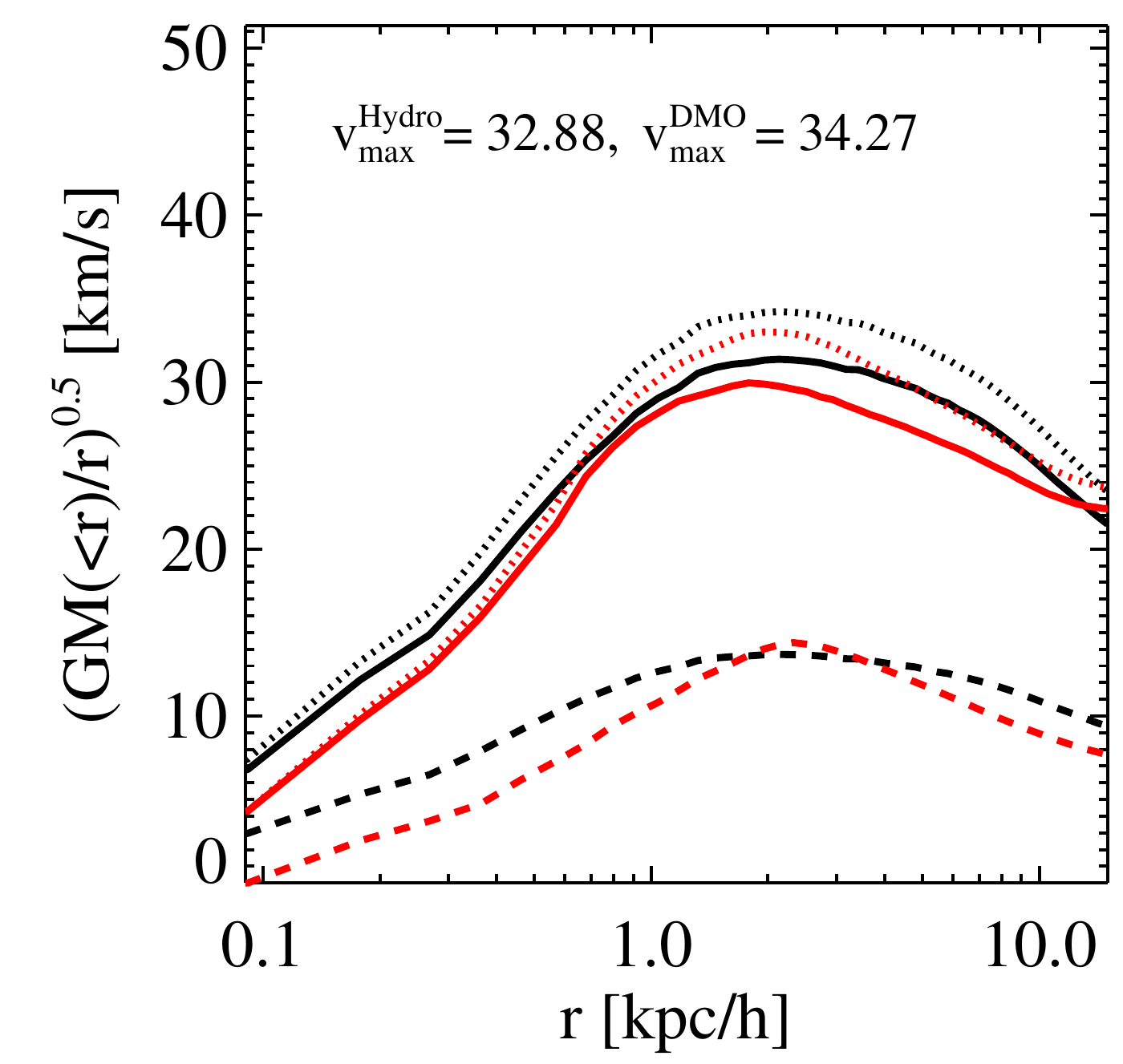}} \\
\resizebox{1.6in}{!}{\includegraphics{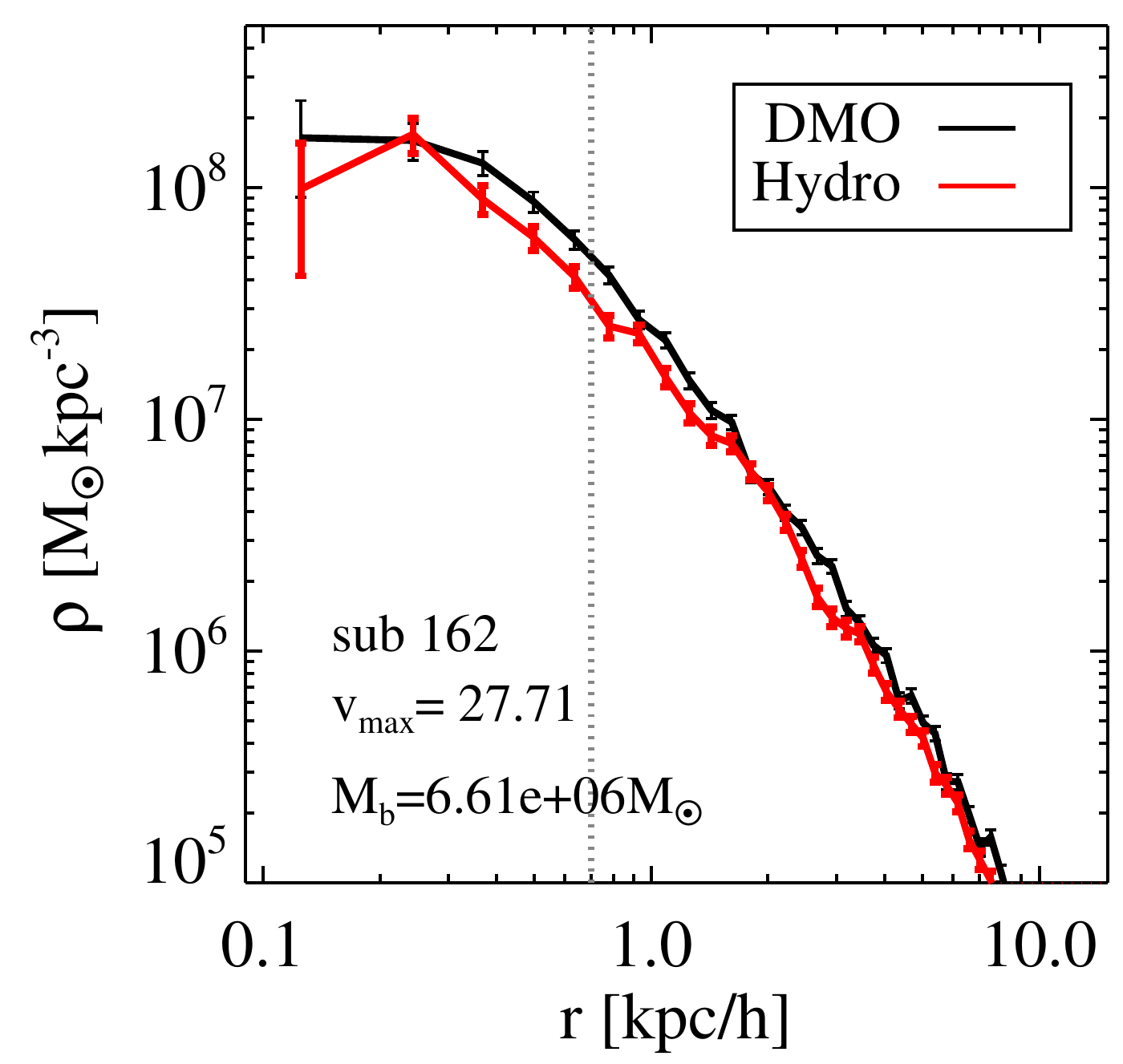}}
\resizebox{1.6in}{!}{\includegraphics{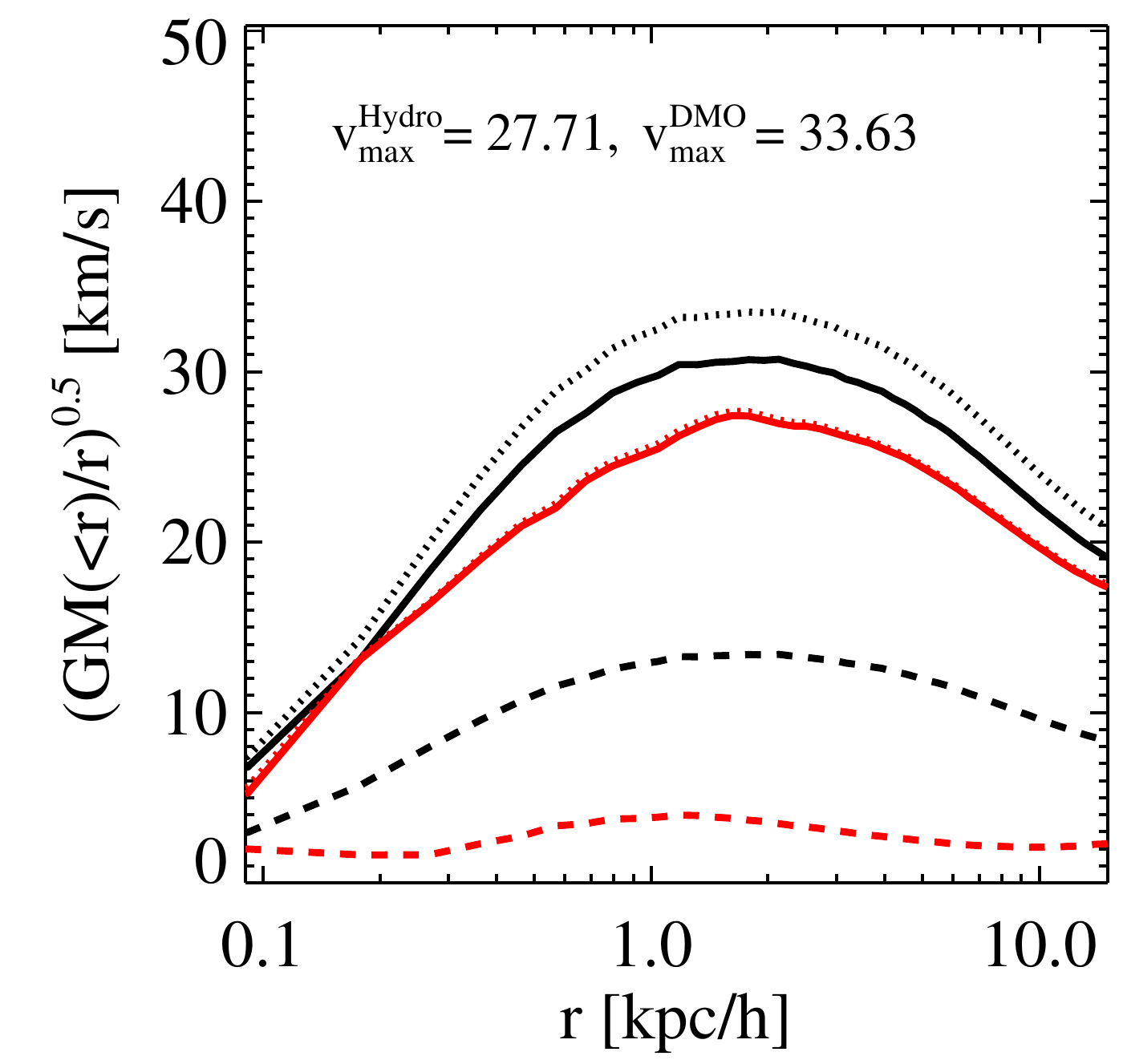}} \\
\end{tabular}
\caption{\label{fig:core_cuspy} A comparison of the dark matter density profiles (left column) and the circular velocity curves (right column) of 6 matched subhalos from the DMO (in black) and Hydro (in red) simulations . As in Figure~\ref{fig:adiabatic_contraction}, the DMO density profiles are multiplied by a factor of $\rm{\frac{\Omega_{m} - \Omega_{b}}{\Omega_{m}}}$, and the dashed vertical line indicates $r=2.8\,\epsilon$. For each subhalo, its ID, $v_{\rm max}$ and baryonic mass $M_{\rm b}$ are listed. In the right panels, the total circular velocity, and contributions from the DM and baryons are represented by the dotted, solid and dashed curves, respectively. The distribution of baryons in the DMO simulation is assumed to follow that of the DM multiplied by a factor of  ${\Omega_{\rm b}}/{\Omega_{\rm m}}$.     
}
\end{center}
\end{figure}	

Similar to the main halo, we identify subhalos in both the DMO and Hydro simulations using the {\small\sc AHF} group finder. We first locate the center of mass from the output of {\small\sc AHF} for each matched object, then compute the spherically averaged density for DM particles of each subhalo based on the particle locations in the original snapshot.  Figure~\ref{fig:core_cuspy} shows the DM density profiles of 6 matched subhalos at $z = 0$ from both simulations, as well as the corresponding circular velocity curves.  These subhalos cover a wide range of total mass, from massive bright dwarf galaxies to ``dark'' subhalos.  The circular velocity of  each subhalo is calculated  as $\sqrt{GM(<r)/r}$, where $M(<r)$ is the enclosed mass within $r$. We further compute the contributions from DM and baryonic components to the rotation curve in order to determine whether the differences in $v_{\rm max}$ from the two simulations are due to highly concentrated baryons or a genuine response of DM to baryonic processes. For the DMO simulation, we assume the distribution of baryons follows that of the DM but differs in mass by a factor of  ${\Omega_{\rm b}}/{\Omega_{\rm m}}$. We note, however, that the  most massive subhalo, Sub~10, contains more baryonic mass in the Hydro simulation than in the DMO simulation, while the least massive subhalos (such as sub 162 and sub 244) contain much less baryonic mass in the Hydro simulation than expected based on the DMO simulation. 

From Figure~\ref{fig:core_cuspy}, the computed density profiles of the subhalos from the Hydro simulation match their counterparts from the DMO simulation quite closely. However, the local distribution of DM is better probed by the rotation curves as they depend sensitively on the mass enclosed within a certain radius. As shown in the right panels of Figure~\ref{fig:core_cuspy}, significant differences are evident between the rotation curves from these two simulations, in particular with respect to the contribution of DM as represented by the solid curves in the right panels. For the first 3 subhalos (Sub~10, 19 and 43), the contribution from the DM in the Hydro simulation is higher than that in the DMO simulation. They contain slightly more dark matter in the Hydro simulation than in the DMO one, showing some mild contraction. In addition, the amount of contraction in these three subhalos varies with their total mass, with Sub~10 showing the strongest contraction and sub 43 the weakest. On the other hand, the other three subhalos, sub 90, 162 and 244, show slightly reduced DM concentrations in the Hydro simulation compared with the DMO one. 

In the inner region, we have not found clear signs of DM cores in these subhalos. However, the absence of DM cores could potentially be due to a limited mass and spatial resolution of the simulations. The gravitational softening length in our simulations is a factor of 2 to 3 larger than in the most recent high resolution simulations focusing on (isolated) dwarf galaxies \citep[][]{Governato2012, Teyssier2013, Madau2014, Onorbe2015}.

\section{The Impact of baryons on the evolution of subhalos}
\label{sec:matchedsubhaloresult}

\subsection{Importance of Individual Physical Processes}
\label{sec:physicalprocesses}

Baryonic processes play a critical role in the formation and evolution of galaxies. In this study we focus on three major mechanisms related to baryons that impact the mass distribution in galaxies: adiabatic contraction, reionization, and tidal disruption. Adiabatic contraction due to gas cooling and condensation leads to an increase of the density in the galaxy center. Reionization not only ionizes and evaporates gas from galaxies, but can also prevent gas accretion from the intergalactic medium (IGM). Tidal truncation from gravitational interactions can result in the removal and redistribution of both dark and baryonic matter components.

The relative impact of these process on galaxy properties and their evolution depends on the galaxy mass. Based on the properties found in Section~3, the subhalos in our simulations can be categorized into three main groups according to their $v_{\rm max}$ (or alternatively mass) at $z = 0$. The first group consists of massive subhalos with $v_{\rm max} > 35~\kms$ (${M_{\rm sub}} > 4\times 10^{9}\, \Msun$), where adiabatic contraction tends to increase the amount of DM within the virial radius. These subhalos usually form stars and are therefore ``bright''. The second group are the least massive ones with $v_{\rm max} < 20~\kms$ (${M_{\rm sub}} < 10^{9}\, \Msun$), and are mostly ``dark'' with little or no star formation. These small subhalos are affected by reionization. The third group are subhalos with intermediate masses with $v_{\rm max} \sim  20~\kms$ -- $35~\kms$. These subhalos show signs of a competition between adiabatic contraction and tidal disruption. While adiabatic contraction is able to increase the $v_{\rm max}$ of these subhalos, they often suffer from strong tidal effects in the Hydro simulation that remove both DM and baryonic mass, thus effectively reducing $v_{\rm max}$ once they are close enough to the central galaxy. In what follows we will show some examples of these evolutionary paths.

\subsubsection{The Role of Adiabatic Contraction}

\begin{figure*}
\begin{center}
\includegraphics[width=\linewidth]{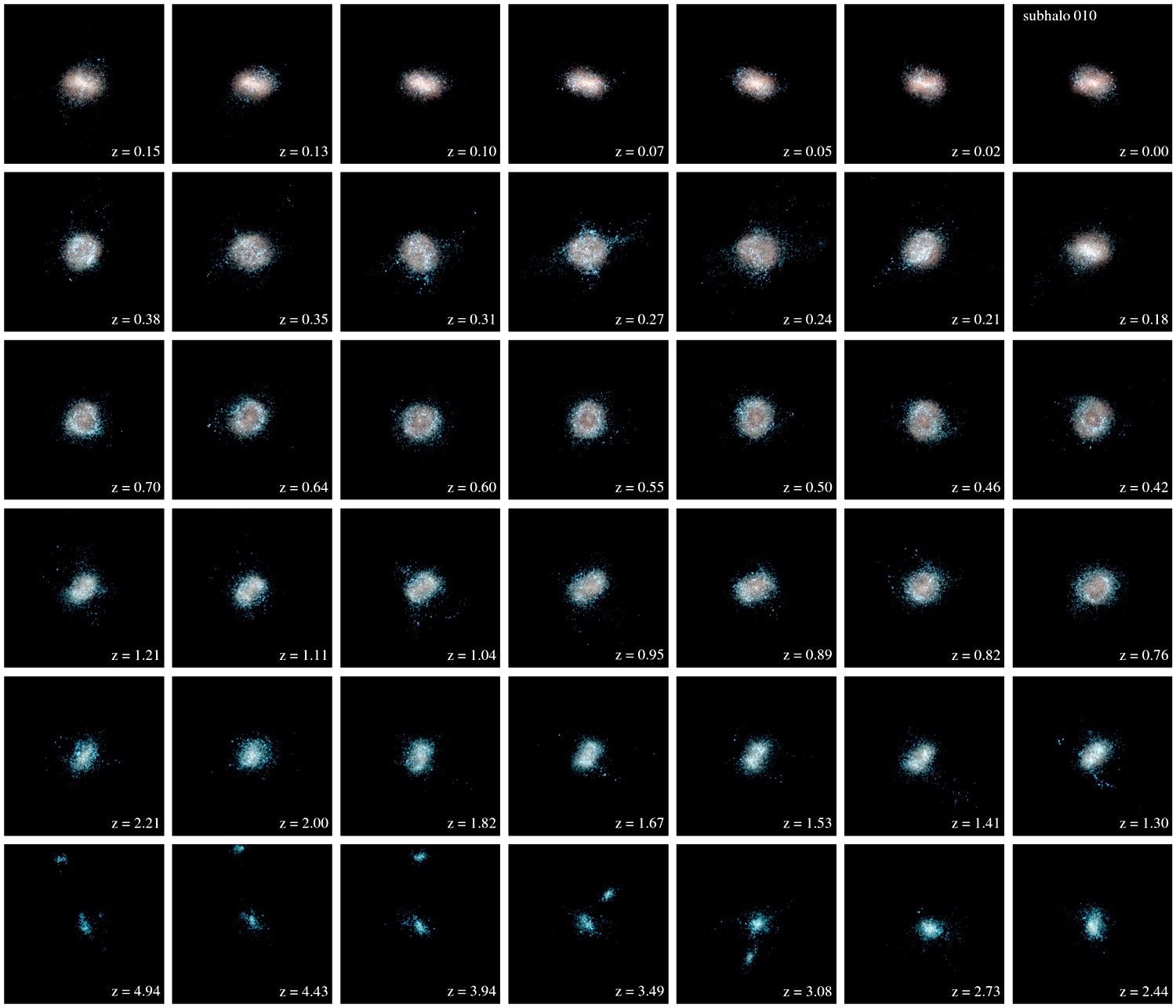}
\includegraphics[width=\linewidth]{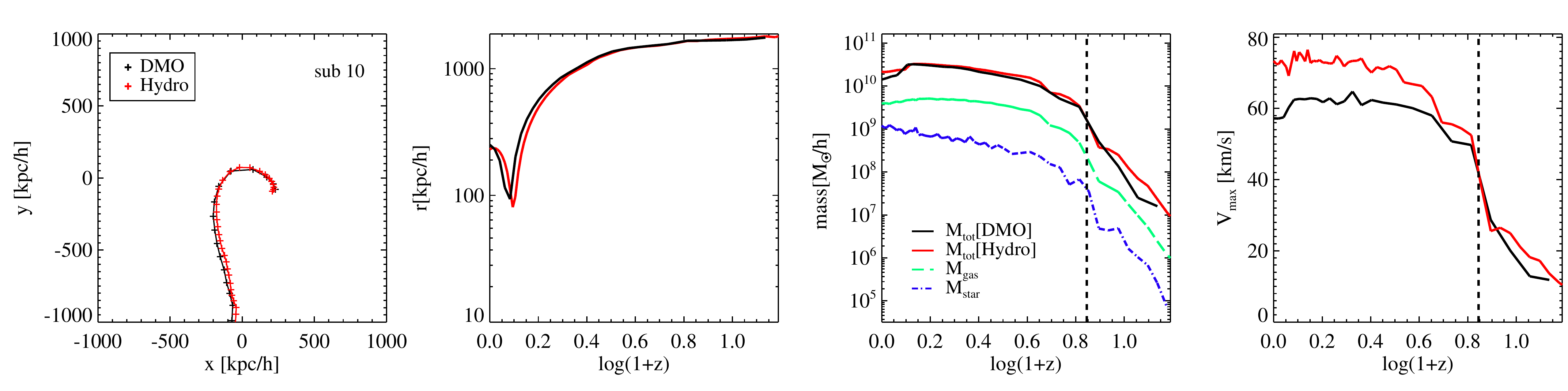}
\caption{\label{fig:sub_10} Time evolution of the most massive subhalo (Sub~10) in the simulations. Top panels: projected density map of the stellar component from redshift $z \sim 5$ to $z=0$. The composite images have used RGB colors mapped from $K$, $B$ and $U$ bands, respectively. The lower panels show (from left to right) the orbit of Sub~10 in the $x$-$y$ plane and its distance to the center of the MW halo, as well as the evolution of its total mass and maximum circular velocity $v_{\rm max}$. As in previous plots, the DMO and Hydro simulations are represented by black and red colors, respectively. In the last two panels, a vertical line represents the end of reionization in the Hydro simulation.}
\end{center}
\end{figure*}

\begin{figure} 
\begin{center} \includegraphics[width=\linewidth]{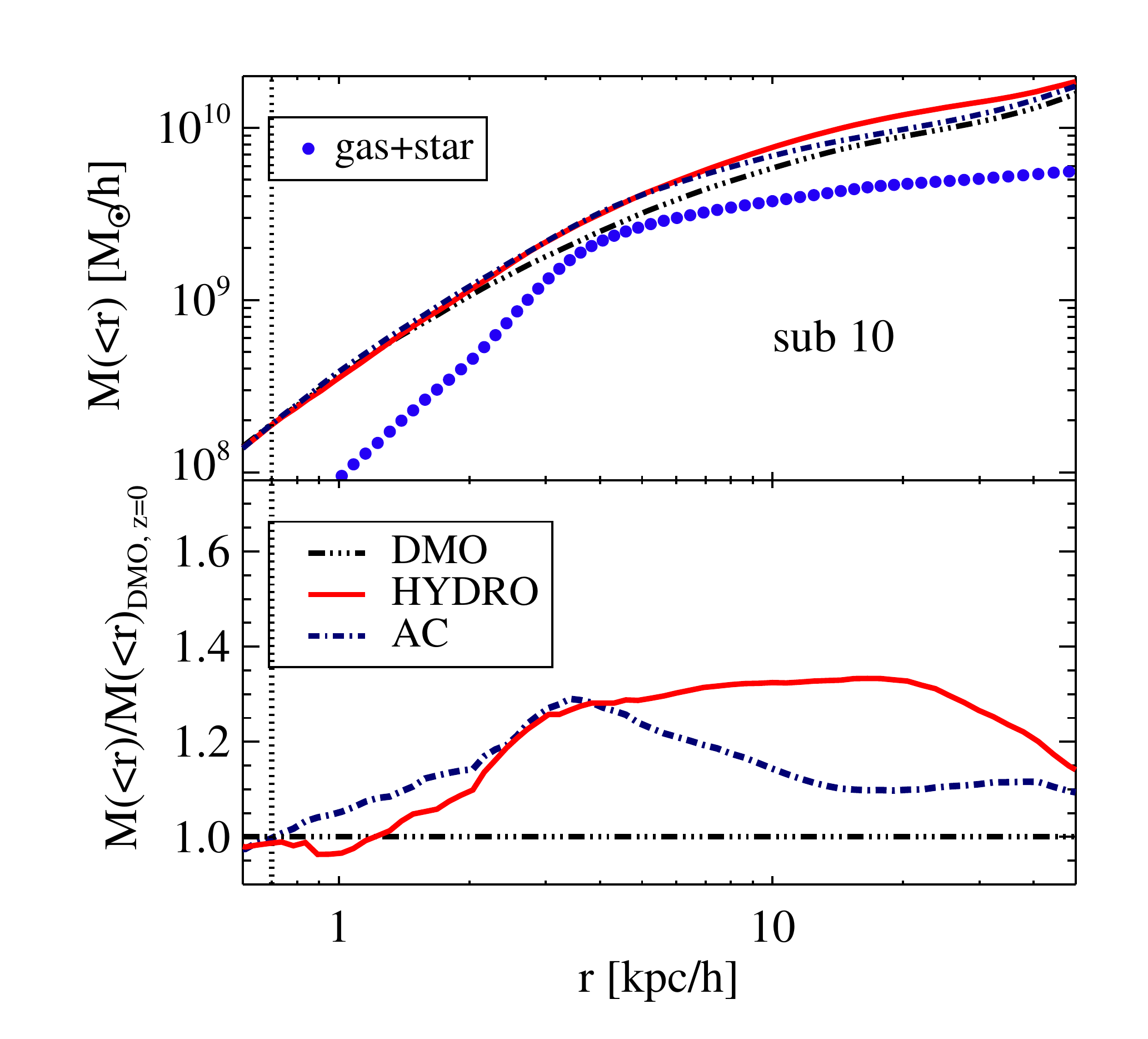}
\includegraphics[width=\linewidth]{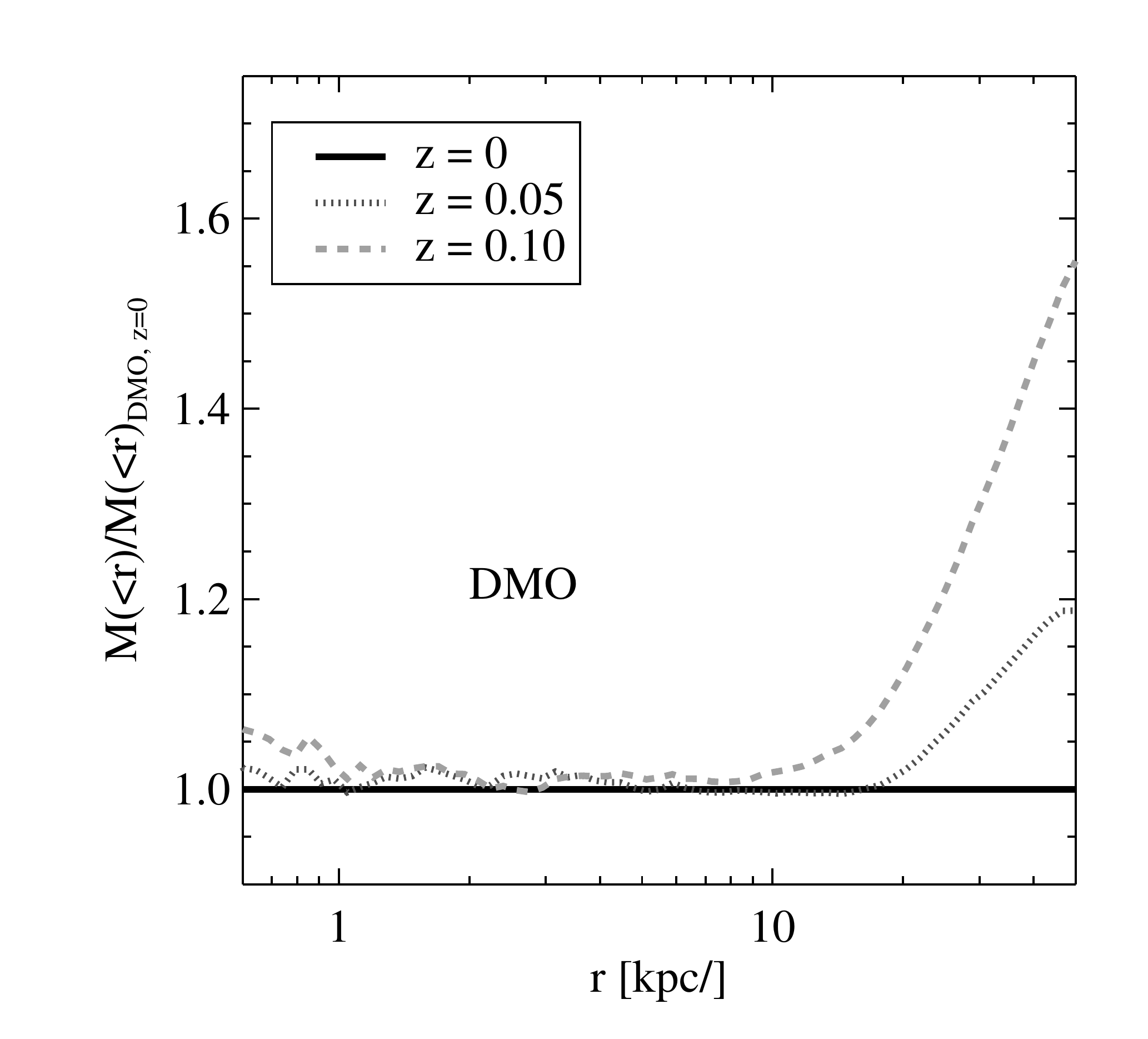}
\caption{\label{fig:adiabatic_contraction_subhalo} 
{\it Top panel:} Comparison of the radially enclosed DM mass profiles of Sub~10 from different simulations (top panel).  
The red solid, black dash-double-dotted and blue dashed-dotted curves represent the Hydro simulation, the DMO simulation and {\small\sc contra} calculation, respectively, and the total enclosed baryonic mass (gas and stars) is also shown (blue filled circles). A factor of $\rm{\frac{\Omega_{m} - \Omega_{b}}{\Omega_{m}}}$ is applied to the density profile from the DMO simulation. This plot shows that Sub~10 retains more mass in the outer region in the Hydro simulation than its DMO counterpart.  {\it Bottom panel:} The radially enclosed DM mass profiles of Sub~10 in the DMO simulation at different evolution times. This subhalo has a pericentric passage at redshift $z = 0.10$ in both simulations. In the DMO simulation, this subhalo experienced substantial mass loss in the outer region which explains the difference between the two mass profiles in the top panel beyond 10 kpc. }
\end{center}
\end{figure}

As we have seen from Figure~\ref{fig:core_cuspy}, three subhalos (Sub~10, 19 and 43) show signs of contracted DM in the Hydro simulation. Subhalo~10 is the most massive  among them, with $v_{\rm max} = 73\,  \kms$ (in the Hydro simulation). Figure~\ref{fig:sub_10} shows the assembly history and dynamical evolution of Sub~10. Not surprisingly, this object is able to fuel star formation continuously from high redshift to the present day, as shown by the ``blue'' stars in the composite images of the figure. Both the Hydro and DMO simulations produce similar  trajectories for Sub~10, which is simply a single fly-by at $z \sim 0.3$. A slight reduction of the total mass and $v_{\rm max}$ is evident in both simulations due to this fly-by; but overall, the effect of baryons on this object is dominated by adiabatic contraction, making it more resilient to disruption.  It builds up mass steadily, and it reaches a total mass of $2.1\times{10^{10}}\, \Msunh$ at $z = 0$ in the Hydro simulation, about $50\%$ more massive than its counterpart in the DMO simulation. 
It is interesting to note that the $v_{\rm max}$ of Sub 10 in the Hydro simulation is substantially larger than that in the DMO simulation throughout the time despite that its total mass is almost identical in the two runs.  This suggests that the contraction of DM 
within Sub 10 has been established at an early stage  if adiabatic contraction is indeed responsible for the enhanced DM 
density profile shown in the top panel of Figure 10. Moreover, in the Hydro simulation there is a non-negligible contribution 
of the baryons to the total gravitational potential of Sub 10 in the inner regions which also helps explaining the different values of $v_{\max}$ 
(see Figure~\ref{fig:adiabatic_contraction_subhalo}).

To demonstrate the effect of adiabatic contraction in subhalos, Figure~\ref{fig:adiabatic_contraction_subhalo} shows a comparison of the radially enclosed DM mass of Sub~10 from both Hydro and DMO simulations against prediction from the {\small\sc contra} code (top panel), and its evolution from the DMO simulation (bottom panel). In the inner region between 1 and 4 kpc, the Hydro simulation and the {\small\sc contra} calculation show similar enhancements of the enclosed DM mass by $\sim 25\%$ due to baryons. Beyond 4 kpc, the enclosed DM mass in the Hydro simulation is consistently higher than the expectation from {\small\sc contra}. 

We find that the discrepancy between the Hydro and DMO simulations is mostly due to tidal removal of the loosely bound material in the outer parts of Sub~10 in the DMO simulation, as evidenced by the change of radial enclosed DM mass with time in Figure~\ref{fig:adiabatic_contraction_subhalo} (bottom panel). Sub~10 experiences continuous mass loss in the outer region after the pericentric passage at redshift $z = 0.10$.  Although mass loss also occurs in Sub~10 in the Hydro simulation, as shown in the mass evolution in Figure~\ref{fig:sub_10}, the amount is much smaller than that in the DMO simulation. These results show that when baryonic effects are included, massive systems similar to Sub~10 become more resilient to tidal disruption since adiabatic contraction and the presence of baryons in the inner regions tend to increase the binding energy.

Similar effects of adiabatic contraction could also help explain the survival of
the bright satellites of galaxy clusters reported for the Illustris Simulation
\citep{Vogelsberger2014a}.   It was found that those galaxies (with
stellar mass $\sim 10^{10}\, \Msunh$, much more massive than Sub~10,
the largest dwarf in our simulation) are more resilient to tidal
disruption in the central cluster regions than satellites in pure
$\textit{N}-$body simulations due to the increased concentration of
DM and stellar components, in agreement with our findings in this
study.

The role of adiabatic contraction becomes progressively less important for lower mass  dark matter halos/subhalos, as we have shown in Figure~\ref{fig:core_cuspy}. In particular, for subhalos with $v_{\rm max} < 35\ \kms$ the total amount of baryons (mostly in the form of  cool gas and stars) no longer plays a substantial gravitational role in these systems. We also note that our  simulations include only a few massive subhalos, so that some scatter in their properties is inevitable. A more accurate estimate of the galaxy mass at which adiabatic contraction turns ineffective will require a much larger sample of (massive) subhalos than the one analyzed in this study.  

\subsubsection{The Role of Reionization}

\begin{figure*} 
\begin{center} \includegraphics[scale=0.33]{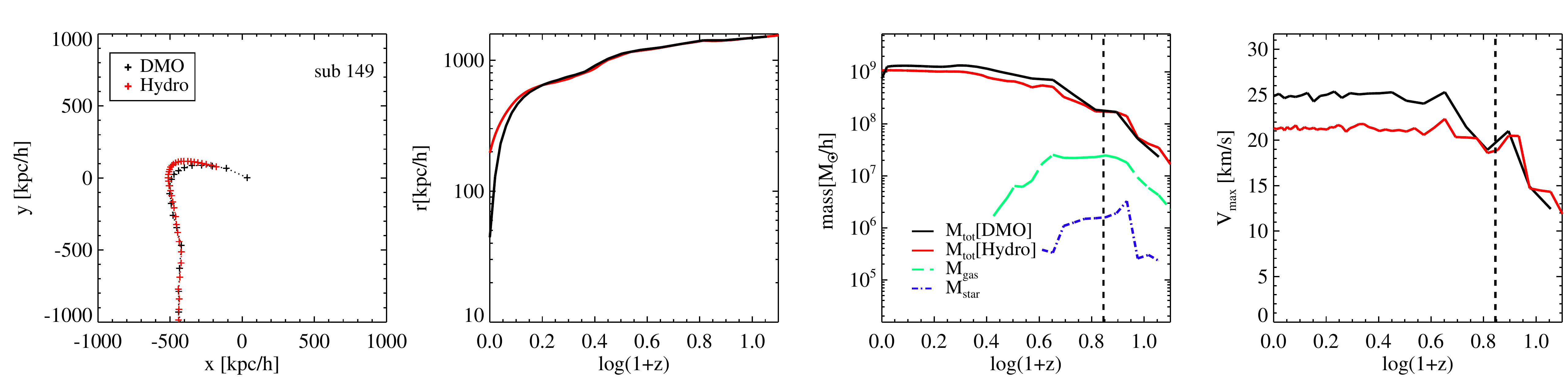} 
\includegraphics[scale=0.33]{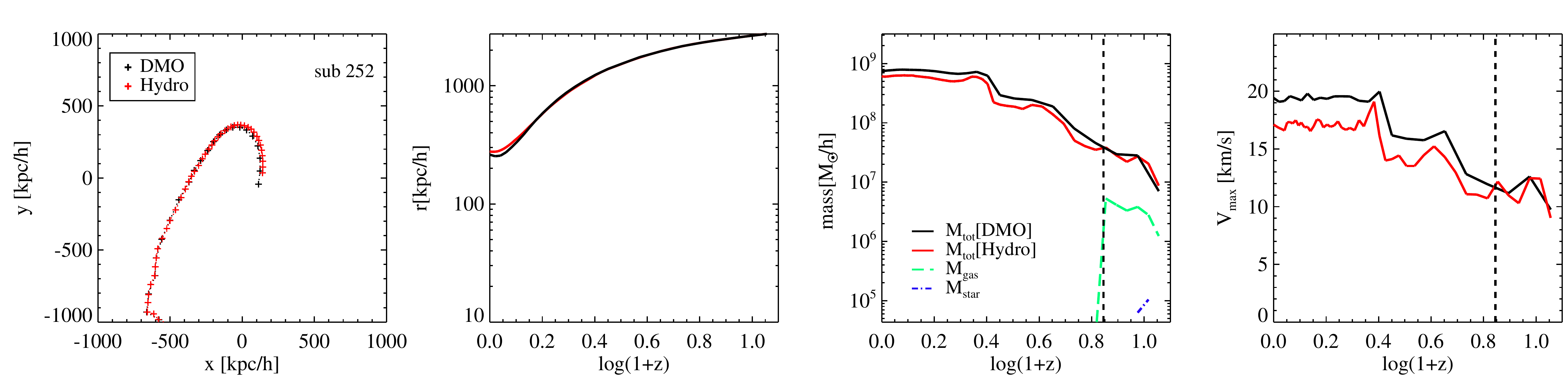} 
\includegraphics[scale=0.33]{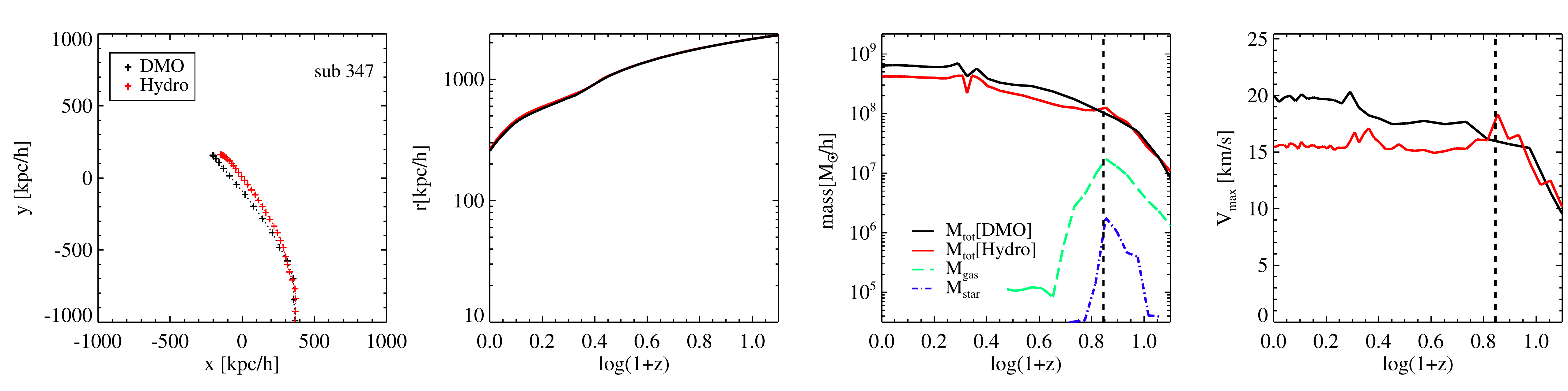} 
\includegraphics[scale=0.33]{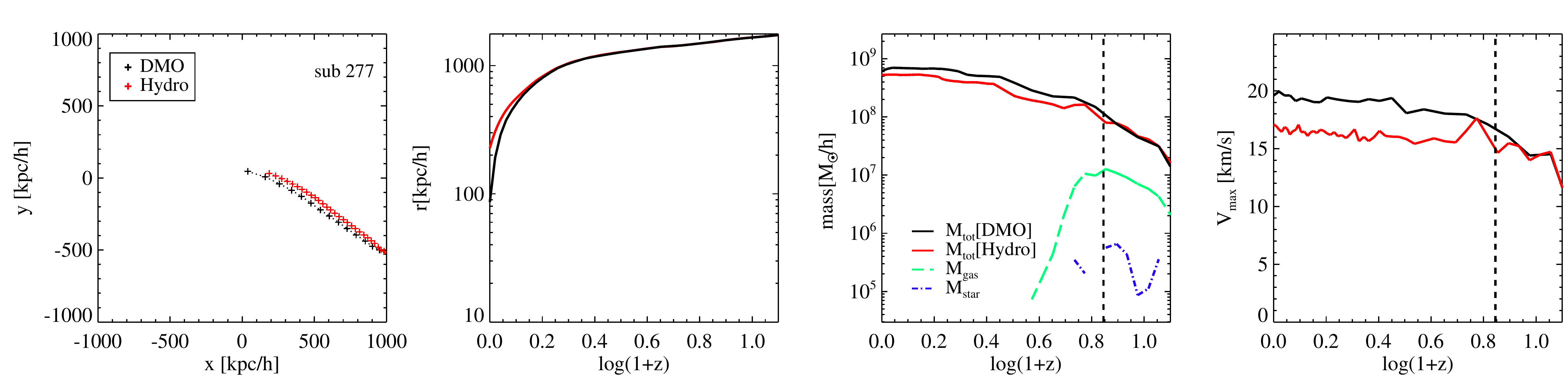} 
\includegraphics[scale=0.33]{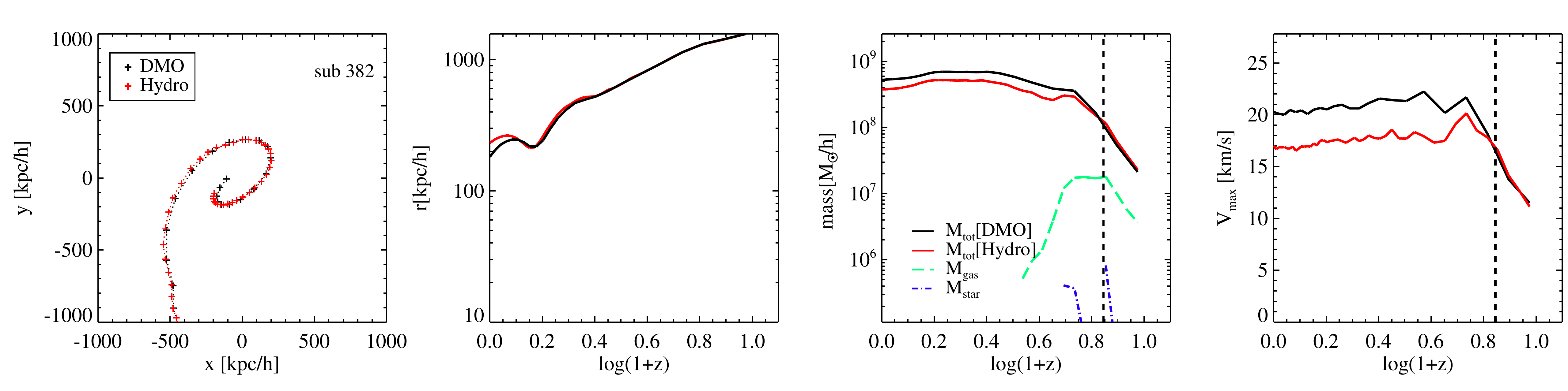} 
\caption{\label{fig:sub_149} The evolution of five low-mass subhalos ($v_{\rm max} < 20~\kms$) from the simulations (namely Sub 149, 252, 347, 277 and 382) in terms of their orbits in the $x$-$y$ planes and their distance to the center of the MW halo at different redshifts (left two columns). Also shown are their growth histories in mass and circular velocity as a function of redshift (right two columns). As in previous plots, the DMO and Hydro simulations are represented by black and red colors, respectively. In the two columns on the right, the total, gas and stellar masses are represented by solid, green dashed-dotted, and blue dotted lines, respectively, and the vertical dashed line indicates the end of reionization at $z=6$.}
\end{center}
\end{figure*}

\begin{figure*}
\begin{center}
\begin{tabular}{cccc}
\resizebox{1.75in}{!}{\includegraphics{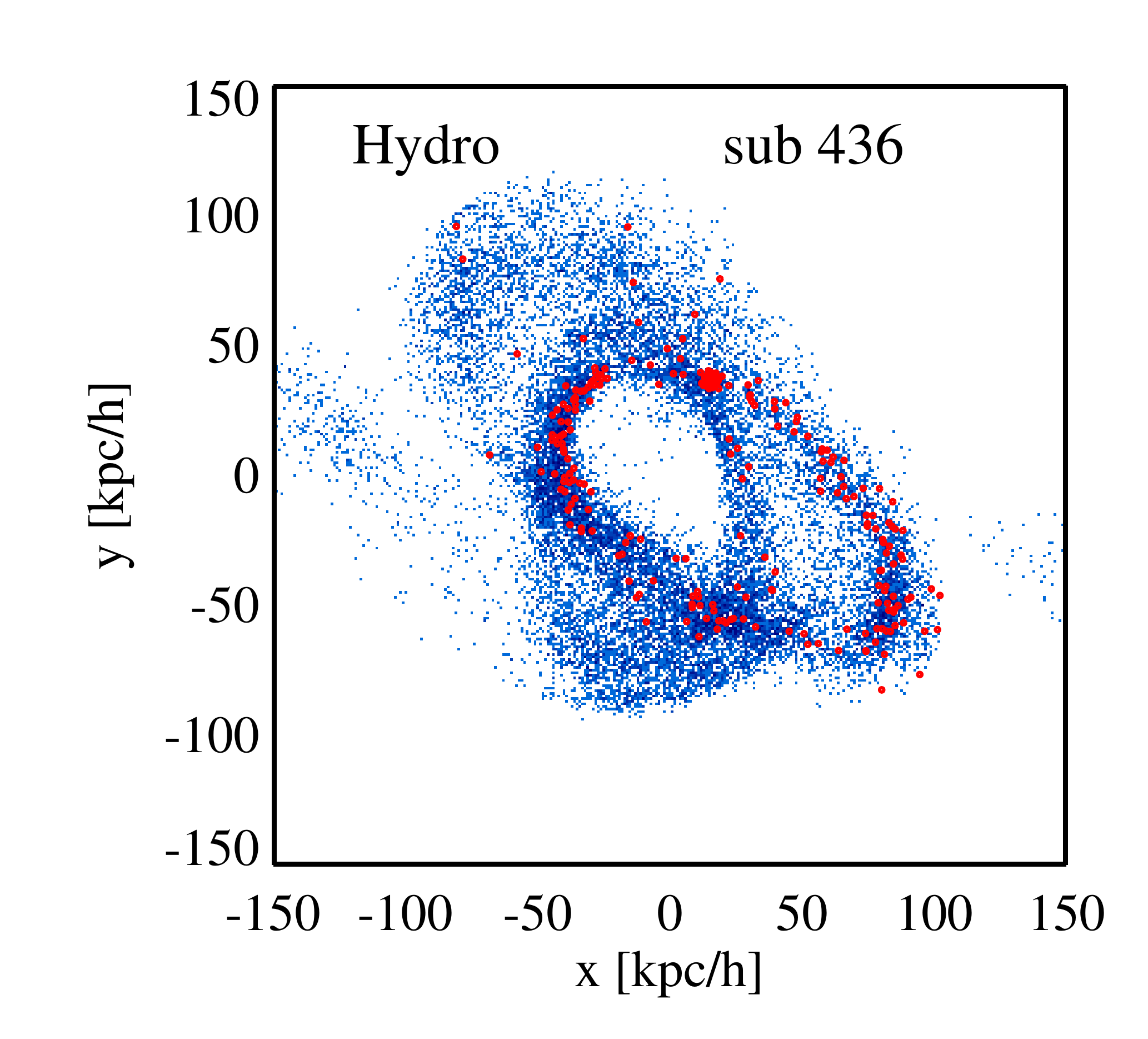}}
\resizebox{1.75in}{!}{\includegraphics{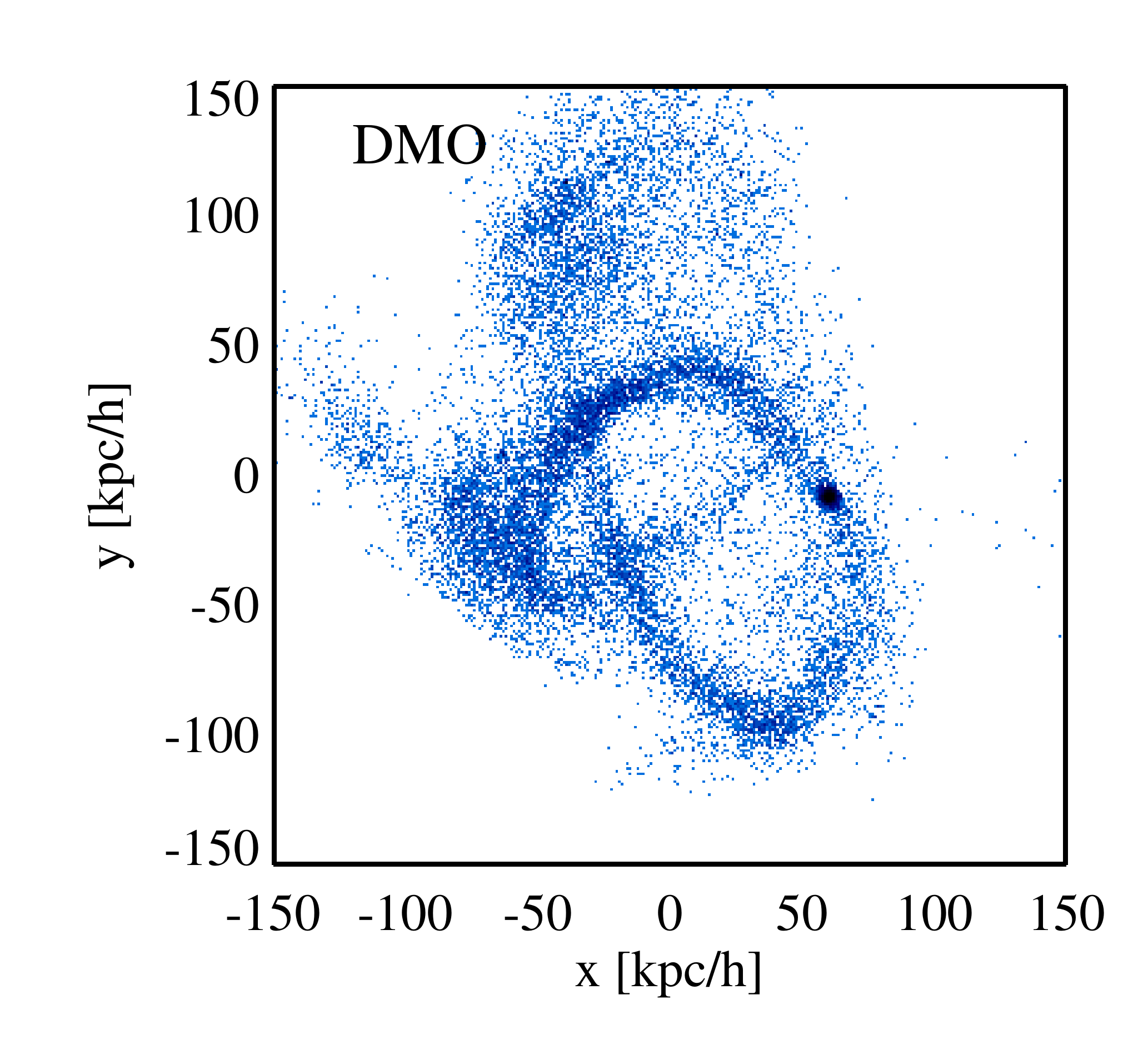}}
\resizebox{1.75in}{!}{\includegraphics{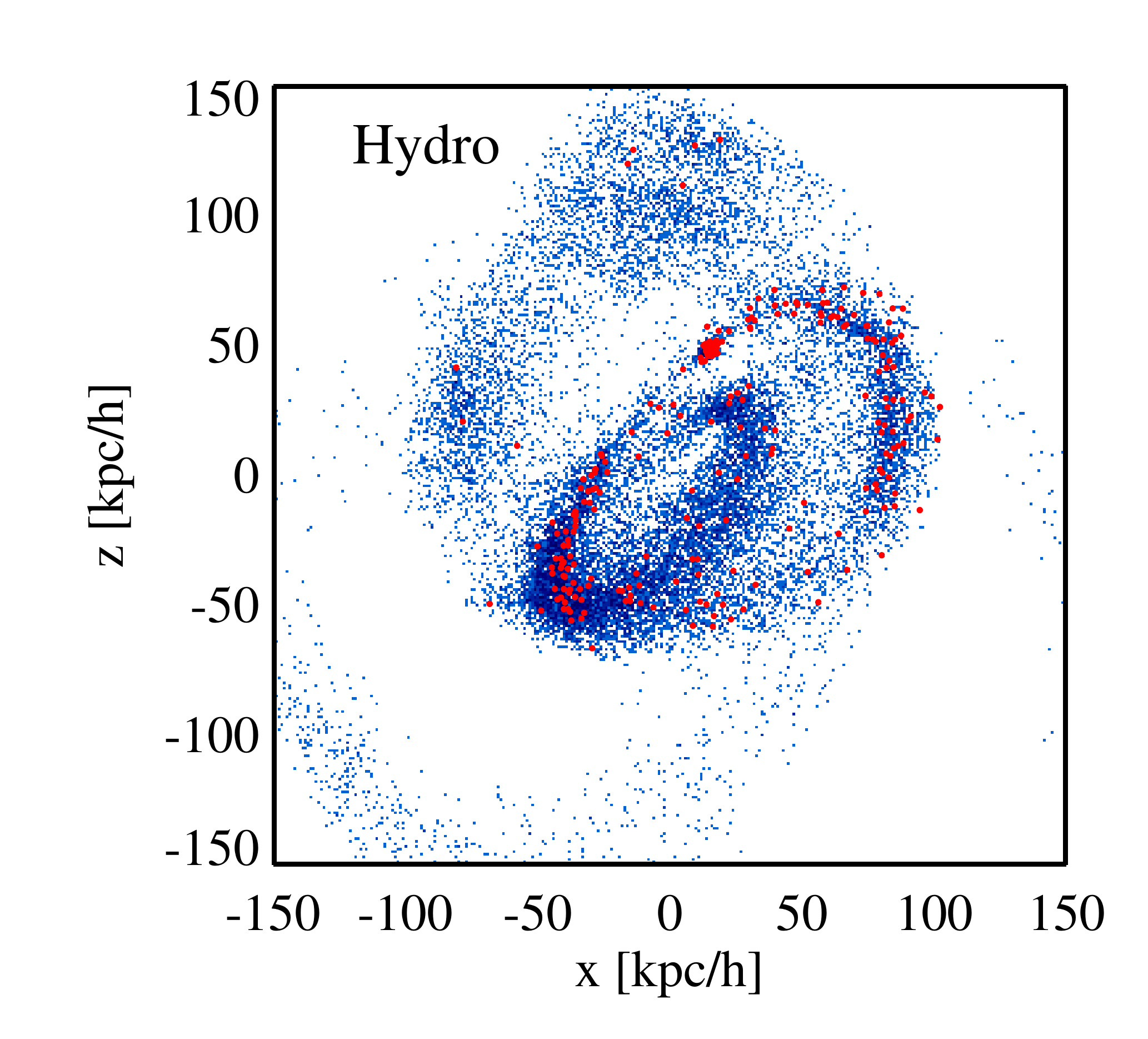}}
\resizebox{1.75in}{!}{\includegraphics{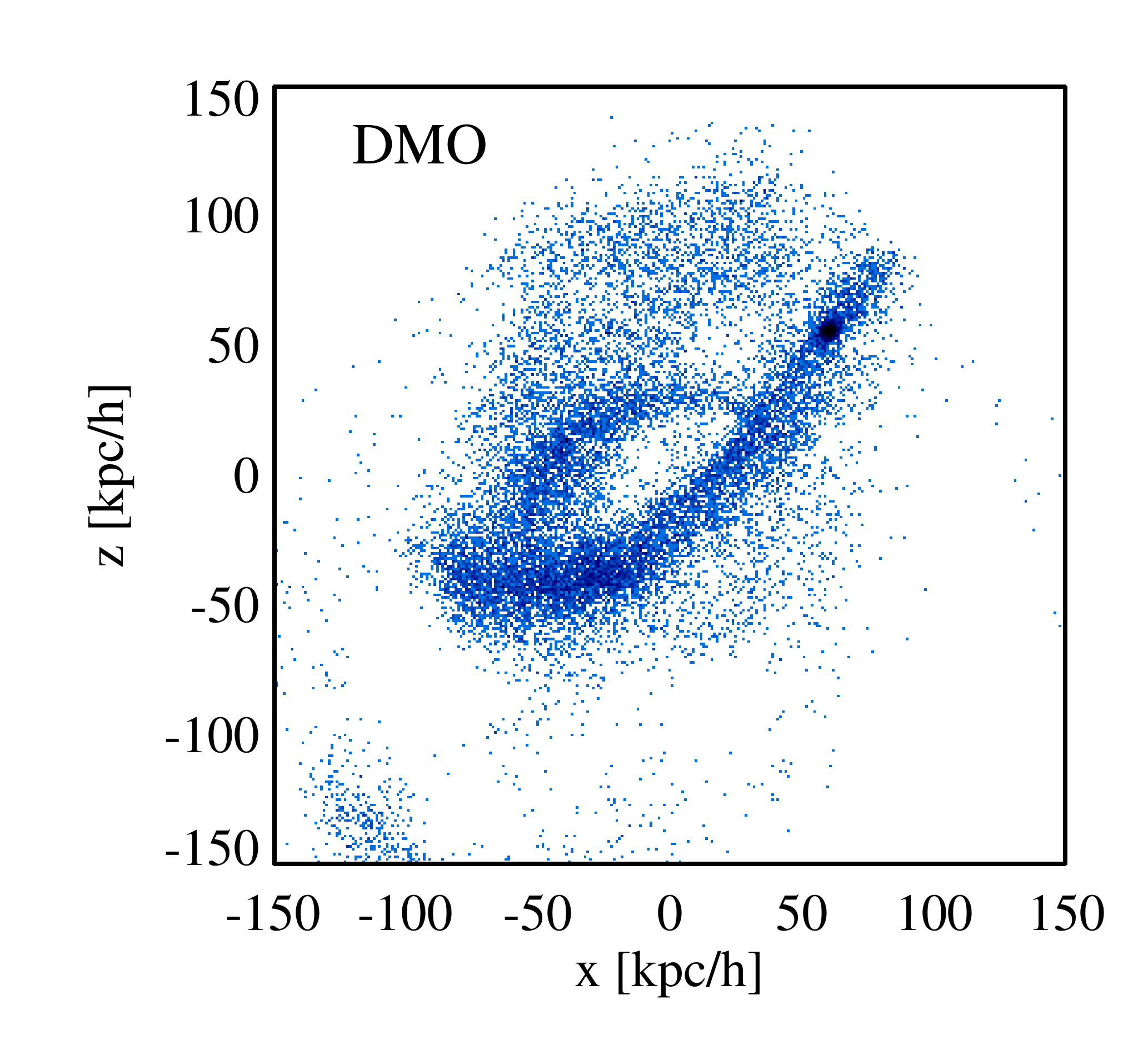}}
\end{tabular}
\begin{tabular}{cccc}
\resizebox{1.75in}{!}{\includegraphics{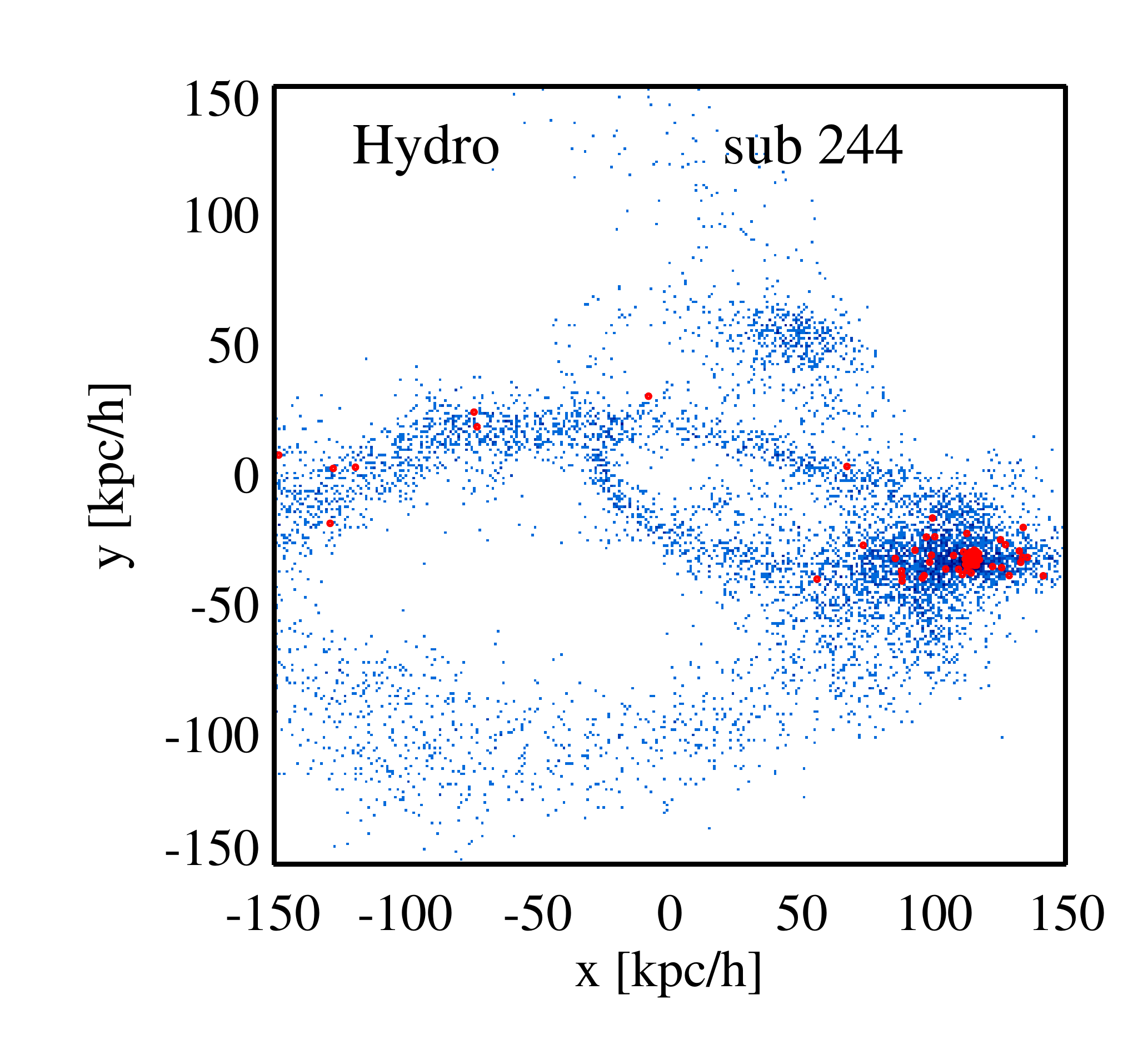}}
\resizebox{1.75in}{!}{\includegraphics{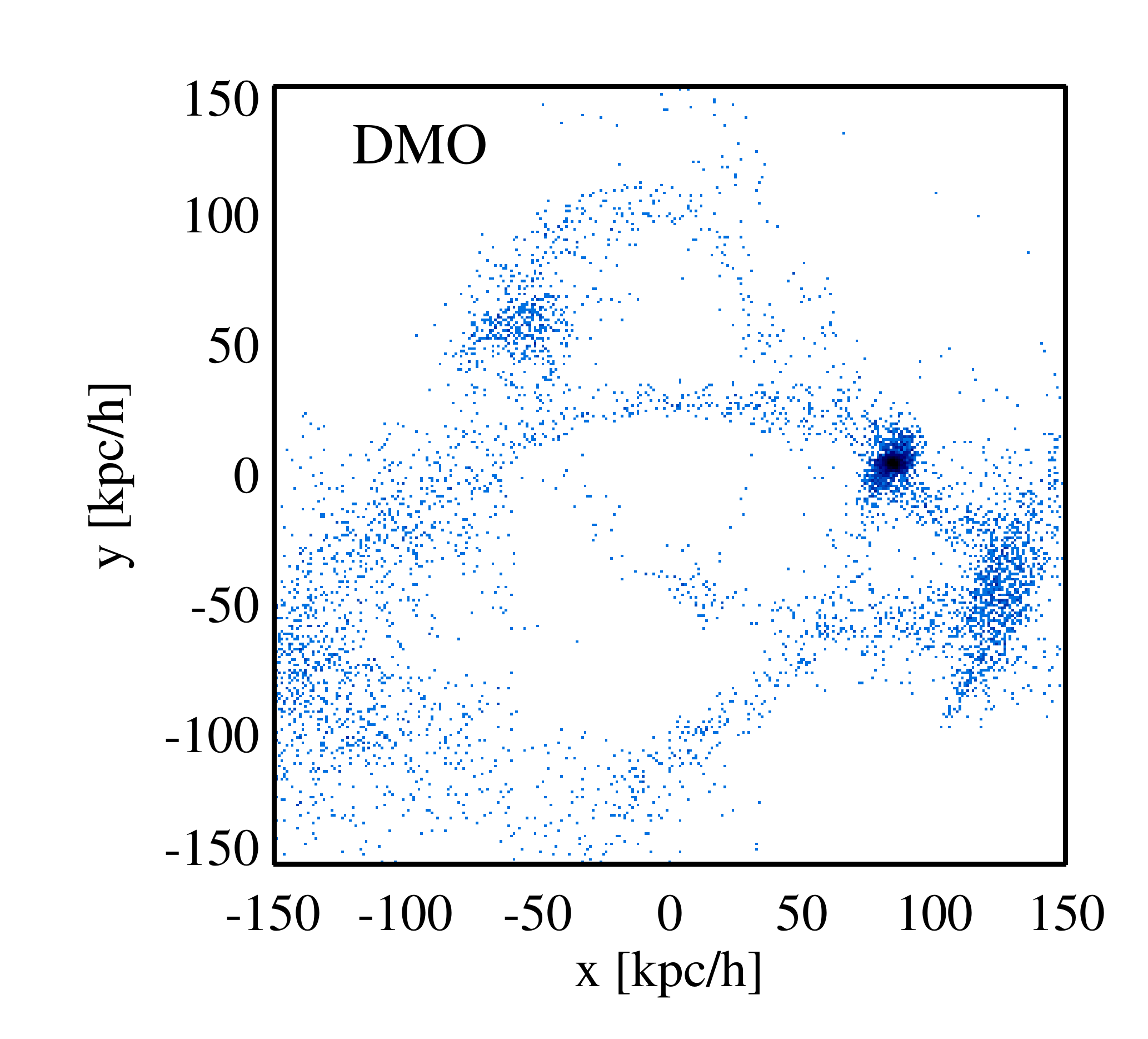}}
\resizebox{1.75in}{!}{\includegraphics{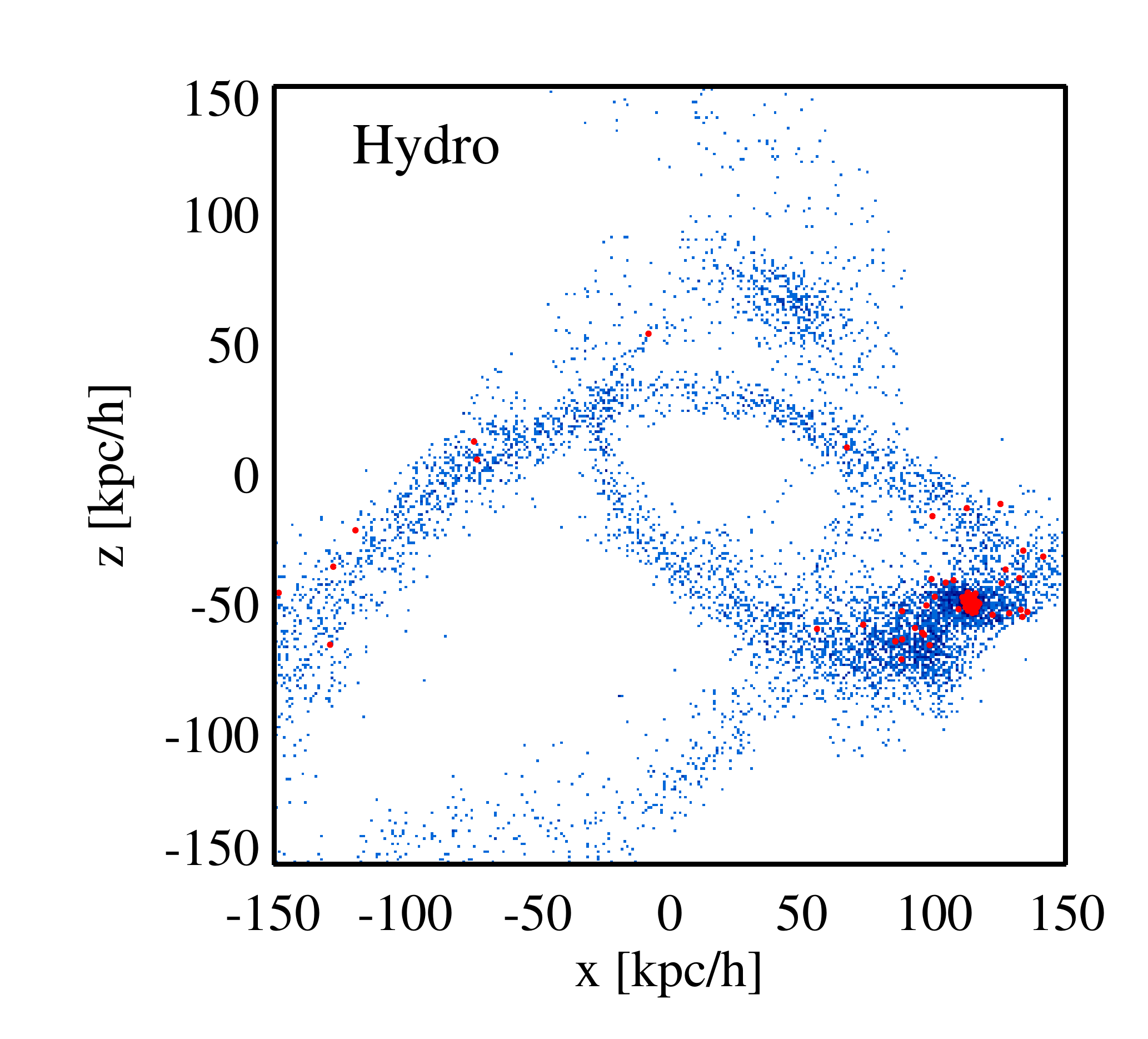}}
\resizebox{1.75in}{!}{\includegraphics{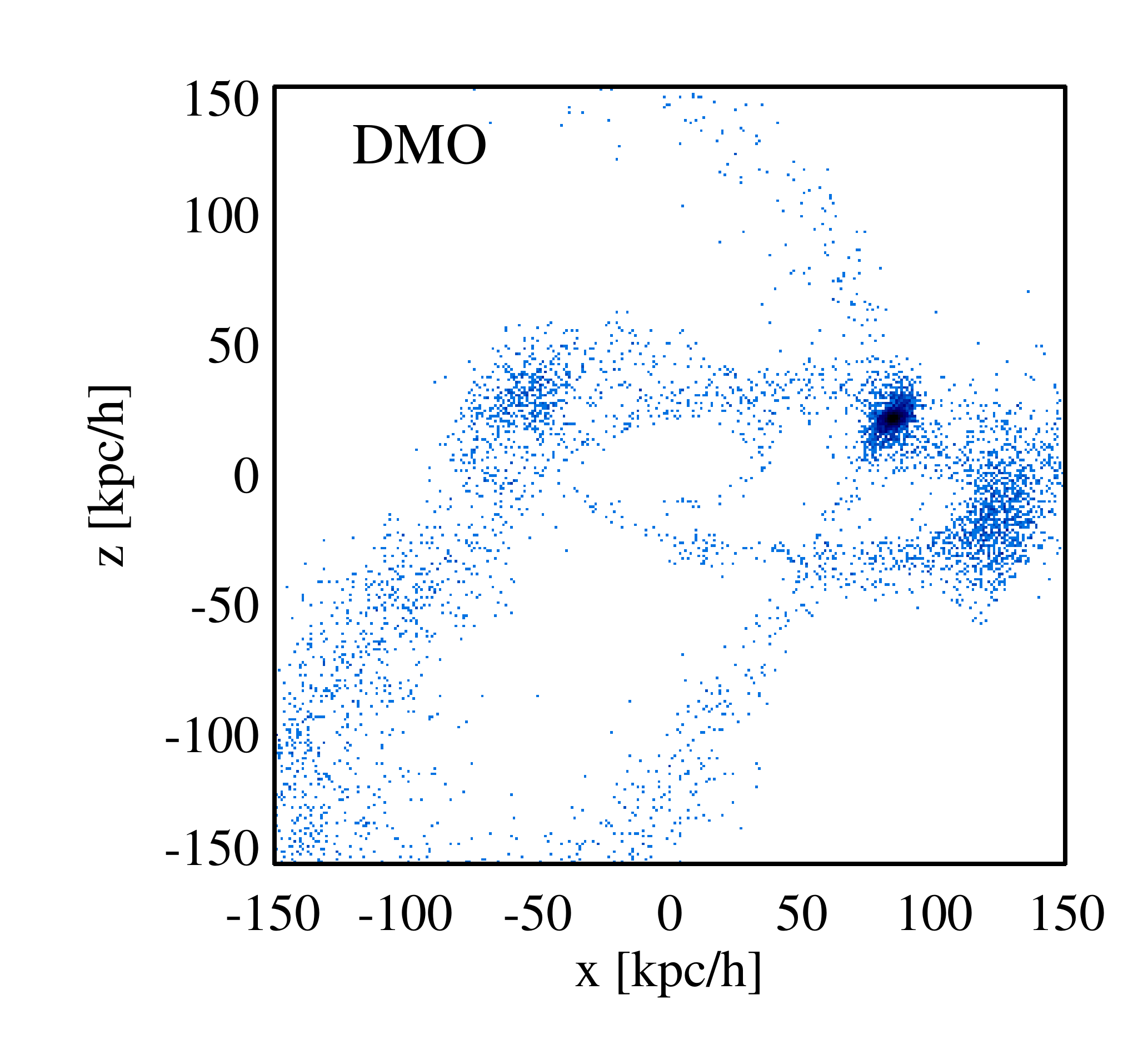}}
\end{tabular}
\caption{\label{fig:tidal_436} Projected maps of the dark matter distribution of two bright satellites (intermediate-mass subhalos with $20\,\kms < v_{\rm max} < 35\,\kms$), namely Sub 436 (top panels) and 244 (bottom panels), at $z = 0$ from both the Hydro and DMO simulations. The left two columns show projections in the $x$-$y$ plane, while the right two columns show the projections in the $x$-$z$ plane. The blue dots represent the dark matter particles, while red dots show stars in the Hydro simulation. The presence of both dark matter and stellar streams in these plots are clear signs of tidal truncation.
}
\end{center}
\end{figure*}

The epoch of reionization is an important landmark event in cosmic history during which photons from young stars or accreting black holes ionize the neutral hydrogen. The latest results of Planck \citep{Planck2015} indicate that the Universe was 50\% reionized at $z \approx 9$, while Gunn-Peterson absorption features in quasar spectra suggest that reionization began as early as $z \sim 14$ and ended at $z \sim 6$ \citep{Fan2006}. 

The majority of the low-mass subhalos in our simulations are unable to form any 
stars due to reionization, thus staying dark. In Figure~\ref{fig:sub_149}, we 
show the evolutionary histories of five of such dark subhalos, Sub 149, 252, 
347, 277 and 382. The trajectories show that they have been recently accreted 
onto the main halo. For each subhalo, the trajectory from the Hydro simulation 
is close to that of the DMO simulation, albeit with some small deviations.  
This is expected since the gravitational potential of the central host is 
modified in the hydrodynamic simulation due to the presence of the stellar disk, 
and the gas ram pressure, which is absent in collisionless $\textit{N-}$body 
simulations,  introduces some additional offsets in orbital phase space.

The growth history of each subhalo shows remarkable differences between the DMO and Hydro simulations after $z = 6$. The subhalo mass from the Hydro simulation is consistently lower than that from the DMO simulation. It is clear that the gas content of the subhalos declines rapidly after reionization, because the shallow gravitational potential of these objects both fails to retain the heated gas and to accrete new gas from the IGM.

A comparison of the growth histories between Figure~\ref{fig:sub_149} and Figure~\ref{fig:sub_10} shows that only the more massive subhalos are able to retain their gas after reionization, likely due to the fact that the densest gas regions in these objects can still reach the critical density needed for self-shielding from the ionizing UV background. These subhalos also have a sufficiently deep gravitational potential to accrete new gas and sustain star formation. These results are consistent with those obtained by \cite{Onorbe2015} for simulations of field dwarf galaxies, and demonstrate that reionization significantly suppresses the formation of low-mass galaxies.

\subsubsection{The Role of Tidal Disruption}

The third group we consider consists of bright satellites (intermediate-mass 
subhalos with $20\,\kms < v_{\rm max} < 35\,\kms$) similar to the dwarf 
spheroidal galaxies near the MW. We have already shown in 
Figure~\ref{fig:sub_number_density} the effect of increased tidal disruption of 
subhalos in the Hydro simulation, which resulted in a lower subhalo abundance in 
the inner region of the main halo. Below we will examine individual subhalos and 
how they are shaped by tidal forces.

In Figure~\ref{fig:tidal_436},  we map the distribution of the DM components of 
two satellites, Sub 436 and 244, at $z = 0$, both in the Hydro and DMO 
simulations. For the Hydro simulation we also examine the stellar distribution. 
The original members of each subhalo are identified at the 
redshift when their $v_{\rm max}$ reached the peak value.  Overall, the 
trajectories of the matched subhalos are similar in the Hydro and DMO 
simulations, but with some subtle differences. For example, the current 
positions of the two subhalos in the DMO simulation lag slightly behind those in 
the Hydro simulation. This may be caused by the deeper potential well of the main halo from a more contracted DM density profile and 
the presence of the stellar disk, which accelerate 
the subhalos close to the host to move at a slightly higher speed in the Hydro simulation than in the DMO 
one.  However, the most striking feature visible in the figure is the presence of tidal 
debris and streams of both DM and stars. This is clear evidence of strong 
gravitational interactions and tidal truncation of these two satellites. 

\begin{figure*}
\begin{center}
\includegraphics[scale=0.33]{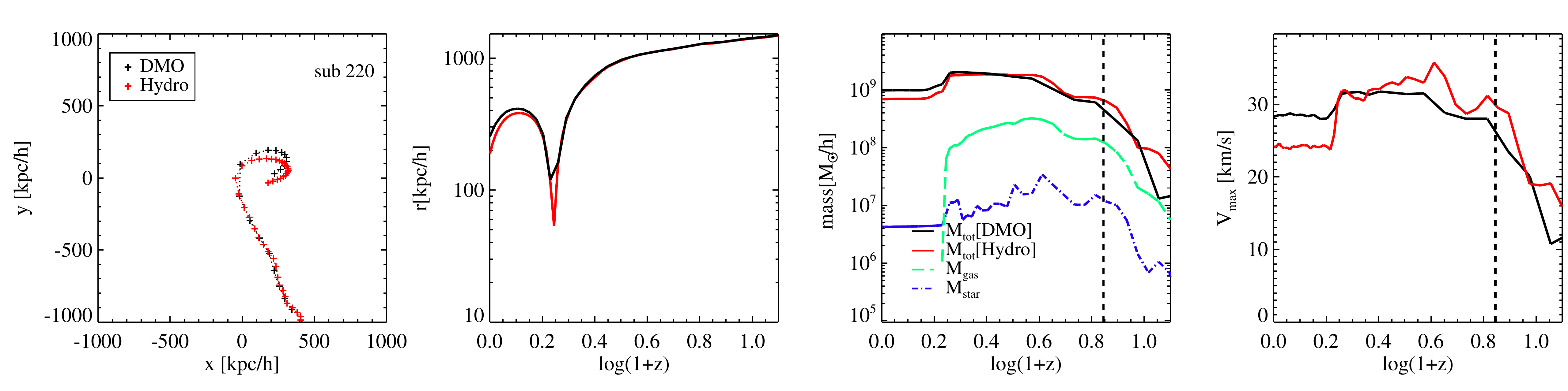}
\includegraphics[scale=0.33]{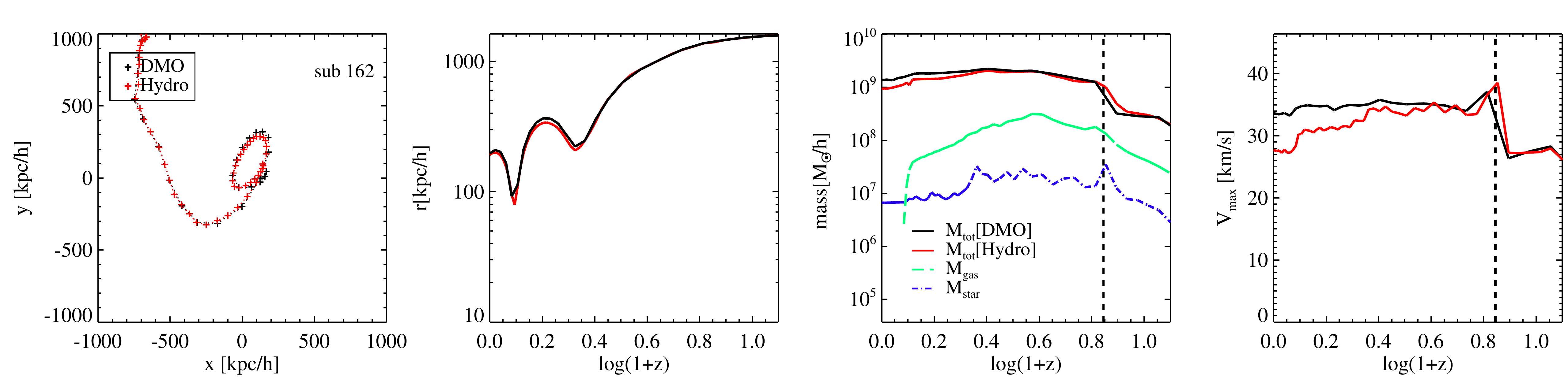}
\includegraphics[scale=0.33]{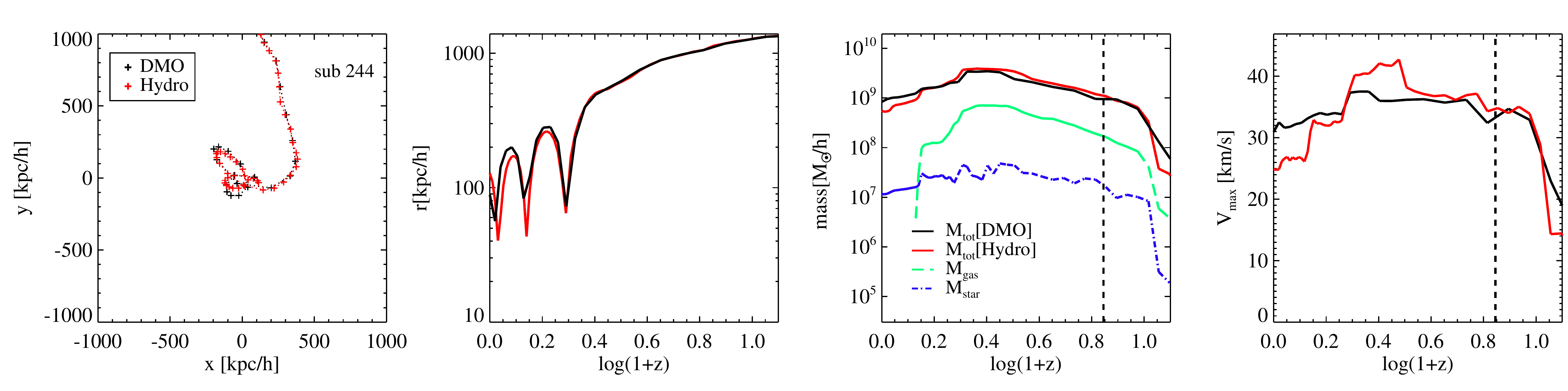}
\includegraphics[scale=0.33]{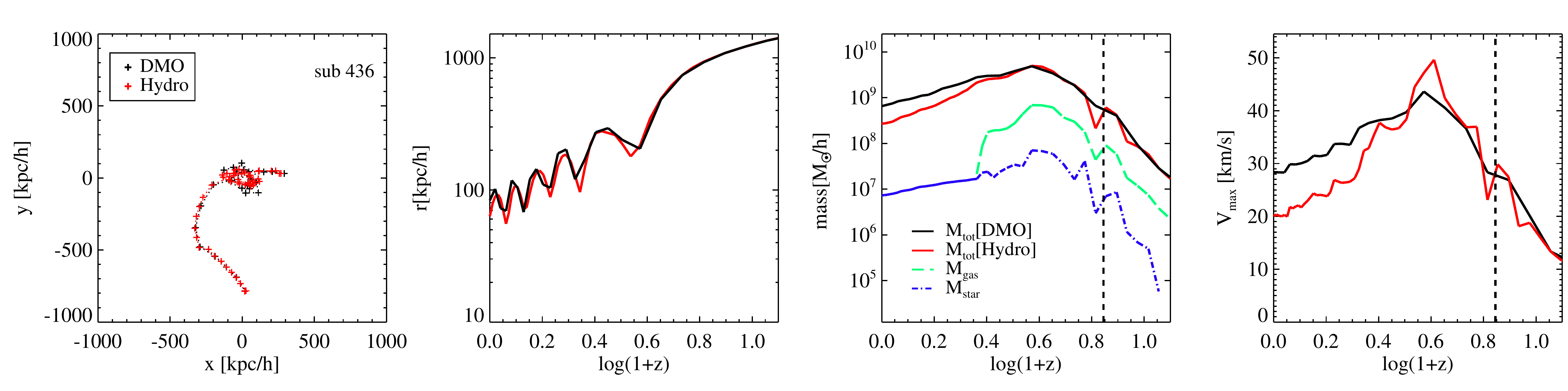}
\caption{\label{fig:sub_436} The evolution of four bright satellites (intermediate-mass subhalos with $20\,\kms < v_{\rm max} < 35\,\kms$) from the simulations (namely Sub 220, 162, 244, and 436) in terms of their orbits in the $x$-$y$ plane and their distance to the center of the MW halo  at different redshifts (left two columns). Their growth histories in mass and circular velocity as a function of redshift are also shown (right two columns). As in previous plots, the DMO and Hydro simulations are represented by black and red colors, respectively. In the two columns on the right, the total, gas and stellar masses are represented by solid, green dashed-dotted, and blue dotted lines, respectively, while the vertical dashed line indicates the end of reionization at $z=6$.}
\end{center}
\end{figure*}

Figure~\ref{fig:sub_436} shows the evolutionary paths and growth histories of four subhalos (Sub 220, 162, 244, 436) that are massive enough to retain gas to fuel star formation in a continuous manner after reionization. The typical stellar mass of these subhalos falls in the range of $10^6 - 10^7\, \Msunh$ at $z = 0$, similar to that of the classical dwarf spheroidal galaxies in the Local Group \citep{Kolchin2012}. These dwarfs have a total mass well above ${10^{9}} \, \Msunh$ before infall to the main host, and they can continue to accrete gas from the IGM after $z = 6$. 

These subhalos experienced strong gravitational interactions with their main host before the final plunges, as shown in their trajectories. These encounters tidally remove both baryons and DM, resulting in a steady reduction of the subhalo mass. The interactions also trigger episodes of starbursts, as shown in the growth curve of the stellar mass, and lead to distinct step-wise fluctuations in the velocity curve. The $v_{\rm max}$ curves experience stronger reductions in the Hydro simulation than in the DMO one. These are characteristic features of tidal forces during pericenter passages of the central galaxy.  These findings are consistent with idealized simulations of tidal disruption of dwarf galaxies by \cite{Penarrubia2008b} and \cite{Arraki2014}.   

In addition, the dwarf galaxies end up gas poor at $z=0$ in the simulation, as 
shown in the evolution curve of the gas mass in Figure~\ref{fig:sub_436}. All 
four dwarfs have experienced an abrupt loss in gas mass during their first 
infall. Gas is loosely bound to the subhalos compared to their stellar and DM 
components. However, tidal forces alone could not completely remove the gas. In 
principle, tidal torques may well funnel some gas into the center of these 
dwarfs and compress it to form a dense component which is difficult to strip 
\citep[][]{Mayer2006}. However, ram-pressure, on the other hand, is able to 
efficiently remove the gas from these subhalos when they pass through the hot 
halo gas of the central galaxy \citep[][]{Mayer2006,Wadepuhl2011, Gatto2013, Arraki2014}. 
Ram-pressure stripping can transform these dwarf galaxies into gas-poor systems, 
and it may also induce other effects. For example, it was suggested by 
\cite{Arraki2014} that  sudden gas loss in a dwarf galaxy due to ram-pressure 
could cause its dark matter to expand adiabatically and thus reduce $v_{\rm 
max}$. However, this effect is small ($< 10\%$ in $v_{\rm max}$) compared to the 
much larger reduction in $v_{\rm max}$ caused by tidal truncation.

\begin{figure*}
\begin{center}
\includegraphics[scale=0.33]{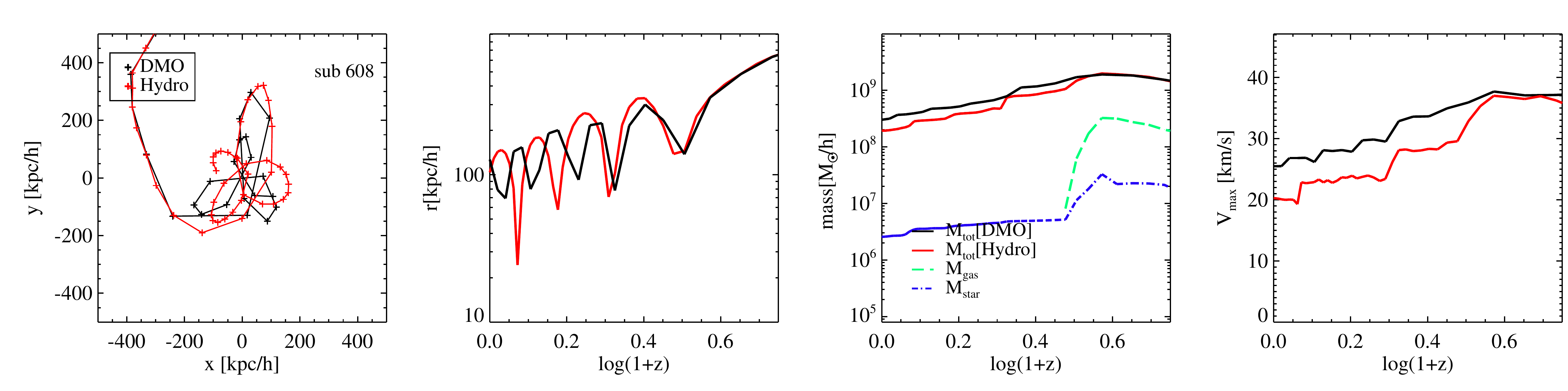}
\includegraphics[scale=0.33]{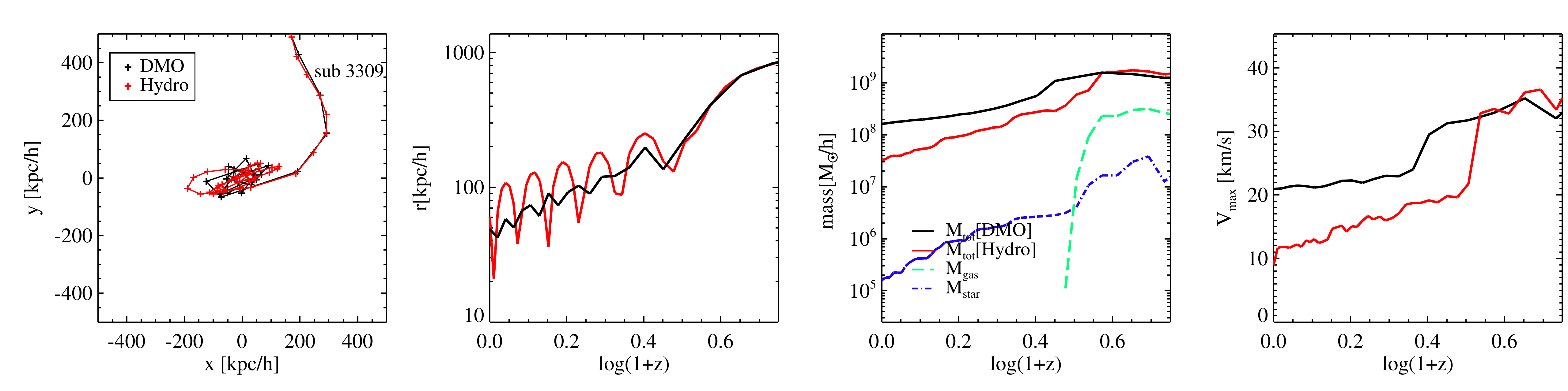}
\includegraphics[scale=0.33]{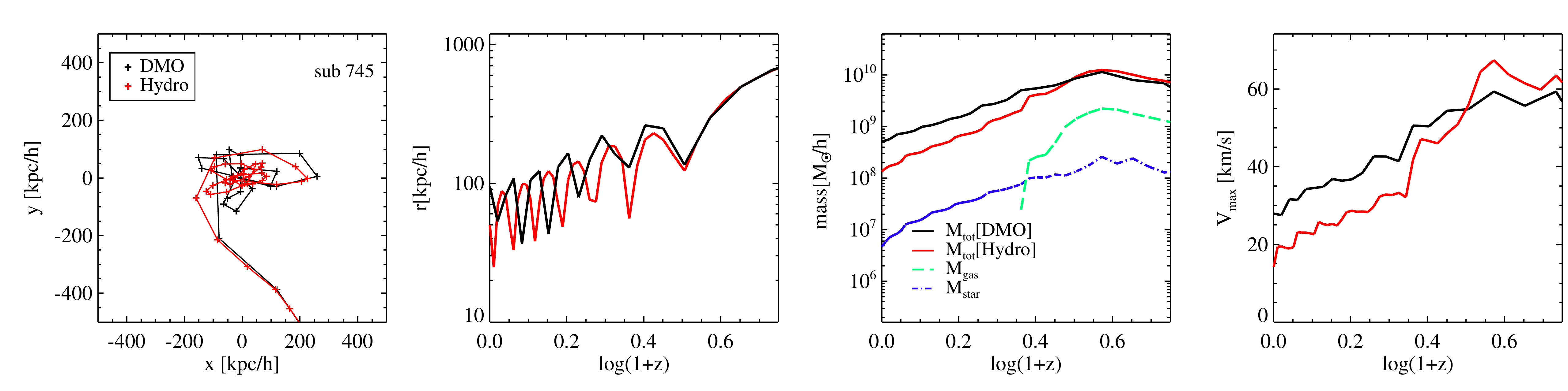}
\caption{\label{fig:sub_3309} Same as Figure~\ref{fig:sub_436}, but for three subhalos (namely Sub 608, 3309 and 745) with closest encounters with the central galaxy at distances comparable to the stellar disk size of $\sim 25$ kpc. As in previous plots, their orbits in the $x$-$y$ plane and their distance to the center of the MW halo at different redshifts are shown in the left two columns, and their growth histories in mass and circular velocity as a function of redshift are shown in  the right two columns. The DMO and Hydro simulations are indicated by black and red colors, the total, gas and stellar masses are represented by solid, green dashed-dotted, and blue dotted lines, respectively. Note that the redshift range $\log(1+z)$ is slightly narrowed compared to Figure~\ref{fig:sub_436} to highlight the temporal evolution of $v_{\rm{max}}$, distance, and mass.}
\end{center}
\end{figure*}

The trends of total mass evolution of subhalos in Figures~\ref{fig:sub_436} show clear tidal disruption caused by the central galaxy. The signature of tidal disruption is also clearly seen in the sharp and distinct decrease of $v_{\rm{max}}$. We show in Figure~\ref{fig:sub_3309} the evolution of three subhalos with even shorter pericentric distances, comparable to the stellar disk size of $\sim 25$ kpc of the central galaxy \citep{Marinacci2014a}. Not surprisingly,  these subhalos experience even more substantial reductions in both mass and $v_{\rm{max}}$ during their close encounters with the host galaxy. In particular, Subhalos 3309 and 745  show the largest reduction in $v_{\rm{max}}$ ($\sim50\%$ of their DMO values) since they have passed the galactic center at much smaller distances  ($\sim 20$  kpc) than the other subhalos. The mass loss in the stellar component is also higher for these three subhalos than those in Figure~\ref{fig:sub_436}.

\begin{figure}
\begin{center} \includegraphics[width=\linewidth]{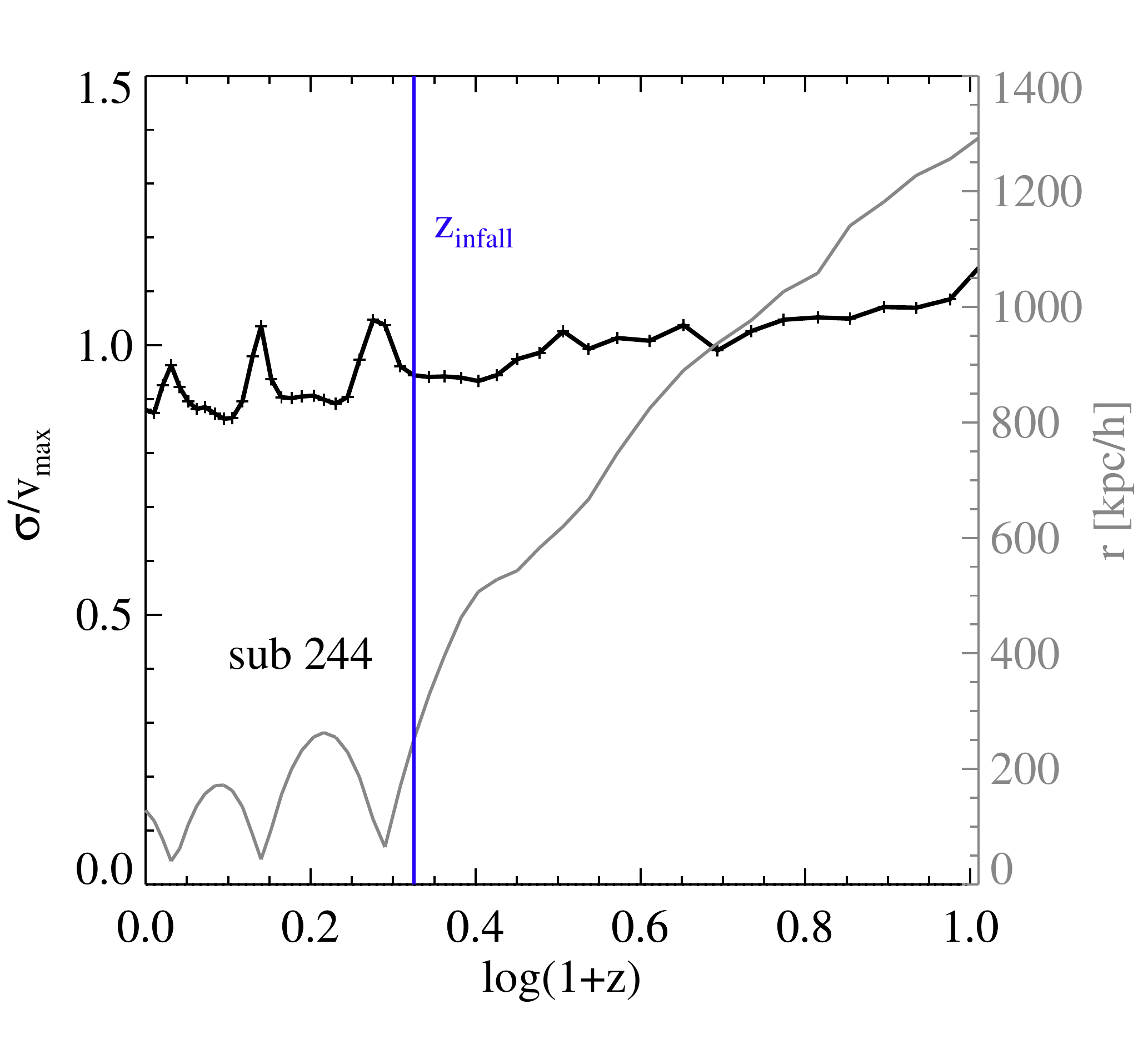}
\caption{\label{fig:tidal_shock}  Effect of tidal shocks on the evolution of Subhalo 244. The black solid curves represent the evolution of $\sigma/v_{\rm max}$ while the gray solid line represents the distance of Sub 244 to the center of the MW halo at different redshifts (in comoving units). Each peak of $\sigma/v_{\rm max}$ corresponds  to a sharp increase of random kinetic energy due to tidal shock heating when the satellite passes through the pericenter. A blue vertical line indicates the infall time of this object at $z = 1.1$, when it has first became a subhalo of the central host. }
\end{center}
\end{figure}

The enhanced tidal disruption rate in the Hydro simulation is likely a combination of several gravitational effects, such as halo shocking from a rapidly varying potential which induces tidal shocks when the objects are on highly eccentric orbits, and disk shocking when they are  passing in the vicinity of the stellar disk.  In comparison, tidal stripping of material is a gentler process that does not increase the kinetic energy within the subhalo, at least when strong resonances are not operating \citep{Donghia2009, Donghia2010a}.  To investigate the impact of tidal shocks on satellites, we follow the evolution of $\sigma/v_{\rm max}$, a ratio between the DM velocity dispersion $\sigma$ and the maximum circular velocity $v_{\rm max}$ a proxy of energy. In Figure~\ref{fig:tidal_shock} we show Subhalo 244 as an example during its infall journey into the main galaxy. Indeed, the internal energy of the subhalo, as indicated by $\sigma/v_{\rm max}$, increases sharply when it passes the pericenter of its trajectory, demonstrating heating from tidal shocks during the close encounter. We have confirmed that the peaks of $\sigma/v_{\rm max}$ are caused by the increase of $\sigma$ when the subhalo passes pericenter. We note, however, that only 64 snapshots were stored for the entire simulation, with the consequence that the time-sampling is not ideal for probing the subhalo trajectory at the time resolution required to clearly disentangle tidal from disk shocking or halo shocking. The discrete time sampling may also likely over-estimate the ``minimum distance" plotted in these figures while the true ``minimum distance" is attained between two snapshots.

\subsubsection{The Origin and Evolution of Bright and Dark Subhalos}

\begin{figure}
\begin{center}
\includegraphics[width=\linewidth]{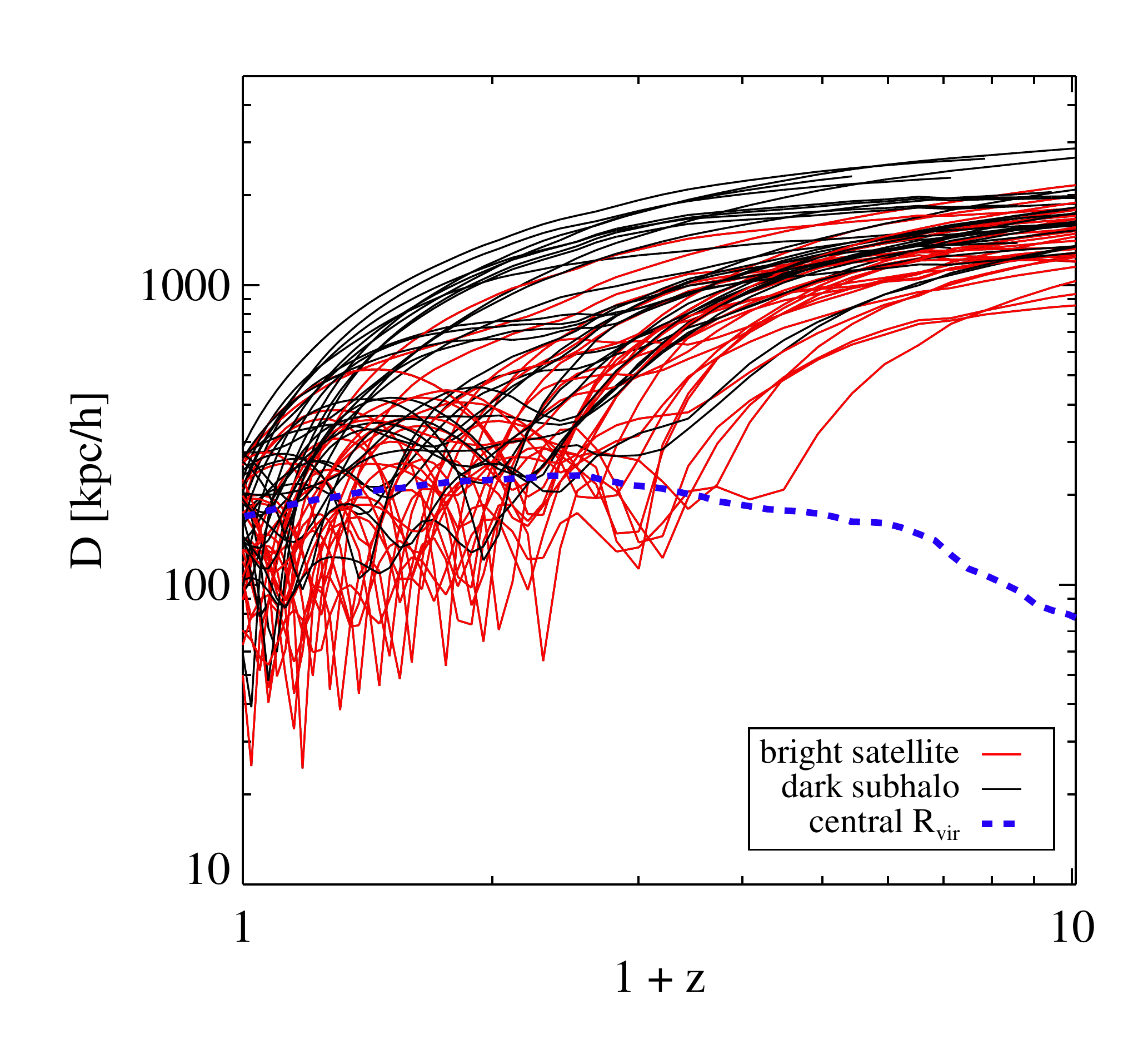}
\caption{\label{fig:subhalo_infall} The distances (in comoving units) to the central galaxy of all subhalos in the mass range of $10^8 - 10^9\, \Msun$ at different times during their infall to the host, taken from the Hydro simulation. The red and black solid curves represent ``bright'' (with stars) and ``dark'' (without stars) subhalos, respectively, while the blue dashed curve indicates the virial radius of the main galaxy at different redshift. }
\end{center}
\end{figure}

\begin{figure}
\begin{center}
\includegraphics[width=\linewidth]{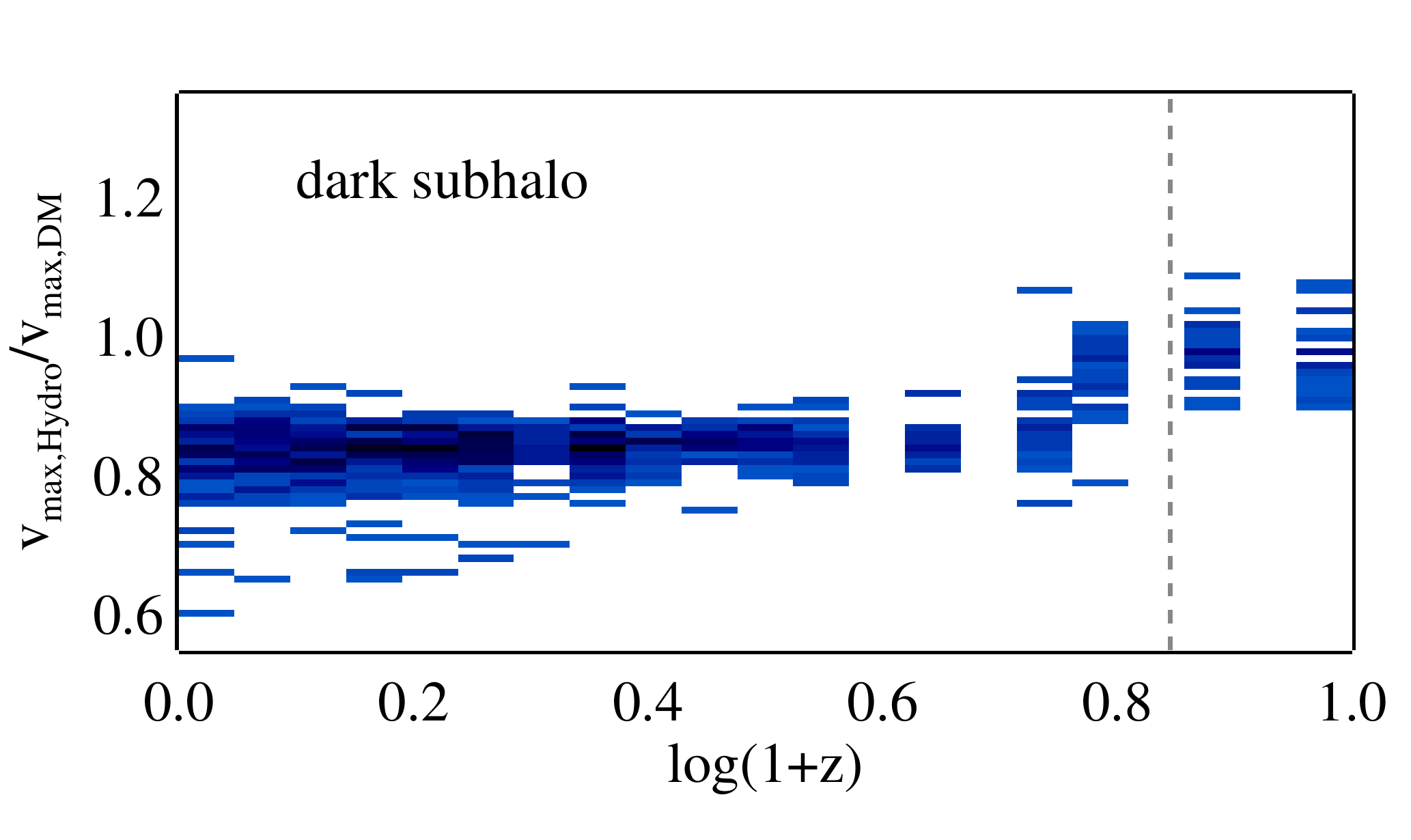}
\includegraphics[width=\linewidth]{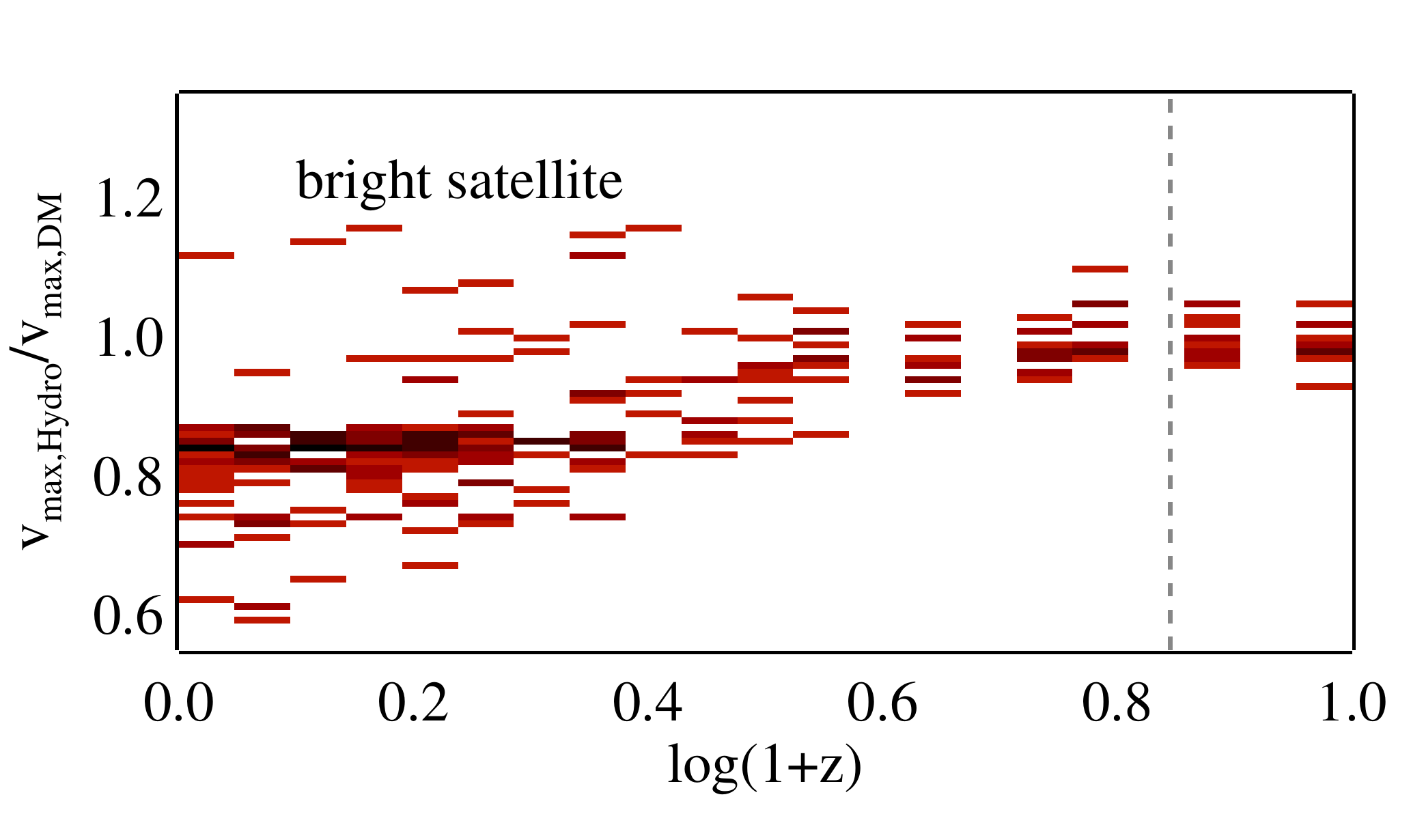}
\caption{\label{fig:vmax_evolution} Evolution of the maximum circular velocity of both dark (top panel) and bright (bottom panel) subhalos in the mass range of $10^8 - 10^9\, \Msun$. These subhalos are identified in the Hydro simulation. In order to compare with the DMO simulation and to identify the effects of baryons, the ratio $v_{\rm {max, Hydro}}/v_{\rm {max, DM}}$ of matched subhalos in both simulations is used. The vertical dashed line represents the end of reionization at redshift $z=6$. The redshift bin size is $0.05$ due to the limited number of output snapshots available for the simulations.}
\end{center}
\end{figure}

To understand the origin of ``bright'' (with stars) and ``dark" (without stars) subhalos in the mass range of $10^{8} - 10^{9}\, \Msunh$, we track their assembly histories in our simulations. In Figure~\ref{fig:subhalo_infall}, we show the distance of each subhalo to the central galaxy during its infall. Interestingly, ``bright'' and ``dark'' subhalos have different accretion paths. On average, the bright satellites are accreted into the host at an earlier redshift than the majority of the dark subhalos, and they typically undergo multiple passages through the main halo. Moreover, we follow the evolution of their $v_{\rm max}$ value and find that bright and dark subhalos have different trends as well, as shown in Figure~\ref{fig:vmax_evolution}. The dark subhalos experience a sharp decline in $v_{\rm max}$ (or mass) shortly after the end of reionization, while the bright satellites do not show such an immediate and dramatic suppression by reionization. In fact, most of them are able to retain the existing gas and replenish some of it long after $z = 6$, thus boosting their mass growth. In both cases, there is a reduction of $v_{\rm max}$ by $\sim 17\%$ on average at $z=0$ in the Hydro simulation compared to the DMO one, highlighting the effects of baryonic processes on mass reduction discussed in the previous sections.

As demonstrated in Figure~\ref{fig:vmax_evolution}, reionization plays an important role in the formation of bright and dark satellites. The impact of reionization on these subhalos is mainly to suppress fresh gas accretion from the IGM, as evidenced by a dip in the curve of the dark subhalos around $\log(1+z) \sim 0.65$ $(z\sim3.5)$. However, a number of the bright satellites in our simulation gain substantial mass even after $z \sim 3$  (see also Figure~\ref{fig:sub_149}). \cite{Ricotti2009} considered a scenario in which the low-mass halos ($v_{\rm max} < 20\,\kms$) in the outer region of the MW are able to accrete low density IGM gas after $z = 3$ and form stars, once the mean temperature of the low-density IGM starts to decrease due to Hubble expansion, similar to what we find here. 

Our simulations show that bright satellites have a different origin and 
evolutionary path from the dark ones. At early times, the bright satellites 
survive better than dark subhalos from reionization and they are able to retain 
gas and form stars afterwards. Moreover, they have an earlier infall time into the 
main galaxy than the dark ones, as also reported by other work 
\citep{Sawala2014b}. Our results suggest that bright satellites may be biased 
tracers of the total subhalo population, as implied by observations of faint 
dwarfs of the Local Group \citep{Weisz2015}.

\section{Discussion}
\label{sec:discussion}

Thanks to significant progress in numerical modeling of galaxy formation  and evolution, recent hydrodynamic simulations are becoming increasingly successful and are now able to reproduce many of the observed properties of galaxies in a more self-consistent manner. However, the complexity of the physical processes and the numerical methods inevitably entail uncertainties in the modeling. In what follows, we will compare our simulations with previous work and discuss the limitations of our model. 

\begin{figure} 
\begin{center} \includegraphics[width=\linewidth]{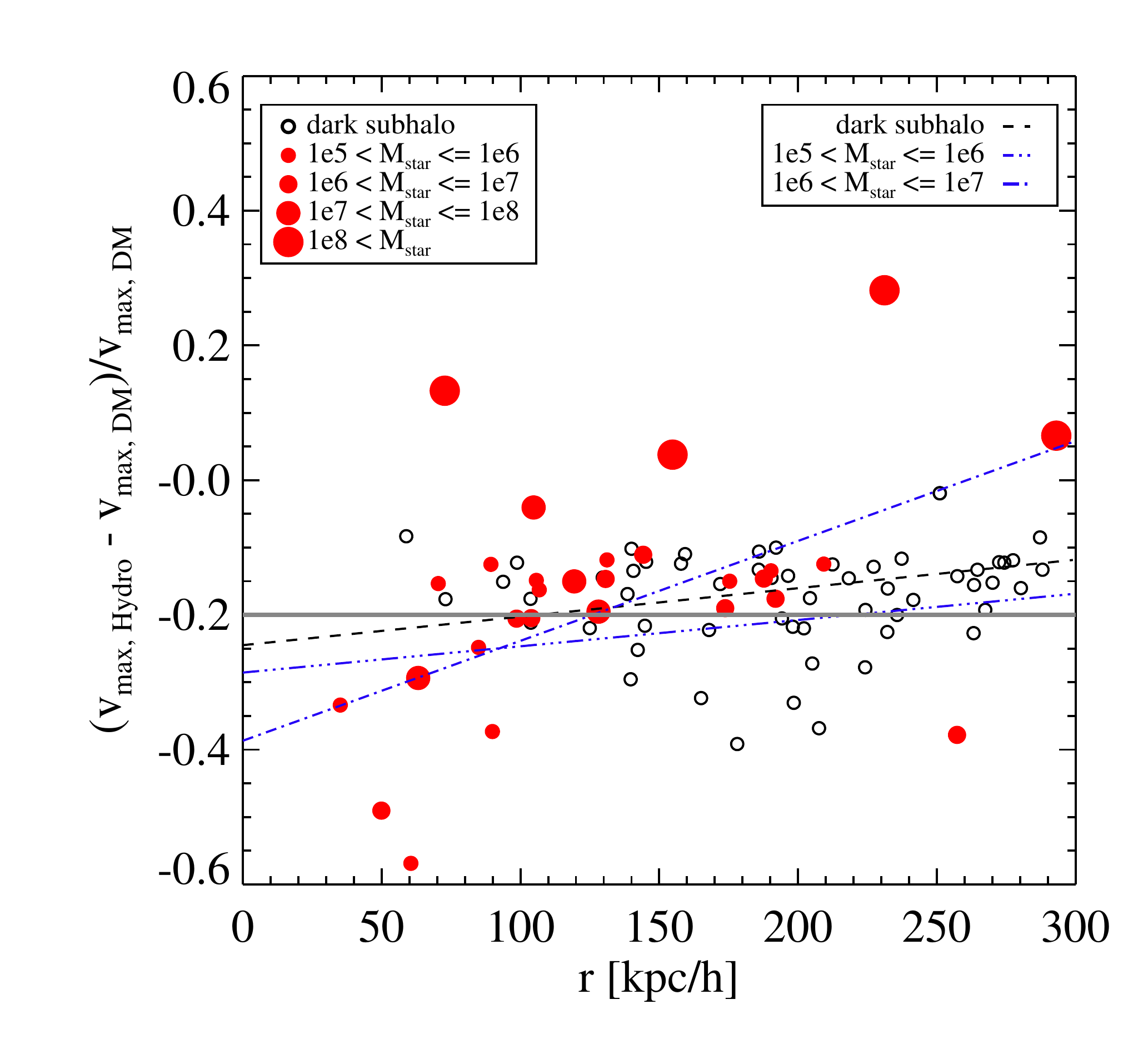} 
\caption{\label{fig:vmax_reduction_distance} The radial distribution of the reduction of $v_{\rm{max}}$ ($\rm{\Delta v_{max}= (v_{max,Hydro} - v_{max,DM})/v_{max,DM}}$) of matched subhalos between Hydro and DMO simulations. The open black circles represent the dark subhalos that contain no stars, while each of the filled red circles represent individual bright satellites with its size proportional to the subhalo mass. The black dashed and the blue lines are simple linear fittings to the dark subhalos, and bright subhalos in the mass ranges $10^5\, \Msun < M_{star}  < 10^6\, \Msun$ and $10^6\, \Msun < M_{star}  < 10^7\, \Msun$, respectively, while the grey horizontal line provides a visual guide of no correlation between $\Delta v_{max}$ and galactic distance $r$. Overall, there is no clear radial dependence of $\Delta v_{max}$ of the subhalos in our simulations. Note the weak $\Delta v_{max} - r$ relation of subhalos with $10^5\, \Msun < M_{star} < 10^6\, \Msun$ is due to two outliers at $r = 50\, \kpch$. Once these two outliers are removed from the sample, the resulted radial dependence is as weak as the ``dark" subhalos and those with $10^6\, \Msun < M_{star}  < 10^7\, \Msun$.}
\end{center}
\end{figure}

\subsection{Dark matter ``cores"}

First of all, it is useful to compare the impact of baryons on the distribution of DM in a MW-size halo found in our study with other works. The enhanced DM concentration in the inner region of the main halo seen in our simulation is consistent with adiabatic contraction. Moreover, we do not find cored dark matter distributions on dwarf galaxy scales in our simulation.  Recently, both \cite{Maccio2012} and \cite{Mollitor2015} have reported flattened DM distributions within 5 kpc from the galactic center in their hydrodynamic simulations of halos with similar mass. This difference cannot be attributed to insufficient numerical resolution as our mass resolution and gravitational softening length, which is finer than in \cite{Maccio2012}, should give reliable results on kpc scales \citep[][]{Power2003, Springel2008}. It is also unlikely to arise from differences in the hydrodynamics or gravity solvers, because grid-based codes such as {\small\sc RAMSES} used in \cite{Mollitor2015} do not have the problem of intrinsic noise in SPH methods \citep[][]{Bauer2012, Zhu2015} and use independent and different numerical methods than employed in \cite{Maccio2012}. The most plausible cause for the difference lies in the feedback models. Here all three simulations have used a similarly large fraction of supernovae energy to drive outflows. However, our outflow model is less bursty than those of \cite{Maccio2012} and \cite{Mollitor2015}, and this has been suggested as an important factor for making cores. We note however that \cite{GarrisonKimmel2013} argued that repeated blowout of gas is not necessarily more effective than a single blowout in reducing central dark matter densities.

The outflow model used here is phenomenological and ties the wind launching 
velocity directly to the properties of subhalos. Once flagged as a wind, 
outflowing particles are temporally decoupled from hydrodynamics to prevent a 
disruption of the sub-resolution ISM phase. Thus, the small-scale creation of 
the wind in a star-forming region and its interaction with the ISM is not 
followed in detail. \cite{DallaVecchia2008} have argued that such a decoupled 
wind is less efficient than a coupled wind in driving strong turbulent motions 
on the scale of dwarf galaxies. A decoupled wind scheme could thus in principle 
miss some of the physical processes needed to generate DM cores, provided random 
bulk motions of gas are indeed able to produce a strongly fluctuating 
gravitational potential, as argued by \cite{Mashchenko2006, Pontzen2012, 
Pontzen2014}.  However, \cite{Sawala2014a} reported that their simulated 
galaxies do not contain cores, even though they use a coupled wind model. This 
indicates that the decoupling feature of the wind model we used is not  
responsible for the absence of DM cores in our run. It thus remains interesting 
to see on what time scales the gravitational potential needs to fluctuate to 
generate DM cores. It is also possible that the delayed cooling mechanism used 
by \cite{Maccio2012} and \cite{Mollitor2015} overestimates the effect of bursty 
supernovae explosions \citep{Agertz2013}.

Despite this difference between our simulation and others that predict DM cores, our results regarding the ``dark'' subhalos and the least luminous dwarfs should not be affected by the wind model, because these objects have the lowest star formation activities and hence an efficient removal of DM is energetically difficult \citep[e.g.][]{Governato2012, GarrisonKimmel2013, Madau2014, DiCintio2014, Onorbe2015}. Indeed, \cite{DiCintio2014} predicted that the most cored density distribution is likely to be found in large halos with $v_{\rm max} = 50\,\kms$, and that DM profiles remain cuspy for dwarf galaxies with the least massive stellar population, in line with some recent hydrodynamic simulations of field dwarf galaxies \citep{Madau2014, Onorbe2015}. The reduction on $v_{\rm max}$ at the low-mass end ($v_{\rm max} < 20\,\kms$, mostly ``dark'') of the cumulative subhalo mass function in our simulation is in good agreement both with other SPH \citep{Zolotov2012, Sawala2014a} and AMR \citep{Mollitor2015} simulations, and we believe that reionization is the primary culprit for the suppressed gas accretion rate from the IGM and the reduced number density of low-mass subhalos.

\subsection{Tidal disruption}
  
It was argued by \cite{Tollerud2014} that tidal forces play a marginal role in resolving the ``too big to fail'' problem as there is a lack of a strong radial dependence in the $r-v_{\rm max}$ relation for the M31 satellite galaxies.  However, we find no strong radial dependence in the reduction of $v_{\rm max}$ in terms of ${(v_{\rm max,Hydro} - v_{\rm max,DMO})/v_{\rm max,DMO}}$ in the Hydro simulation, as shown in Figure~\ref{fig:vmax_reduction_distance}. Overall, the distribution of ${(v_{\rm max,Hydro} - v_{\rm max,DMO})/v_{\rm max,DMO}}$ is rather flat, close to $-0.2$, as a function of distance $r$. In this Figure, we further divide the subhalos into several subgroups according to their stellar mass and fit the results with simple linear functions. Only the subhalo sample with $10^5\, {\Msun} < M_{\rm star} < 10^6\, \Msun$ shows a clear radial dependence. However, this strong radial trend is dominated by two subhalos. If we remove the two outliers around $50\, \kpch$, the radial trend disappears accordingly. For subhalos with $10^6{\Msun} < M_{\rm star} < 10^7 \Msun$, which is in the same mass range as the dSphs in \cite{Tollerud2014}, there is no clear radial dependence for the reduction of $v_{\rm max}$. Hence, we conclude that a lack of a strong radial dependence cannot refute the role of tidal disruption.

\cite{Tollerud2014} attributed the reduction of the number of subhalos to 
supernova feedback. However, we find that this plays a minor role in our 
simulation,  especially for the subhalos with an intermediate $v_{\rm{max}}$ (see Fig.~14), in which 
the (maximum) circular velocity is  always larger in the Hydro case than in the DMO one before those objects are 
accreted into the central galaxy where stronger tidal disruption in the Hydro 
simulation comes into play.  In our simulations, the tension between $\Lambda$CDM 
and observations of dwarf galaxies leading to the so-called ``too big to fail'' 
problem is largely alleviated by stronger tidal disruption, caused by an 
enhanced DM concentration and the stellar disk of the central galaxy \citep[See also][]{Romano2010}. The 
problem is further mitigated by the fact that these subhalos are accreted by the 
host at a much earlier time than the average ``dark'' subhalos as shown in 
Figure~\ref{fig:subhalo_infall}. Thus, they have experienced a more extended 
tidal influence from their host galaxy compared to the average subhalo. On the 
other hand, if the host halo in our simulation is slightly less massive, similar 
to the halo masses used by other groups \cite[e.g.][]{Guedes2011, Sawala2014a}, 
one may not have a ``too big to fail'' problem at all to begin with 
\citep{Wang2012}. However, it is unclear how tidal disruption or similar 
environmental effects would work for the recently reported ``too big to fail'' 
problem of field dwarf galaxies \citep{GarrisonKimmel2014, Papastergis2015, 
Klypin2014}.

\subsection{Mass reduction}

\begin{figure}
\begin{center}
\includegraphics[width=0.98\linewidth]{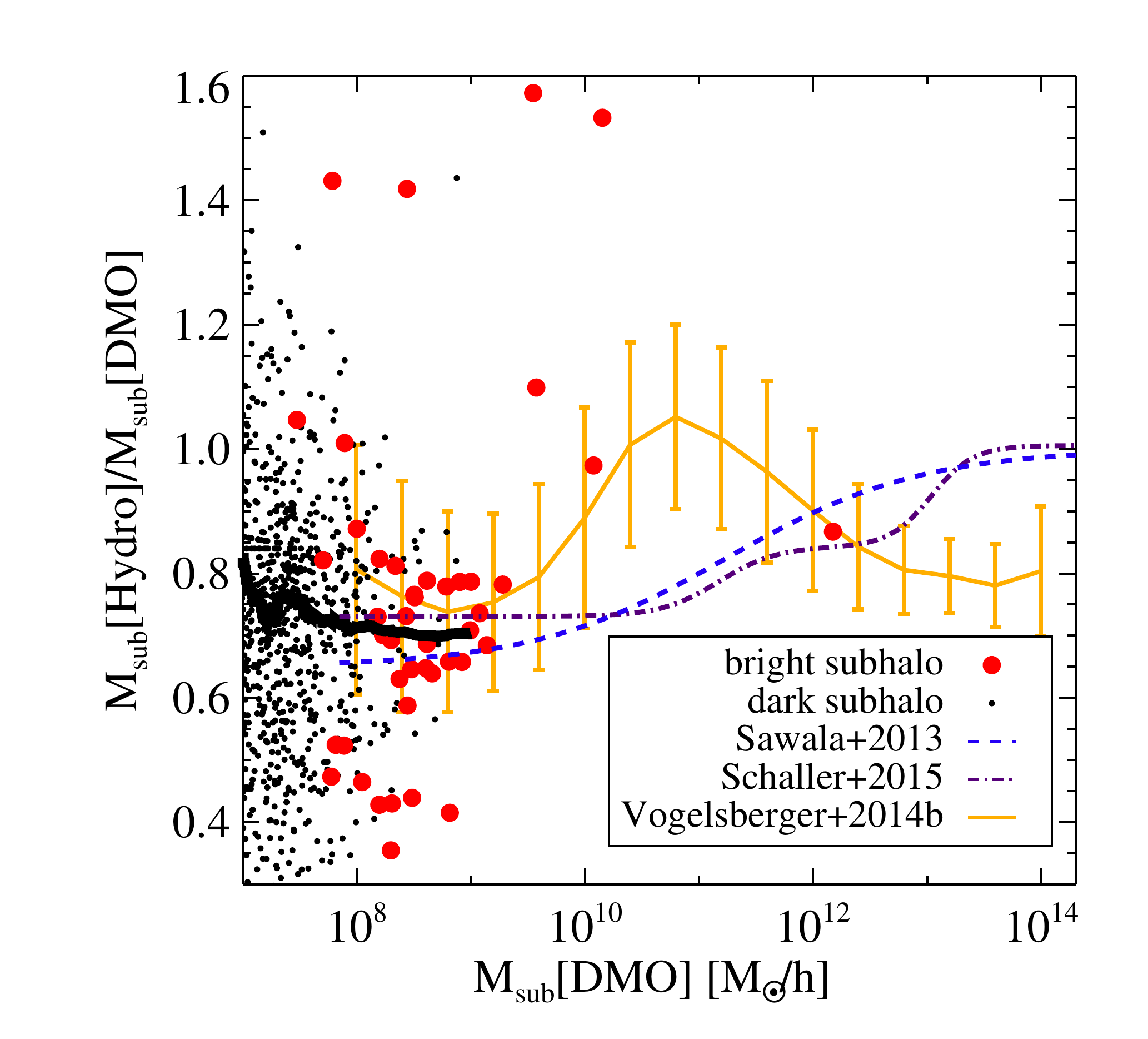}
\includegraphics[width=0.98\linewidth]{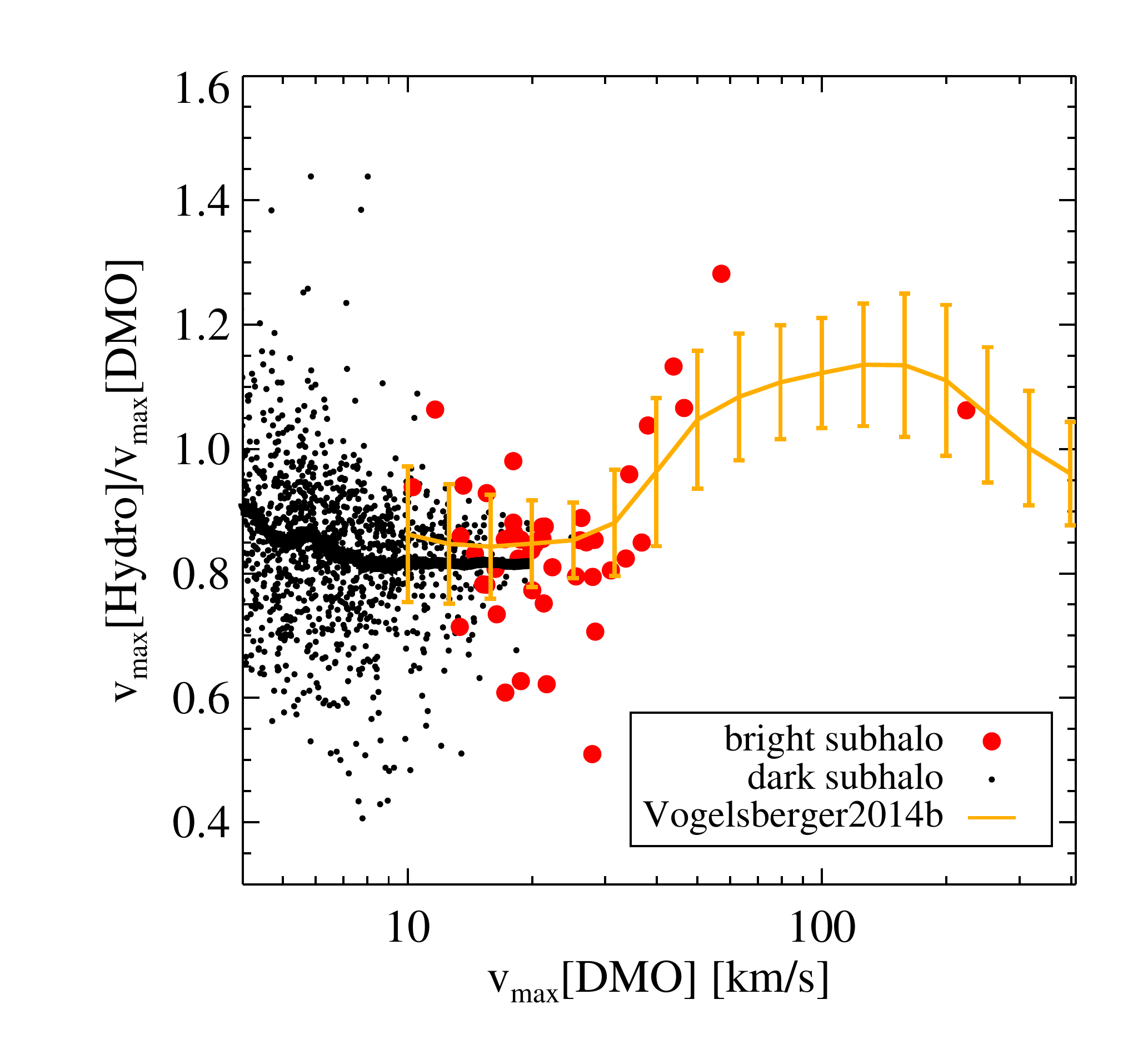}
\caption{\label{fig:mass_reduction_comparison} 
A comparison of our work with previous studies of the reduction of mass and 
and $v_{\rm{max}}$. Top panel: Ratio of $\rm{M_{sub}[Hydro]}/\rm{M_{sub}[DMO]}$ as a 
function of subhalo mass $\rm{M_{sub}[DMO]}$ from our simulations and fitting 
relations from \protect\cite{Sawala2013}, \protect\cite{Schaller2015} and 
\protect\cite{Vogelsberger2014b}. The filled symbols indicate our simulation 
data (red for bright satellites, black for dark subhalos), while 
the relations from \protect\cite{Sawala2013} ({\sc gimic} simulation) and 
\protect\cite{Schaller2015} ({\sc eagle}  simulation) are shown as  
long-dashed and dashed-dotted lines, respectively. We include all the matched 
subhalos between the Illustris full physics run 
and Illustris-Dark without further separating them into subsamples as in 
\protect\citet[][yellow line]{Vogelsberger2014b}. Scatter from the 
fitting relations, shown as 1-$\sigma$ error bars, is computed within each 
mass bin. Bottom panel: Ratio of $v_{\rm{max}}\rm{[Hydro]}$/$v_{\rm{max}}\rm{[DMO]}$ as a 
function of $v_{\rm{max}}$[DMO]. The scatter of this relation, as indicated by 
1-$\sigma$ error bars, is smaller than the top panel for Illustris. In 
both panels, a moving average of 200 data points is shown (black thick solid 
curve) to highlight the overall trend in the low-mass range.}
\end{center}
\end{figure}

A number of previous works have examined the effects of baryons on halos in 
different mass ranges (e.g., \citealt{Sawala2013, Schaller2015}). In order to 
compare with these studies, we show in  Figure~\ref{fig:mass_reduction_comparison} 
the reduction of mass and $v_{\rm{max}}$  from different simulations. 
The top panel compares the total mass reduction in the halos/subhalos as a function of 
subhalo mass from our simulations with the fitting relations from \cite{Sawala2013} ({\sc gimic} simulation) and 
\cite{Schaller2015} ({\sc eagle}  simulation), as well as all subhalos from the 
matched catalog between the Illustris and Illustris-Dark simulations 
\citep[][]{Nelson2015}\footnote{Raw data from each simulation are available from 
\url{http://www.illustris-project.org/data/}.}.  To allow for a comparison with our results we extrapolated the fitting relations
of \cite{Sawala2013} and \cite{Schaller2015} down to $10^8\,\Msun$. 
Given the form of these relations this is equivalent to using a constant  value of 0.65 and 0.73, respectively, for subhalos less massive than $\sim 10^8\,\Msun$.
Under this assumption, the plot shows that all studies are in good agreement for the mass 
reduction of the central MW-size galaxy, as well as the low-mass end 
($<5\times10^{9}\Msun$) where most of the dark subhalos and bright satellites in our 
simulations are located.  There is substantial scatter in $\rm{M_{200,Hydro}/M_{200,DMO}}$, which is
both evident in the data points as well as the error bars of the Illustris results. 
We show a moving average of 200 data  points with a thick solid black curve to highlight the overall trends in the low-mass 
range covered by the high resolution simulation used in this study.

However, significant differences are present in the mass range between 
$5\times10^{9}\Msun$ and $10^{12}\Msun$. The mass ratio given by 
\cite{Sawala2013} or \cite{Schaller2015} is evidently lower than ours. Although 
we suffer from small-number statistics in our simulation, the mass excess in 
the Hydro simulation appears to be a real signal as also shown by a much larger 
galaxy sample drawn from the Illustris matched subhaloes. Considering that the 
universal baryonic mass fraction is $\sim16\%$, galaxies in \cite{Sawala2013} or 
\cite{Schaller2015} actually have less mass in the dark matter component in their 
hydrodynamic simulations. This led them to conclude that supernova feedback,  
which is strongly operating in this mass range, has also removed some DM, thus 
producing lighter halos. However, we do not see DM mass reduction in 
this mass range in our simulation. As we have found in our earlier discussions
on Sub~10 ($v_{\rm{max}} = 73 \kms$), we actually observe an increased dark matter mass 
in this regime due to adiabatic contraction. 
	
We note that the mass values returned by halo finders do have some uncertainties, 
which is evident in the large scatter of the raw data in the upper panel of 
Figure~\ref{fig:mass_reduction_comparison}, while $v_{\rm{max}}$ is not as severely affected 
\citep[e.g.][]{Behroozi2015}. In the lower panel of 
Figure~\ref{fig:mass_reduction_comparison}, we compare the ratio of 
$v_{\rm{max}}\rm{[Hydro]}$/$v_{\rm{max}}\rm{[DMO]}$ from our simulations with that of
the Illustris simulations. Both works show good agreement over the common $v_{max}$ 
range, with subhalos above $v_{\rm{max}}\sim35\kms$ 
($\rm{M_{sub}} > 5\times10^{9}\Msun$) having higher  $v_{\rm{max}}$ in the hydrodynamic 
simulation than in their DMO counterpart. Similar to the top panel,  
we show a moving average of 200 data points at the low $v_{\rm max}$ end 
with a thick solid black curve to highlight the overall 
behavior, which is consistent with the Illustris relation.

Since our simulations employ the same code and essentially the same physical models as the 
Illustris simulations, the differences between our study and previous ones by 
\cite{Sawala2013} and \cite{Schaller2015} may owe to
different implementation of feedback 
processes or different hydrodynamical methods used in these 
works. These comparisons thus highlight the need for comprehensive and systematic 
investigations similar to the Aquila Comparison Project, which is beyond the 
scope of this paper, but we plan to pursue it in future work.

\subsection{Implications for dark matter detections}          
$N-$body simulations are routinely used to model direct/indirect signals from 
Galactic dark matter structures. However, all the recent studies agree that 
baryons have significant impacts on dark matter distributions. The interplay 
between baryonic and DM distributions has a direct impact on current attempts to 
indirectly measure DM in the MW as well as in other galaxies. Traditionally, DM 
substructures are thought to be responsible for the observed radio flux-ratio 
anomaly in gravitational lensing \citep[e.g.][]{Mao1998}. Recent studies by 
\cite{Xu2010, Xu2015} based on the $N-$body simulations by \cite{Springel2008} 
and \cite{Gao2012} concluded that DM substructures alone can not be the whole 
reason for radio flux-ratio anomalies. \cite{Xu2015} suggested that a 
substantial improvement over the existing modeling of strong lensing would be to 
consider substructures in baryonic simulations that take the impact of baryonic 
processes on DM into account. Similarly, accurate predictions of the DM 
annihilation rate \citep{fermidwarf2013} sensitively depend on the DM 
distribution in the dwarf galaxies. The uncertainties in the DM distributions 
enter in factors describing the clumping of the DM density (proportional to the 
density squared) along the line of sight \citep[see][]{GeringerSameth2015}. For 
example, \cite{GomezVargas2013} have shown that the cross section estimated from 
$\gamma-$ray photons from the galactic center can be affected by adiabatic 
contraction of DM by more than an order of magnitude. So far, these studies are 
based on inferences from high resolution $\textit{N}-$body simulations. 
Hydrodynamical simulations change this picture and offer a more accurate and 
self-consistent physical modeling for analyzing direct and indirect detection 
experiments.

\section{Conclusions}
\label{sec:conclusion}

The availability of hydrodynamical simulations allows us to reassess some of the 
well-known potential issues of $\Lambda$CDM that have been identified with pure 
\textit{N}-body simulations. In particular, hydrodynamical simulations are able 
to self-consistently model the impact of baryonic processes on the DM 
distribution, a point which has previously often been ignored in galaxy 
formation studies based on DM-only simulations.

In this work, we analyze a high-resolution cosmological hydrodynamic simulation 
on a moving mesh and compare it to its DM-only counterpart in order to study the 
DM distribution in a MW-like galaxy and its subhalos. This simulation uses 
essentially the same physical model employed in the Illustris simulation, and hence is 
consistent with a globally successful model for explaining the observed galaxy 
population. Moreover, the properties of the simulated galaxy are well converged 
with resolution, making our results numerically robust. Here we summarize our 
main findings:

\begin{itemize}

\item We identify three physical processes induced by baryons that shape 
the overall distribution of DM in the main halo and its subhalos depending on 
mass: (1) adiabatic contraction due to gas cooling and condensation increases 
the DM concentration in the inner region of the main halo and in massive 
subhalos with $v_{\rm max} > 35~\kms$ (${M_{\rm sub}} > 4\times10^{9}\, 
\Msunh$), making them more spherical in the inner regions; (2) 
reionization plays a critical role in 
the formation and evolution of low-mass halos with $v_{\rm max} < 20~\kms$ 
(${M_{\rm sub}} < 10^{9}\, \Msunh$) by removing the gas from the halo and 
suppressing new gas accretion from the IGM, making them ``dark'' with little or 
no star formation; (3)  strong tidal forces in the Hydro simulation effectively 
remove the stellar and DM components of intermediate-mass subhalos in 
the range of $v_{\rm max} \sim 20~\kms$ -- $35~\kms$ during their infall to the main galaxy, leaving behind tidal debris 
and streams of DM and stars.

\item As a result of these major effects from baryons, the total number of subhalos in a MW-like galaxy and their total mass are significantly reduced in the hydrodynamic simulation compared to the DM-only one. Our results are in good agreement with observations of dwarf satellites in the MW galaxy, suggesting a viable solution to  long-standing problems such as the ``missing satellites'' and ``too big to fail'' issues that arose from pure $\textit{N-}$body simulations. 

\item The ranking of subhalos based on either their peak mass or their present-day mass is modified in the Hydro simulation compared with the DMO run. A large fraction of subhalos with a peak mass below ${10^{9}\Msunh}$ are found to host no stars. These findings suggest that the assumption of a monotonic relation between stellar mass and peak mass as commonly used in abundance matching is not strictly valid, and that effects of baryonic processes should be included in this modeling as well.
	
\end{itemize}
	
Interestingly, our high-resolution hydrodynamic simulation does not produce cored DM distributions as observationally suggested in some low surface brightness galaxies, and unlike some other recent simulation models with very bursty feedback modes.  More sophisticated hydrodynamic simulations are needed to further study this puzzling problem, and to improve the realism of the inclusion of feedback processes to determine whether they can indeed provide a solution to small-scale tensions identified in the $\Lambda$CDM cosmogony.

\section*{Acknowledgements}
We thank Steinn Sigurdsson, Priyamvada Natarajan, Annalisa Pillepich, Arthur Kosowsky, 
Eddie Chua and Dylan Nelson for their suggestions and helpful discussions.
We also thank an anonymous referee for valuable comments that have helped improve the manuscript.   
YL acknowledges support from NSF grants AST-0965694, AST-1009867 and AST-1412719. 
LH acknowledges support from NSF grant AST-1312095 and NASA grant NNX12AC67G. 
VS acknowledges support by the DFG Research Centre SFB-881 
`The Milky Way System' through project A1, by the European Research Council under 
ERC-StG grant EXAGAL-308037, and by the Klaus Tschira Foundation. We acknowledge the Institute For 
CyberScience at The Pennsylvania State University for providing computational 
resources and services that have contributed to the research results reported in 
this paper. The Institute for Gravitation and the Cosmos is supported by the 
Eberly College of Science and the Office of the Senior Vice President for 
Research at the Pennsylvania State University.

\bibliographystyle{mn2efixed}

\label{lastpage}

\end{document}

%% file: JournalAbbr.tex
\def\aj{AJ}%
\def\araa{ARA\&A}%
\def\apj{ApJ}%
\def\apjl{ApJ}%
\def\apjs{ApJS}%
\def\aap{A\&A}%
\def\mnras{MNRAS}%
\def\prd{Phys.~Rev.~D}%
\def\nat{Nature}%
\def\jcap{JCAP}

%% file: ms.bbl
\begin{thebibliography}{154}
\expandafter\ifx\csname natexlab\endcsname\relax\def\natexlab#1{#1}\fi

\bibitem[{{Addison} {et~al}\mbox{.}(2013){Addison}, {Hinshaw}, \&
  {Halpern}}]{Addison2013}
{Addison} G.~E., {Hinshaw} G., {Halpern} M., 2013, \mnras, 436, 1674

\bibitem[{{Agertz} {et~al}\mbox{.}(2013){Agertz}, {Kravtsov}, {Leitner}, \&
  {Gnedin}}]{Agertz2013}
{Agertz} O., {Kravtsov} A.~V., {Leitner} S.~N., {Gnedin} N.~Y., 2013, \apj,
  770, 25

\bibitem[{{Agertz} {et~al}\mbox{.}(2007){Agertz}, {Moore}, {Stadel}, {Potter},
  {Miniati}, {Read}, {Mayer}, {Gawryszczak}, {Kravtsov}, {Nordlund}, {Pearce},
  {Quilis}, {Rudd}, {Springel}, {Stone}, {Tasker}, {Teyssier}, {Wadsley}, \&
  {Walder}}]{Agertz2007}
{Agertz} O. {et~al.}, 2007, \mnras, 380, 963

\bibitem[{{Agertz} {et~al}\mbox{.}(2011){Agertz}, {Teyssier}, \&
  {Moore}}]{Agertz2011}
{Agertz} O., {Teyssier} R., {Moore} B., 2011, \mnras, 410, 1391

\bibitem[{{Allen} {et~al}\mbox{.}(2011){Allen}, {Evrard}, \&
  {Mantz}}]{Allen2011}
{Allen} S.~W., {Evrard} A.~E., {Mantz} A.~B., 2011, \araa, 49, 409

\bibitem[{{Amorisco} \& {Evans}(2012)}]{Amorisco2012}
{Amorisco} N.~C., {Evans} N.~W., 2012, \mnras, 419, 184

\bibitem[{{Arraki} {et~al}\mbox{.}(2014){Arraki}, {Klypin}, {More}, \&
  {Trujillo-Gomez}}]{Arraki2014}
{Arraki} K.~S., {Klypin} A., {More} S., {Trujillo-Gomez} S., 2014, \mnras, 438,
  1466

\bibitem[{{Aumer} {et~al}\mbox{.}(2013){Aumer}, {White}, {Naab}, \&
  {Scannapieco}}]{Aumer2013}
{Aumer} M., {White} S.~D.~M., {Naab} T., {Scannapieco} C., 2013, \mnras, 434,
  3142

\bibitem[{{Barnes} \& {White}(1984)}]{Barnes1984}
{Barnes} J., {White} S.~D.~M., 1984, \mnras, 211, 753

\bibitem[{{Barnes} \& {Hernquist}(1996)}]{Barnes1996}
{Barnes} J.~E., {Hernquist} L., 1996, \apj, 471, 115

\bibitem[{{Barnes} \& {Hernquist}(1991)}]{Barnes1991}
{Barnes} J.~E., {Hernquist} L.~E., 1991, \apjl, 370, L65

\bibitem[{{Bauer} \& {Springel}(2012)}]{Bauer2012}
{Bauer} A., {Springel} V., 2012, \mnras, 423, 2558

\bibitem[{{Behroozi} {et~al}\mbox{.}(2015){Behroozi}, {Knebe}, {Pearce},
  {Elahi}, {Han}, {Lux}, {Mao}, {Muldrew}, {Potter}, \&
  {Srisawat}}]{Behroozi2015}
{Behroozi} P. {et~al.}, 2015, \mnras, 454, 3020

\bibitem[{{Blumenthal} {et~al}\mbox{.}(1986){Blumenthal}, {Faber}, {Flores}, \&
  {Primack}}]{Blumenthal1986}
{Blumenthal} G.~R., {Faber} S.~M., {Flores} R., {Primack} J.~R., 1986, \apj,
  301, 27

\bibitem[{{Bosch} {et~al}\mbox{.}(2014){Bosch}, {Jiang}, {Hearin}, {Campbell},
  {Watson}, \& {Padmanabhan}}]{vandenBosch2014b}
{Bosch} F.~C.~v.~d., {Jiang} F., {Hearin} A., {Campbell} D., {Watson} D.,
  {Padmanabhan} N., 2014, \mnras, 445, 1713

\bibitem[{{Boylan-Kolchin} {et~al}\mbox{.}(2011){Boylan-Kolchin}, {Bullock}, \&
  {Kaplinghat}}]{Kolchin2011}
{Boylan-Kolchin} M., {Bullock} J.~S., {Kaplinghat} M., 2011, \mnras, 415, L40

\bibitem[{{Boylan-Kolchin} {et~al}\mbox{.}(2012){Boylan-Kolchin}, {Bullock}, \&
  {Kaplinghat}}]{Kolchin2012}
{Boylan-Kolchin} M., {Bullock} J.~S., {Kaplinghat} M., 2012, \mnras, 422, 1203

\bibitem[{{Brooks} {et~al}\mbox{.}(2013){Brooks}, {Kuhlen}, {Zolotov}, \&
  {Hooper}}]{Brooks2013}
{Brooks} A.~M., {Kuhlen} M., {Zolotov} A., {Hooper} D., 2013, \apj, 765, 22

\bibitem[{{Brooks} \& {Zolotov}(2014)}]{Brooks2014}
{Brooks} A.~M., {Zolotov} A., 2014, \apj, 786, 87

\bibitem[{{Bryan} {et~al}\mbox{.}(2013){Bryan}, {Kay}, {Duffy}, {Schaye},
  {Dalla Vecchia}, \& {Booth}}]{Bryan2013}
{Bryan} S.~E., {Kay} S.~T., {Duffy} A.~R., {Schaye} J., {Dalla Vecchia} C.,
  {Booth} C.~M., 2013, \mnras, 429, 3316

\bibitem[{{Clowe} {et~al}\mbox{.}(2006){Clowe}, {Brada{\v c}}, {Gonzalez},
  {Markevitch}, {Randall}, {Jones}, \& {Zaritsky}}]{Clowe2006}
{Clowe} D., {Brada{\v c}} M., {Gonzalez} A.~H., {Markevitch} M., {Randall}
  S.~W., {Jones} C., {Zaritsky} D., 2006, \apjl, 648, L109

\bibitem[{{Conselice}(2014)}]{Conselice2014}
{Conselice} C.~J., 2014, \araa, 52, 291

\bibitem[{{Correa} {et~al}\mbox{.}(2015){Correa}, {Wyithe}, {Schaye}, \&
  {Duffy}}]{Correa2015}
{Correa} C.~A., {Wyithe} J.~S.~B., {Schaye} J., {Duffy} A.~R., 2015, \mnras,
  450, 1514

\bibitem[{{Dalla Vecchia} \& {Schaye}(2008)}]{DallaVecchia2008}
{Dalla Vecchia} C., {Schaye} J., 2008, \mnras, 387, 1431

\bibitem[{{Dav{\'e}} {et~al}\mbox{.}(2001){Dav{\'e}}, {Spergel}, {Steinhardt},
  \& {Wandelt}}]{Dave2001}
{Dav{\'e}} R., {Spergel} D.~N., {Steinhardt} P.~J., {Wandelt} B.~D., 2001,
  \apj, 547, 574

\bibitem[{{de Blok}(2010)}]{deBlok2010}
{de Blok} W.~J.~G., 2010, Advances in Astronomy, 2010, 5

\bibitem[{{Di Cintio} {et~al}\mbox{.}(2014){Di Cintio}, {Brook}, {Dutton},
  {Macci{\`o}}, {Stinson}, \& {Knebe}}]{DiCintio2014}
{Di Cintio} A., {Brook} C.~B., {Dutton} A.~A., {Macci{\`o}} A.~V., {Stinson}
  G.~S., {Knebe} A., 2014, \mnras, 441, 2986

\bibitem[{{Di Matteo} {et~al}\mbox{.}(2005){Di Matteo}, {Springel}, \&
  {Hernquist}}]{Dimatteo2005}
{Di Matteo} T., {Springel} V., {Hernquist} L., 2005, \nat, 433, 604

\bibitem[{{D'Onghia} {et~al}\mbox{.}(2009){D'Onghia}, {Besla}, {Cox}, \&
  {Hernquist}}]{Donghia2009}
{D'Onghia} E., {Besla} G., {Cox} T.~J., {Hernquist} L., 2009, \nat, 460, 605

\bibitem[{{D'Onghia} {et~al}\mbox{.}(2010{\natexlab{a}}){D'Onghia}, {Springel},
  {Hernquist}, \& {Keres}}]{DOnghia2010}
{D'Onghia} E., {Springel} V., {Hernquist} L., {Keres} D., 2010{\natexlab{a}},
  \apj, 709, 1138

\bibitem[{{D'Onghia} {et~al}\mbox{.}(2010{\natexlab{b}}){D'Onghia},
  {Vogelsberger}, {Faucher-Giguere}, \& {Hernquist}}]{Donghia2010a}
{D'Onghia} E., {Vogelsberger} M., {Faucher-Giguere} C.-A., {Hernquist} L.,
  2010{\natexlab{b}}, \apj, 725, 353

\bibitem[{{Elbert} {et~al}\mbox{.}(2014){Elbert}, {Bullock}, {Garrison-Kimmel},
  {Rocha}, {O{\~n}orbe}, \& {Peter}}]{Elbert2014}
{Elbert} O.~D., {Bullock} J.~S., {Garrison-Kimmel} S., {Rocha} M., {O{\~n}orbe}
  J., {Peter} A.~H.~G., 2014, ArXiv e-prints

\bibitem[{{Evans} {et~al}\mbox{.}(2009){Evans}, {An}, \& {Walker}}]{Evans2009}
{Evans} N.~W., {An} J., {Walker} M.~G., 2009, \mnras, 393, L50

\bibitem[{{Fan} {et~al}\mbox{.}(2006){Fan}, {Strauss}, {Becker}, {White},
  {Gunn}, {Knapp}, {Richards}, {Schneider}, {Brinkmann}, \&
  {Fukugita}}]{Fan2006}
{Fan} X. {et~al.}, 2006, \aj, 132, 117

\bibitem[{{Faucher-Gigu{\`e}re} {et~al}\mbox{.}(2009){Faucher-Gigu{\`e}re},
  {Lidz}, {Zaldarriaga}, \& {Hernquist}}]{Faucher2009}
{Faucher-Gigu{\`e}re} C.-A., {Lidz} A., {Zaldarriaga} M., {Hernquist} L., 2009,
  \apj, 703, 1416

\bibitem[{{Frenk} \& {White}(2012)}]{Frenk2012}
{Frenk} C.~S., {White} S.~D.~M., 2012, Annalen der Physik, 524, 507

\bibitem[{{Frieman} {et~al}\mbox{.}(2008){Frieman}, {Turner}, \&
  {Huterer}}]{Frieman2008}
{Frieman} J.~A., {Turner} M.~S., {Huterer} D., 2008, \araa, 46, 385

\bibitem[{{Gao} {et~al}\mbox{.}(2012){Gao}, {Navarro}, {Frenk}, {Jenkins},
  {Springel}, \& {White}}]{Gao2012}
{Gao} L., {Navarro} J.~F., {Frenk} C.~S., {Jenkins} A., {Springel} V., {White}
  S.~D.~M., 2012, \mnras, 425, 2169

\bibitem[{{Garrison-Kimmel}
  {et~al}\mbox{.}(2014{\natexlab{a}}){Garrison-Kimmel}, {Boylan-Kolchin},
  {Bullock}, \& {Kirby}}]{GarrisonKimmel2014}
{Garrison-Kimmel} S., {Boylan-Kolchin} M., {Bullock} J.~S., {Kirby} E.~N.,
  2014{\natexlab{a}}, \mnras, 444, 222

\bibitem[{{Garrison-Kimmel}
  {et~al}\mbox{.}(2014{\natexlab{b}}){Garrison-Kimmel}, {Boylan-Kolchin},
  {Bullock}, \& {Lee}}]{GarrisonKimmel2014b}
{Garrison-Kimmel} S., {Boylan-Kolchin} M., {Bullock} J.~S., {Lee} K.,
  2014{\natexlab{b}}, \mnras, 438, 2578

\bibitem[{{Garrison-Kimmel} {et~al}\mbox{.}(2013){Garrison-Kimmel}, {Rocha},
  {Boylan-Kolchin}, {Bullock}, \& {Lally}}]{GarrisonKimmel2013}
{Garrison-Kimmel} S., {Rocha} M., {Boylan-Kolchin} M., {Bullock} J.~S., {Lally}
  J., 2013, \mnras, 433, 3539

\bibitem[{{Gatto} {et~al}\mbox{.}(2013){Gatto}, {Fraternali}, {Read},
  {Marinacci}, {Lux}, \& {Walch}}]{Gatto2013}
{Gatto} A., {Fraternali} F., {Read} J.~I., {Marinacci} F., {Lux} H., {Walch}
  S., 2013, \mnras, 433, 2749

\bibitem[{{Genel} {et~al}\mbox{.}(2014){Genel}, {Vogelsberger}, {Springel},
  {Sijacki}, {Nelson}, {Snyder}, {Rodriguez-Gomez}, {Torrey}, \&
  {Hernquist}}]{Genel2014}
{Genel} S. {et~al.}, 2014, \mnras, 445, 175

\bibitem[{{Geringer-Sameth} {et~al}\mbox{.}(2015){Geringer-Sameth},
  {Koushiappas}, \& {Walker}}]{GeringerSameth2015}
{Geringer-Sameth} A., {Koushiappas} S.~M., {Walker} M.~G., 2015, \prd, 91,
  083535

\bibitem[{{Gilmore} {et~al}\mbox{.}(2007){Gilmore}, {Wilkinson}, {Wyse},
  {Kleyna}, {Koch}, {Evans}, \& {Grebel}}]{Gilmore2007}
{Gilmore} G., {Wilkinson} M.~I., {Wyse} R.~F.~G., {Kleyna} J.~T., {Koch} A.,
  {Evans} N.~W., {Grebel} E.~K., 2007, \apj, 663, 948

\bibitem[{{Gnedin} {et~al}\mbox{.}(2011){Gnedin}, {Ceverino}, {Gnedin},
  {Klypin}, {Kravtsov}, {Levine}, {Nagai}, \& {Yepes}}]{Gnedin2011}
{Gnedin} O.~Y., {Ceverino} D., {Gnedin} N.~Y., {Klypin} A.~A., {Kravtsov}
  A.~V., {Levine} R., {Nagai} D., {Yepes} G., 2011, ArXiv e-prints

\bibitem[{{Gnedin} {et~al}\mbox{.}(2004){Gnedin}, {Kravtsov}, {Klypin}, \&
  {Nagai}}]{Gnedin2004}
{Gnedin} O.~Y., {Kravtsov} A.~V., {Klypin} A.~A., {Nagai} D., 2004, \apj, 616,
  16

\bibitem[{{G{\'o}mez-Vargas} {et~al}\mbox{.}(2013){G{\'o}mez-Vargas},
  {S{\'a}nchez-Conde}, {Huh}, {Peir{\'o}}, {Prada}, {Morselli}, {Klypin},
  {Cerde{\~n}o}, {Mambrini}, \& {Mu{\~n}oz}}]{GomezVargas2013}
{G{\'o}mez-Vargas} G.~A. {et~al.}, 2013, \jcap, 10, 29

\bibitem[{{Governato} {et~al}\mbox{.}(2012){Governato}, {Zolotov}, {Pontzen},
  {Christensen}, {Oh}, {Brooks}, {Quinn}, {Shen}, \& {Wadsley}}]{Governato2012}
{Governato} F. {et~al.}, 2012, \mnras, 422, 1231

\bibitem[{{Guedes} {et~al}\mbox{.}(2011){Guedes}, {Callegari}, {Madau}, \&
  {Mayer}}]{Guedes2011}
{Guedes} J., {Callegari} S., {Madau} P., {Mayer} L., 2011, \apj, 742, 76

\bibitem[{{Guo} \& {White}(2014)}]{Guo2014}
{Guo} Q., {White} S., 2014, \mnras, 437, 3228

\bibitem[{{Guo} {et~al}\mbox{.}(2010){Guo}, {White}, {Li}, \&
  {Boylan-Kolchin}}]{Guo2010}
{Guo} Q., {White} S., {Li} C., {Boylan-Kolchin} M., 2010, \mnras, 404, 1111

\bibitem[{{Hinshaw} {et~al}\mbox{.}(2013){Hinshaw}, {Larson}, {Komatsu},
  {Spergel}, {Bennett}, {Dunkley}, {Nolta}, {Halpern}, {Hill}, {Odegard},
  {Page}, {Smith}, {Weiland}, {Gold}, {Jarosik}, {Kogut}, {Limon}, {Meyer},
  {Tucker}, {Wollack}, \& {Wright}}]{Hinshaw2013}
{Hinshaw} G. {et~al.}, 2013, \apjs, 208, 19

\bibitem[{{Hopkins}(2013)}]{Hopkins2013}
{Hopkins} P.~F., 2013, \mnras, 428, 2840

\bibitem[{{Hopkins} {et~al}\mbox{.}(2014){Hopkins}, {Kere{\v s}}, {O{\~n}orbe},
  {Faucher-Gigu{\`e}re}, {Quataert}, {Murray}, \& {Bullock}}]{Hopkins2014b}
{Hopkins} P.~F., {Kere{\v s}} D., {O{\~n}orbe} J., {Faucher-Gigu{\`e}re} C.-A.,
  {Quataert} E., {Murray} N., {Bullock} J.~S., 2014, \mnras, 445, 581

\bibitem[{{Hu} {et~al}\mbox{.}(2014){Hu}, {Naab}, {Walch}, {Moster}, \&
  {Oser}}]{Hu2014}
{Hu} C.-Y., {Naab} T., {Walch} S., {Moster} B.~P., {Oser} L., 2014, \mnras,
  443, 1173

\bibitem[{{Kennedy} {et~al}\mbox{.}(2014){Kennedy}, {Frenk}, {Cole}, \&
  {Benson}}]{Kennedy2014}
{Kennedy} R., {Frenk} C., {Cole} S., {Benson} A., 2014, \mnras, 442, 2487

\bibitem[{{Klypin} {et~al}\mbox{.}(2014){Klypin}, {Karachentsev}, {Makarov}, \&
  {Nasonova}}]{Klypin2014}
{Klypin} A., {Karachentsev} I., {Makarov} D., {Nasonova} O., 2014, ArXiv
  e-prints

\bibitem[{{Klypin} {et~al}\mbox{.}(1999){Klypin}, {Kravtsov}, {Valenzuela}, \&
  {Prada}}]{Klypin1999}
{Klypin} A., {Kravtsov} A.~V., {Valenzuela} O., {Prada} F., 1999, \apj, 522, 82

\bibitem[{{Knollmann} \& {Knebe}(2009)}]{Knollmann2009}
{Knollmann} S.~R., {Knebe} A., 2009, \apjs, 182, 608

\bibitem[{{Komatsu} {et~al}\mbox{.}(2011){Komatsu}, {Smith}, {Dunkley},
  {Bennett}, {Gold}, {Hinshaw}, {Jarosik}, {Larson}, {Nolta}, {Page},
  {Spergel}, {Halpern}, {Hill}, {Kogut}, {Limon}, {Meyer}, {Odegard}, {Tucker},
  {Weiland}, {Wollack}, \& {Wright}}]{Komatsu2011}
{Komatsu} E. {et~al.}, 2011, \apjs, 192, 18

\bibitem[{{Kravtsov}(2010)}]{Kravtsov2009}
{Kravtsov} A., 2010, Advances in Astronomy, 2010, 8

\bibitem[{{Kravtsov} \& {Borgani}(2012)}]{Kravtsov2012}
{Kravtsov} A.~V., {Borgani} S., 2012, \araa, 50, 353

\bibitem[{{Kravtsov} {et~al}\mbox{.}(2004){Kravtsov}, {Gnedin}, \&
  {Klypin}}]{Kravtsov2004}
{Kravtsov} A.~V., {Gnedin} O.~Y., {Klypin} A.~A., 2004, \apj, 609, 482

\bibitem[{{Macci{\`o}} {et~al}\mbox{.}(2012){Macci{\`o}}, {Stinson}, {Brook},
  {Wadsley}, {Couchman}, {Shen}, {Gibson}, \& {Quinn}}]{Maccio2012}
{Macci{\`o}} A.~V., {Stinson} G., {Brook} C.~B., {Wadsley} J., {Couchman}
  H.~M.~P., {Shen} S., {Gibson} B.~K., {Quinn} T., 2012, \apjl, 744, L9

\bibitem[{{Madau} {et~al}\mbox{.}(2014){Madau}, {Shen}, \&
  {Governato}}]{Madau2014}
{Madau} P., {Shen} S., {Governato} F., 2014, \apjl, 789, L17

\bibitem[{{Mao} \& {Schneider}(1998)}]{Mao1998}
{Mao} S., {Schneider} P., 1998, \mnras, 295, 587

\bibitem[{{Marinacci} {et~al}\mbox{.}(2014{\natexlab{a}}){Marinacci}, {Pakmor},
  \& {Springel}}]{Marinacci2014a}
{Marinacci} F., {Pakmor} R., {Springel} V., 2014{\natexlab{a}}, \mnras, 437,
  1750

\bibitem[{{Marinacci} {et~al}\mbox{.}(2014{\natexlab{b}}){Marinacci}, {Pakmor},
  {Springel}, \& {Simpson}}]{Marinacci2014b}
{Marinacci} F., {Pakmor} R., {Springel} V., {Simpson} C.~M.,
  2014{\natexlab{b}}, \mnras, 442, 3745

\bibitem[{{Martinez}(2013)}]{Martinez2013}
{Martinez} G.~D., 2013, ArXiv e-prints

\bibitem[{{Mashchenko} {et~al}\mbox{.}(2006){Mashchenko}, {Couchman}, \&
  {Wadsley}}]{Mashchenko2006}
{Mashchenko} S., {Couchman} H.~M.~P., {Wadsley} J., 2006, \nat, 442, 539

\bibitem[{{Mashchenko} {et~al}\mbox{.}(2008){Mashchenko}, {Wadsley}, \&
  {Couchman}}]{Mashchenko2008}
{Mashchenko} S., {Wadsley} J., {Couchman} H.~M.~P., 2008, Science, 319, 174

\bibitem[{{Mayer} {et~al}\mbox{.}(2006){Mayer}, {Mastropietro}, {Wadsley},
  {Stadel}, \& {Moore}}]{Mayer2006}
{Mayer} L., {Mastropietro} C., {Wadsley} J., {Stadel} J., {Moore} B., 2006,
  \mnras, 369, 1021

\bibitem[{{Mihos} \& {Hernquist}(1996)}]{Mihos1996}
{Mihos} J.~C., {Hernquist} L., 1996, \apj, 464, 641

\bibitem[{{Mollitor} {et~al}\mbox{.}(2015){Mollitor}, {Nezri}, \&
  {Teyssier}}]{Mollitor2015}
{Mollitor} P., {Nezri} E., {Teyssier} R., 2015, \mnras, 447, 1353

\bibitem[{{Moore} {et~al}\mbox{.}(1999){Moore}, {Ghigna}, {Governato}, {Lake},
  {Quinn}, {Stadel}, \& {Tozzi}}]{Moore1999}
{Moore} B., {Ghigna} S., {Governato} F., {Lake} G., {Quinn} T., {Stadel} J.,
  {Tozzi} P., 1999, \apjl, 524, L19

\bibitem[{{Moster} {et~al}\mbox{.}(2013){Moster}, {Naab}, \&
  {White}}]{Moster2013}
{Moster} B.~P., {Naab} T., {White} S.~D.~M., 2013, \mnras, 428, 3121

\bibitem[{{Navarro} {et~al}\mbox{.}(1996){Navarro}, {Frenk}, \&
  {White}}]{Navarro1996}
{Navarro} J.~F., {Frenk} C.~S., {White} S.~D.~M., 1996, \apj, 462, 563

\bibitem[{{Navarro} {et~al}\mbox{.}(1997){Navarro}, {Frenk}, \&
  {White}}]{Navarro1997}
{Navarro} J.~F., {Frenk} C.~S., {White} S.~D.~M., 1997, \apj, 490, 493

\bibitem[{{Nelson} {et~al}\mbox{.}(2015){Nelson}, {Pillepich}, {Genel},
  {Vogelsberger}, {Springel}, {Torrey}, {Rodriguez-Gomez}, {Sijacki}, {Snyder},
  {Griffen}, {Marinacci}, {Blecha}, {Sales}, {Xu}, \& {Hernquist}}]{Nelson2015}
{Nelson} D. {et~al.}, 2015, ArXiv e-prints

\bibitem[{{O{\~n}orbe} {et~al}\mbox{.}(2015){O{\~n}orbe}, {Boylan-Kolchin},
  {Bullock}, {Hopkins}, {Ker{\v e}s}, {Faucher-Gigu{\`e}re}, {Quataert}, \&
  {Murray}}]{Onorbe2015}
{O{\~n}orbe} J., {Boylan-Kolchin} M., {Bullock} J.~S., {Hopkins} P.~F., {Ker{\v
  e}s} D., {Faucher-Gigu{\`e}re} C.-A., {Quataert} E., {Murray} N., 2015, ArXiv
  e-prints

\bibitem[{{Ogiya} \& {Burkert}(2015)}]{Ogiya2015}
{Ogiya} G., {Burkert} A., 2015, \mnras, 446, 2363

\bibitem[{{Okamoto}(2013)}]{Okamoto2013}
{Okamoto} T., 2013, \mnras, 428, 718

\bibitem[{{Okamoto} \& {Frenk}(2009)}]{Okamoto2009}
{Okamoto} T., {Frenk} C.~S., 2009, \mnras, 399, L174

\bibitem[{{Okamoto} {et~al}\mbox{.}(2010){Okamoto}, {Frenk}, {Jenkins}, \&
  {Theuns}}]{Okamoto2010}
{Okamoto} T., {Frenk} C.~S., {Jenkins} A., {Theuns} T., 2010, \mnras, 406, 208

\bibitem[{{Okamoto} {et~al}\mbox{.}(2008){Okamoto}, {Gao}, \&
  {Theuns}}]{Okamoto2008}
{Okamoto} T., {Gao} L., {Theuns} T., 2008, \mnras, 390, 920

\bibitem[{{Onions} {et~al}\mbox{.}(2012){Onions}, {Knebe}, {Pearce}, {Muldrew},
  {Lux}, {Knollmann}, {Ascasibar}, {Behroozi}, {Elahi}, {Han}, {Maciejewski},
  {Merch{\'a}n}, {Neyrinck}, {Ruiz}, {Sgr{\'o}}, {Springel}, \&
  {Tweed}}]{Onions2012}
{Onions} J. {et~al.}, 2012, \mnras, 423, 1200

\bibitem[{{Papastergis} {et~al}\mbox{.}(2015){Papastergis}, {Giovanelli},
  {Haynes}, \& {Shankar}}]{Papastergis2015}
{Papastergis} E., {Giovanelli} R., {Haynes} M.~P., {Shankar} F., 2015, \aap,
  574, A113

\bibitem[{{Pe{\~n}arrubia} {et~al}\mbox{.}(2010){Pe{\~n}arrubia}, {Benson},
  {Walker}, {Gilmore}, {McConnachie}, \& {Mayer}}]{Penarrubia2010}
{Pe{\~n}arrubia} J., {Benson} A.~J., {Walker} M.~G., {Gilmore} G.,
  {McConnachie} A.~W., {Mayer} L., 2010, \mnras, 406, 1290

\bibitem[{{Pe{\~n}arrubia} {et~al}\mbox{.}(2008{\natexlab{a}}){Pe{\~n}arrubia},
  {McConnachie}, \& {Navarro}}]{Penarrubia2008}
{Pe{\~n}arrubia} J., {McConnachie} A.~W., {Navarro} J.~F., 2008{\natexlab{a}},
  \apj, 672, 904

\bibitem[{{Pe{\~n}arrubia} {et~al}\mbox{.}(2008{\natexlab{b}}){Pe{\~n}arrubia},
  {Navarro}, \& {McConnachie}}]{Penarrubia2008b}
{Pe{\~n}arrubia} J., {Navarro} J.~F., {McConnachie} A.~W., 2008{\natexlab{b}},
  \apj, 673, 226

\bibitem[{{Perlmutter} {et~al}\mbox{.}(1999){Perlmutter}, {Aldering},
  {Goldhaber}, {Knop}, {Nugent}, {Castro}, {Deustua}, {Fabbro}, {Goobar},
  {Groom}, {Hook}, {Kim}, {Kim}, {Lee}, {Nunes}, {Pain}, {Pennypacker},
  {Quimby}, {Lidman}, {Ellis}, {Irwin}, {McMahon}, {Ruiz-Lapuente}, {Walton},
  {Schaefer}, {Boyle}, {Filippenko}, {Matheson}, {Fruchter}, {Panagia},
  {Newberg}, {Couch}, \& {Project}}]{Perlmutter1999}
{Perlmutter} S. {et~al.}, 1999, \apj, 517, 565

\bibitem[{{Pillepich} {et~al}\mbox{.}(2014){Pillepich}, {Kuhlen}, {Guedes}, \&
  {Madau}}]{Pillepich2014}
{Pillepich} A., {Kuhlen} M., {Guedes} J., {Madau} P., 2014, \apj, 784, 161

\bibitem[{{Planck Collaboration} {et~al}\mbox{.}(2015){Planck Collaboration},
  {Ade}, {Aghanim}, {Arnaud}, {Ashdown}, {Aumont}, {Baccigalupi}, {Banday},
  {Barreiro}, {Bartlett}, \& et~al.}]{Planck2015}
{Planck Collaboration} {et~al.}, 2015, ArXiv e-prints

\bibitem[{{Polisensky} \& {Ricotti}(2014)}]{Polisensky2014}
{Polisensky} E., {Ricotti} M., 2014, \mnras, 437, 2922

\bibitem[{{Pontzen} \& {Governato}(2012)}]{Pontzen2012}
{Pontzen} A., {Governato} F., 2012, \mnras, 421, 3464

\bibitem[{{Pontzen} \& {Governato}(2014)}]{Pontzen2014}
{Pontzen} A., {Governato} F., 2014, \nat, 506, 171

\bibitem[{{Power} {et~al}\mbox{.}(2003){Power}, {Navarro}, {Jenkins}, {Frenk},
  {White}, {Springel}, {Stadel}, \& {Quinn}}]{Power2003}
{Power} C., {Navarro} J.~F., {Jenkins} A., {Frenk} C.~S., {White} S.~D.~M.,
  {Springel} V., {Stadel} J., {Quinn} T., 2003, \mnras, 338, 14

\bibitem[{{Puchwein} \& {Springel}(2013)}]{Puchwein2013}
{Puchwein} E., {Springel} V., 2013, \mnras, 428, 2966

\bibitem[{{Pujol} {et~al}\mbox{.}(2014){Pujol}, {Gazta{\~n}aga}, {Giocoli},
  {Knebe}, {Pearce}, {Skibba}, {Ascasibar}, {Behroozi}, {Elahi}, {Han}, {Lux},
  {Muldrew}, {Neyrinck}, {Onions}, {Potter}, \& {Tweed}}]{Pujol2014}
{Pujol} A. {et~al.}, 2014, \mnras, 438, 3205

\bibitem[{{Read} \& {Hayfield}(2012)}]{Read2012}
{Read} J.~I., {Hayfield} T., 2012, \mnras, 422, 3037

\bibitem[{{Ricotti}(2009)}]{Ricotti2009}
{Ricotti} M., 2009, \mnras, 392, L45

\bibitem[{{Riess} {et~al}\mbox{.}(1998){Riess}, {Filippenko}, {Challis},
  {Clocchiatti}, {Diercks}, {Garnavich}, {Gilliland}, {Hogan}, {Jha},
  {Kirshner}, {Leibundgut}, {Phillips}, {Reiss}, {Schmidt}, {Schommer},
  {Smith}, {Spyromilio}, {Stubbs}, {Suntzeff}, \& {Tonry}}]{Riess1998}
{Riess} A.~G. {et~al.}, 1998, \aj, 116, 1009

\bibitem[{{Romano-D{\'{\i}}az} {et~al}\mbox{.}(2010){Romano-D{\'{\i}}az},
  {Shlosman}, {Heller}, \& {Hoffman}}]{Romano2010}
{Romano-D{\'{\i}}az} E., {Shlosman} I., {Heller} C., {Hoffman} Y., 2010, \apj,
  716, 1095

\bibitem[{{Rubin} {et~al}\mbox{.}(1980){Rubin}, {Ford}, \&
  {.~Thonnard}}]{Rubin1980}
{Rubin} V.~C., {Ford} W.~K.~J., {.~Thonnard} N., 1980, \apj, 238, 471

\bibitem[{{Ryden} \& {Gunn}(1987)}]{Ryden1987}
{Ryden} B.~S., {Gunn} J.~E., 1987, \apj, 318, 15

\bibitem[{{Sawala} {et~al}\mbox{.}(2013){Sawala}, {Frenk}, {Crain}, {Jenkins},
  {Schaye}, {Theuns}, \& {Zavala}}]{Sawala2013}
{Sawala} T., {Frenk} C.~S., {Crain} R.~A., {Jenkins} A., {Schaye} J., {Theuns}
  T., {Zavala} J., 2013, \mnras, 431, 1366

\bibitem[{{Sawala} {et~al}\mbox{.}(2014{\natexlab{a}}){Sawala}, {Frenk},
  {Fattahi}, {Navarro}, {Bower}, {Crain}, {Dalla Vecchia}, {Furlong}, {Helly},
  {Jenkins}, {Oman}, {Schaller}, {Schaye}, {Theuns}, {Trayford}, \&
  {White}}]{Sawala2014a}
{Sawala} T. {et~al.}, 2014{\natexlab{a}}, ArXiv e-prints

\bibitem[{{Sawala} {et~al}\mbox{.}(2014{\natexlab{b}}){Sawala}, {Frenk},
  {Fattahi}, {Navarro}, {Theuns}, {Bower}, {Crain}, {Furlong}, {Jenkins},
  {Schaller}, \& {Schaye}}]{Sawala2014b}
{Sawala} T. {et~al.}, 2014{\natexlab{b}}, ArXiv e-prints

\bibitem[{{Sawala} {et~al}\mbox{.}(2012){Sawala}, {Scannapieco}, \&
  {White}}]{Sawala2012}
{Sawala} T., {Scannapieco} C., {White} S., 2012, \mnras, 420, 1714

\bibitem[{{Scannapieco} {et~al}\mbox{.}(2012){Scannapieco}, {Wadepuhl},
  {Parry}, {Navarro}, {Jenkins}, {Springel}, {Teyssier}, {Carlson}, {Couchman},
  {Crain}, {Dalla Vecchia}, {Frenk}, {Kobayashi}, {Monaco}, {Murante},
  {Okamoto}, {Quinn}, {Schaye}, {Stinson}, {Theuns}, {Wadsley}, {White}, \&
  {Woods}}]{Scannapieco2012}
{Scannapieco} C. {et~al.}, 2012, \mnras, 423, 1726

\bibitem[{{Schaller} {et~al}\mbox{.}(2015){Schaller}, {Frenk}, {Bower},
  {Theuns}, {Jenkins}, {Schaye}, {Crain}, {Furlong}, {Dalla Vecchia}, \&
  {McCarthy}}]{Schaller2015}
{Schaller} M. {et~al.}, 2015, \mnras, 451, 1247

\bibitem[{{Schaye} {et~al}\mbox{.}(2015){Schaye}, {Crain}, {Bower}, {Furlong},
  {Schaller}, {Theuns}, {Dalla Vecchia}, {Frenk}, {McCarthy}, {Helly},
  {Jenkins}, {Rosas-Guevara}, {White}, {Baes}, {Booth}, {Camps}, {Navarro},
  {Qu}, {Rahmati}, {Sawala}, {Thomas}, \& {Trayford}}]{Schaye2015}
{Schaye} J. {et~al.}, 2015, \mnras, 446, 521

\bibitem[{{Schneider} {et~al}\mbox{.}(2014){Schneider}, {Anderhalden},
  {Macci{\`o}}, \& {Diemand}}]{Schneider2014}
{Schneider} A., {Anderhalden} D., {Macci{\`o}} A.~V., {Diemand} J., 2014,
  \mnras, 441, L6

\bibitem[{{Sijacki} {et~al}\mbox{.}(2007){Sijacki}, {Springel}, {Di Matteo}, \&
  {Hernquist}}]{Sijacki2007}
{Sijacki} D., {Springel} V., {Di Matteo} T., {Hernquist} L., 2007, \mnras, 380,
  877

\bibitem[{{Sijacki} {et~al}\mbox{.}(2012){Sijacki}, {Vogelsberger}, {Kere{\v
  s}}, {Springel}, \& {Hernquist}}]{Sijacki2012}
{Sijacki} D., {Vogelsberger} M., {Kere{\v s}} D., {Springel} V., {Hernquist}
  L., 2012, \mnras, 424, 2999

\bibitem[{{Somerville} \& {Dav{\'e}}(2014)}]{Somerville2015}
{Somerville} R.~S., {Dav{\'e}} R., 2014, ArXiv e-prints

\bibitem[{{Spergel} {et~al}\mbox{.}(2007){Spergel}, {Bean}, {Dor{\'e}},
  {Nolta}, {Bennett}, {Dunkley}, {Hinshaw}, {Jarosik}, {Komatsu}, {Page},
  {Peiris}, {Verde}, {Halpern}, {Hill}, {Kogut}, {Limon}, {Meyer}, {Odegard},
  {Tucker}, {Weiland}, {Wollack}, \& {Wright}}]{Spergel2007}
{Spergel} D.~N. {et~al.}, 2007, \apjs, 170, 377

\bibitem[{{Springel}(2010)}]{Springel2010}
{Springel} V., 2010, \mnras, 401, 791

\bibitem[{{Springel} {et~al}\mbox{.}(2005{\natexlab{a}}){Springel}, {Di
  Matteo}, \& {Hernquist}}]{Springel2005a}
{Springel} V., {Di Matteo} T., {Hernquist} L., 2005{\natexlab{a}}, \mnras, 361,
  776

\bibitem[{{Springel} {et~al}\mbox{.}(2006){Springel}, {Frenk}, \&
  {White}}]{Springel2006}
{Springel} V., {Frenk} C.~S., {White} S.~D.~M., 2006, \nat, 440, 1137

\bibitem[{{Springel} \& {Hernquist}(2003)}]{Springel2003}
{Springel} V., {Hernquist} L., 2003, \mnras, 339, 289

\bibitem[{{Springel} {et~al}\mbox{.}(2008){Springel}, {Wang}, {Vogelsberger},
  {Ludlow}, {Jenkins}, {Helmi}, {Navarro}, {Frenk}, \& {White}}]{Springel2008}
{Springel} V. {et~al.}, 2008, \mnras, 391, 1685

\bibitem[{{Springel} {et~al}\mbox{.}(2004){Springel}, {White}, \&
  {Hernquist}}]{Springel2004}
{Springel} V., {White} S.~D.~M., {Hernquist} L., 2004, in IAU Symposium, Vol.
  220, Dark Matter in Galaxies, {Ryder} S., {Pisano} D., {Walker} M., {Freeman}
  K., eds., p. 421

\bibitem[{{Springel} {et~al}\mbox{.}(2005{\natexlab{b}}){Springel}, {White},
  {Jenkins}, {Frenk}, {Yoshida}, {Gao}, {Navarro}, {Thacker}, {Croton},
  {Helly}, {Peacock}, {Cole}, {Thomas}, {Couchman}, {Evrard}, {Colberg}, \&
  {Pearce}}]{Springel2005}
{Springel} V. {et~al.}, 2005{\natexlab{b}}, \nat, 435, 629

\bibitem[{{Springel} {et~al}\mbox{.}(2001){Springel}, {White}, {Tormen}, \&
  {Kauffmann}}]{Springel2001}
{Springel} V., {White} S.~D.~M., {Tormen} G., {Kauffmann} G., 2001, \mnras,
  328, 726

\bibitem[{{Stinson} {et~al}\mbox{.}(2013){Stinson}, {Brook}, {Macci{\`o}},
  {Wadsley}, {Quinn}, \& {Couchman}}]{Stinson2013}
{Stinson} G.~S., {Brook} C., {Macci{\`o}} A.~V., {Wadsley} J., {Quinn} T.~R.,
  {Couchman} H.~M.~P., 2013, \mnras, 428, 129

\bibitem[{{Strigari} {et~al}\mbox{.}(2010){Strigari}, {Frenk}, \&
  {White}}]{Strigari2010}
{Strigari} L.~E., {Frenk} C.~S., {White} S.~D.~M., 2010, \mnras, 408, 2364

\bibitem[{{Tegmark} {et~al}\mbox{.}(2006){Tegmark}, {Eisenstein}, {Strauss},
  {Weinberg}, {Blanton}, {Frieman}, {Fukugita}, {Gunn}, {Hamilton}, {Knapp},
  {Nichol}, {Ostriker}, {Padmanabhan}, {Percival}, {Schlegel}, {Schneider},
  {Scoccimarro}, {Seljak}, {Seo}, {Swanson}, {Szalay}, {Vogeley}, {Yoo},
  {Zehavi}, {Abazajian}, {Anderson}, {Annis}, {Bahcall}, {Bassett}, {Berlind},
  {Brinkmann}, {Budavari}, {Castander}, {Connolly}, {Csabai}, {Doi},
  {Finkbeiner}, {Gillespie}, {Glazebrook}, {Hennessy}, {Hogg}, {Ivezi{\'c}},
  {Jain}, {Johnston}, {Kent}, {Lamb}, {Lee}, {Lin}, {Loveday}, {Lupton},
  {Munn}, {Pan}, {Park}, {Peoples}, {Pier}, {Pope}, {Richmond}, {Rockosi},
  {Scranton}, {Sheth}, {Stebbins}, {Stoughton}, {Szapudi}, {Tucker}, {vanden
  Berk}, {Yanny}, \& {York}}]{Tegmark2006}
{Tegmark} M. {et~al.}, 2006, \prd, 74, 123507

\bibitem[{{Teyssier} {et~al}\mbox{.}(2013){Teyssier}, {Pontzen}, {Dubois}, \&
  {Read}}]{Teyssier2013}
{Teyssier} R., {Pontzen} A., {Dubois} Y., {Read} J.~I., 2013, \mnras, 429, 3068

\bibitem[{{The Fermi-LAT Collaboration} {et~al}\mbox{.}(2013){The Fermi-LAT
  Collaboration}, {:}, {Ackermann}, {Albert}, {Anderson}, {Baldini}, {Ballet},
  {Barbiellini}, {Bastieri}, {Bechtol}, {Bellazzini}, {Bissaldi}, {Bloom},
  {Bonamente}, {Bouvier}, {Brandt}, {Bregeon}, {Brigida}, {Bruel}, {Buehler},
  {Buson}, {Caliandro}, {Cameron}, {Caragiulo}, {Caraveo}, {Cecchi}, {Charles},
  {Chekhtman}, {Chiang}, {Ciprini}, {Claus}, {Cohen-Tanugi}, {Conrad},
  {D'Ammando}, {de Angelis}, {Dermer}, {Digel}, {Silva}, {Drell},
  {Drlica-Wagner}, {Essig}, {Favuzzi}, {Ferrara}, {Franckowiak}, {Fukazawa},
  {Funk}, {Fusco}, {Gargano}, {Gasparrini}, {Giglietto}, {Giroletti},
  {Godfrey}, {Gomez-Vargas}, {Grenier}, {Guiriec}, {Gustafsson}, {Hayashida},
  {Hays}, {Hewitt}, {Hughes}, {Jogler}, {Kamae}, {Kn{\"o}dlseder}, {Kocevski},
  {Kuss}, {Larsson}, {Latronico}, {Llena Garde}, {Longo}, {Loparco},
  {Lovellette}, {Lubrano}, {Martinez}, {Mayer}, {Mazziotta}, {Michelson},
  {Mitthumsiri}, {Mizuno}, {Moiseev}, {Monzani}, {Morselli}, {Moskalenko},
  {Murgia}, {Nemmen}, {Nuss}, {Ohsugi}, {Orlando}, {Ormes}, {Perkins}, {Piron},
  {Pivato}, {Porter}, {Rain{\`o}}, {Rando}, {Razzano}, {Razzaque}, {Reimer},
  {Reimer}, {Ritz}, {S{\`a}nchez-Conde}, {Sehgal}, {Sgr{\`o}}, {Siskind},
  {Spinelli}, {Strigari}, {Suson}, {Tajima}, {Takahashi}, {Thayer}, {Tibaldo},
  {Tinivella}, {Torres}, {Uchiyama}, {Usher}, {Vandenbroucke}, {Vianello},
  {Vitale}, {Werner}, {Winer}, {Wood}, {Wood}, {Zaharijas}, \&
  {Zimmer}}]{fermidwarf2013}
{The Fermi-LAT Collaboration} {et~al.}, 2013, ArXiv e-prints

\bibitem[{{Tollerud} {et~al}\mbox{.}(2014){Tollerud}, {Boylan-Kolchin}, \&
  {Bullock}}]{Tollerud2014}
{Tollerud} E.~J., {Boylan-Kolchin} M., {Bullock} J.~S., 2014, \mnras, 440, 3511

\bibitem[{{Torrey} {et~al}\mbox{.}(2012){Torrey}, {Vogelsberger}, {Sijacki},
  {Springel}, \& {Hernquist}}]{Torrey2012}
{Torrey} P., {Vogelsberger} M., {Sijacki} D., {Springel} V., {Hernquist} L.,
  2012, \mnras, 427, 2224

\bibitem[{{Velliscig} {et~al}\mbox{.}(2014){Velliscig}, {van Daalen}, {Schaye},
  {McCarthy}, {Cacciato}, {Le Brun}, \& {Dalla Vecchia}}]{Velliscig2014}
{Velliscig} M., {van Daalen} M.~P., {Schaye} J., {McCarthy} I.~G., {Cacciato}
  M., {Le Brun} A.~M.~C., {Dalla Vecchia} C., 2014, \mnras, 442, 2641

\bibitem[{{Vera-Ciro} {et~al}\mbox{.}(2013){Vera-Ciro}, {Helmi}, {Starkenburg},
  \& {Breddels}}]{VeraCiro2013}
{Vera-Ciro} C.~A., {Helmi} A., {Starkenburg} E., {Breddels} M.~A., 2013,
  \mnras, 428, 1696

\bibitem[{{Vera-Ciro} {et~al}\mbox{.}(2011){Vera-Ciro}, {Sales}, {Helmi},
  {Frenk}, {Navarro}, {Springel}, {Vogelsberger}, \& {White}}]{Vera-Ciro2011}
{Vera-Ciro} C.~A., {Sales} L.~V., {Helmi} A., {Frenk} C.~S., {Navarro} J.~F.,
  {Springel} V., {Vogelsberger} M., {White} S.~D.~M., 2011, \mnras, 416, 1377

\bibitem[{{Vikhlinin} {et~al}\mbox{.}(2006){Vikhlinin}, {Kravtsov}, {Forman},
  {Jones}, {Markevitch}, {Murray}, \& {Van Speybroeck}}]{Vikhlinin2006}
{Vikhlinin} A., {Kravtsov} A., {Forman} W., {Jones} C., {Markevitch} M.,
  {Murray} S.~S., {Van Speybroeck} L., 2006, \apj, 640, 691

\bibitem[{{Vogelsberger} {et~al}\mbox{.}(2013){Vogelsberger}, {Genel},
  {Sijacki}, {Torrey}, {Springel}, \& {Hernquist}}]{Vogelsberger2013}
{Vogelsberger} M., {Genel} S., {Sijacki} D., {Torrey} P., {Springel} V.,
  {Hernquist} L., 2013, \mnras, 436, 3031

\bibitem[{{Vogelsberger} {et~al}\mbox{.}(2014{\natexlab{a}}){Vogelsberger},
  {Genel}, {Springel}, {Torrey}, {Sijacki}, {Xu}, {Snyder}, {Bird}, {Nelson},
  \& {Hernquist}}]{Vogelsberger2014a}
{Vogelsberger} M. {et~al.}, 2014{\natexlab{a}}, \nat, 509, 177

\bibitem[{{Vogelsberger} {et~al}\mbox{.}(2014{\natexlab{b}}){Vogelsberger},
  {Genel}, {Springel}, {Torrey}, {Sijacki}, {Xu}, {Snyder}, {Nelson}, \&
  {Hernquist}}]{Vogelsberger2014b}
{Vogelsberger} M. {et~al.}, 2014{\natexlab{b}}, \mnras, 444, 1518

\bibitem[{{Vogelsberger} {et~al}\mbox{.}(2012){Vogelsberger}, {Zavala}, \&
  {Loeb}}]{Vogelsberger2012}
{Vogelsberger} M., {Zavala} J., {Loeb} A., 2012, \mnras, 423, 3740

\bibitem[{{Vogelsberger} {et~al}\mbox{.}(2014{\natexlab{c}}){Vogelsberger},
  {Zavala}, {Simpson}, \& {Jenkins}}]{Vogelsberger2014c}
{Vogelsberger} M., {Zavala} J., {Simpson} C., {Jenkins} A., 2014{\natexlab{c}},
  \mnras, 444, 3684

\bibitem[{{Wadepuhl} \& {Springel}(2011)}]{Wadepuhl2011}
{Wadepuhl} M., {Springel} V., 2011, \mnras, 410, 1975

\bibitem[{{Wang} {et~al}\mbox{.}(2012){Wang}, {Frenk}, {Navarro}, {Gao}, \&
  {Sawala}}]{Wang2012}
{Wang} J., {Frenk} C.~S., {Navarro} J.~F., {Gao} L., {Sawala} T., 2012, \mnras,
  424, 2715

\bibitem[{{Weisz} {et~al}\mbox{.}(2015){Weisz}, {Dolphin}, {Skillman},
  {Holtzman}, {Gilbert}, {Dalcanton}, \& {Williams}}]{Weisz2015}
{Weisz} D.~R., {Dolphin} A.~E., {Skillman} E.~D., {Holtzman} J., {Gilbert}
  K.~M., {Dalcanton} J.~J., {Williams} B.~F., 2015, \apj, 804, 136

\bibitem[{{Xu} {et~al}\mbox{.}(2015){Xu}, {Sluse}, {Gao}, {Wang}, {Frenk},
  {Mao}, {Schneider}, \& {Springel}}]{Xu2015}
{Xu} D., {Sluse} D., {Gao} L., {Wang} J., {Frenk} C., {Mao} S., {Schneider} P.,
  {Springel} V., 2015, \mnras, 447, 3189

\bibitem[{{Xu} {et~al}\mbox{.}(2010){Xu}, {Mao}, {Cooper}, {Wang}, {Gao},
  {Frenk}, \& {Springel}}]{Xu2010}
{Xu} D.~D., {Mao} S., {Cooper} A.~P., {Wang} J., {Gao} L., {Frenk} C.~S.,
  {Springel} V., 2010, \mnras, 408, 1721

\bibitem[{{Young}(1980)}]{Young1980}
{Young} P., 1980, \apj, 242, 1232

\bibitem[{{Yurin} \& {Springel}(2015)}]{Yurin2015}
{Yurin} D., {Springel} V., 2015, \mnras, 452, 2343

\bibitem[{{Zemp} {et~al}\mbox{.}(2011){Zemp}, {Gnedin}, {Gnedin}, \&
  {Kravtsov}}]{Zemp2011}
{Zemp} M., {Gnedin} O.~Y., {Gnedin} N.~Y., {Kravtsov} A.~V., 2011, \apjs, 197,
  30

\bibitem[{{Zemp} {et~al}\mbox{.}(2012){Zemp}, {Gnedin}, {Gnedin}, \&
  {Kravtsov}}]{Zemp2012}
{Zemp} M., {Gnedin} O.~Y., {Gnedin} N.~Y., {Kravtsov} A.~V., 2012, \apj, 748,
  54

\bibitem[{{Zhu} {et~al}\mbox{.}(2015){Zhu}, {Hernquist}, \& {Li}}]{Zhu2015}
{Zhu} Q., {Hernquist} L., {Li} Y., 2015, \apj, 800, 6

\bibitem[{{Zolotov} {et~al}\mbox{.}(2012){Zolotov}, {Brooks}, {Willman},
  {Governato}, {Pontzen}, {Christensen}, {Dekel}, {Quinn}, {Shen}, \&
  {Wadsley}}]{Zolotov2012}
{Zolotov} A. {et~al.}, 2012, \apj, 761, 71

\bibitem[{{Zwicky}(1937)}]{Zwicky1937}
{Zwicky} F., 1937, \apj, 86, 217

\end{thebibliography}
